\PassOptionsToPackage{numbers,sort&compress}{natbib}
\documentclass[onecolumn,numbers]{els-mrw} 
\usepackage{amsmath,amssymb,amsfonts,amsthm,makeidx,graphicx}
\usepackage{txfonts}
\usepackage{helvet}
\usepackage{tabularx}
\usepackage{array}
\usepackage{booktabs}
\usepackage{graphicx}
\usepackage[colorlinks=true,
            linkcolor=blue,
            citecolor=blue,
            urlcolor=blue]{hyperref}

\begin{document}

%\begin{frontmatter}

\chapter{Precision Muon Physics}\label{}    
%\title{Precision Muon Physics}

\author[1]{Tim Gorringe}

\address[1]{Department of Physics and Astronomy, University of Kentucky, Lexington, KY, 40506, USA}

%\articletag{Chapter Article tagline: update of previous edition,, reprint..}

\maketitle

\begin{abstract}

The muon has become a versatile tool for fundamental physics. Measurements of muon
decay determine the Fermi constant and establish the chiral structure of weak interactions.
Measurements with stored muons and bound muons enable precision determinations of the muon charge, mass, and magnetic moment.
These studies also provide far-reaching tests of bound-state QED and the
Standard Model, as well as unique sensitivity to physics beyond our current understanding of the particles, forces, and laws of nature.

We survey the landscape of precision muon physics, emphasizing both the physics reach and the experimental breadth. We highlight the range of techniques,
from charged-particle timing, tracking, and polarimetry to storage-ring experiments, microwave resonance methods, and Doppler-free laser spectroscopy. We also describe the interplay between experimental work and theoretical developments in quantum electrodynamics, quantum chromodynamics, and effective field theory.
\end{abstract}

 \begin{keywords}

muon, muonium, precision measurements, fundamental constants, quantum electrodynamics, weak interactions, Standard Model

\end{keywords}

\section*{Key points}
\begin{itemize}

\item
Muons provide unique opportunities for measurements of fundamental constants, tests of theories describing the elementary interactions, and searches for new 
particles and forces.

\item
Precision measurements of muon decay observables determine the Fermi constant $G_F$ and establish the chiral structure of the weak interaction.

\item 
Muonium spectroscopy enables precision determinations of muon properties --- charge, mass and magnetic moment --- and stringent tests of bound-state quantum electrodynamics.

\item
Storage ring experiments enable a precision determination of the muon anomalous magnetic moment and a stringent test of the Standard Model.

\item 
Muon experiments have played an important role in  the development and the validation of the Standard Model and driven advances in experimental techniques and theoretical methods.
\end{itemize}

%\tableofcontents
%\input{Introduction_draft1.tex} 
%\input{history_draft1}
%\input{MuonDecay_draft1}  
%\input{MuonMichel_draft1}
%\input{Muonium_draft1} 
%\input{Muon-g2-draft1}

\section{Introduction}

Nearly a century has passed since the muon was identified as the penetrating component of the Earth's cosmic radiation. 
At the time, the electron, proton, and neutron
were widely considered the fundamental constituents of matter.
Atomic binding was understood as long-range electromagnetic interactions mediated by photon exchange, 
while nuclear binding was attributed to short-range strong interactions mediated by heavy meson exchange.
With a mass of about 200~$m_e$, the muon fell in the expected mass  range of the heavy meson. 

However, other properties of muons were found to be incompatible with this identity.
Most strikingly, muons
interacted with nuclei
far more weakly than expected for a mediator of the strong interaction. 
Over time, the muon was recognized as a point-like, spin-1/2 fermion with electron-like electromagnetic and weak
interactions.
Fifty years after the electron’s discovery, a heavier counterpart was unexpectedly found.

In retrospect, the discovery of the muon 
was the first indication  of the generational structure of the Standard Model (SM).
The first-generation particles constitute ordinary matter,
while  the second- and third-generation particles are short-lived, heavier relatives.
Within the SM framework, the three generations possess identical gauge interactions,
but acquire different masses through their different couplings to the Higgs field.

Despite the
apparent redundancy of SM generations,
the muon has become
an important tool in fundamental 
physics.
Its large mass compared with electrons enhances sensitivity to short-distance physics, while its small mass compared with hadrons permits clean theoretical interpretation of its interactions. Its long lifetime permits the transport, stopping, and storage of muons, as well as formation of hydrogen-like atoms. The parity-violating nature of the $\pi \rightarrow \mu \rightarrow e$ decay sequence
provides a means to polarize and analyze the muon spin.

This chapter describes the achievements of precision muon physics.
Measurements of the muon lifetime and its decay observables determine the strength and  structure of the weak interaction. Experiments with stored muons and bound muons 
enable measurements of the muon charge, mass, and  magnetic moment, and provide rigorous tests of both QED and the Standard Model. Muon experiments have also motivated theoretical advances in quantum electrodynamics,  quantum chromodynamics, and effective field theory.

The chapter also describes the diverse  methods and innovative technologies that underpin muon experiments. These range from state-of-the-art timing and tracking detectors to storage rings, microwave techniques, and laser spectroscopy.
Experiments are carried
out at facilities worldwide,
including the TRIUMF Cyclotron Facility in Canada, the J-PARC Muon Science Facility in Japan, the Paul Scherrer Institute (PSI) in Switzerland, and
Fermilab in the United States. 

For further reading we suggest a number of recent reviews covering muon g-2 \citep{KeshavarziKhawYoshioka2022, HertzogHoferichter2026}, muonium spectroscopy \citep{OhayonBurkleyCrivelli2021,ChenSunTang2026}, and broader precision muon physics topics \citep{EchenardPetrov2026,
GorringeHertzog2015}.

\subsection{History of muon experiments}
\label{sec:muonrole}

Contemporary muon experiments are built upon earlier experimental discoveries. They revealed the muon’s instability, established the $\mu \rightarrow e \nu \bar{\nu}$ decay mode, and demonstrated the handedness and universality of weak interactions. They also identified the formation of muonium ($\mu^+ e^-$) atoms 
by positive muons and muonic ($\mu^-Z$) atoms 
by negative muons. In this section we summarize the experimental foundations of precision muon physics (see Table \ref{tab:muon_milestones}).

The discovery of the muon as a new particle is usually credited to %Anderson and Neddermeyer 
\citet{Anderson:1936zz} 
and %Street and Stevenson 
\citet{Street:1937me}.\footnote{They observed both positive muons and negative muons and posited that muons 
carry a  single unit of electronic charge. Their efforts concluded many years of work on understanding the nature of  penetrating particles in cosmic rays.}  Both experiments used cloud chambers to image the trajectories of penetrating charged particles in sea-level cosmic ray showers. The discrimination of muons from lighter electrons and heavier protons was accomplished using absorbing materials, magnetic fields and techniques based on ionizing power, energy loss, and magnetic rigidity.

The first evidence for the decay of the muon was a single cloud chamber photograph recorded by %Williams and Roberts in 1940 
\citet{Williams:1940}. The photograph showed the decay of a muon into an electron on a microsecond timescale.
The authors described their event as a transformation rather than decay and their result as intriguing but tentative. 

Soon, an indirect determination 
of the muon lifetime, $ 2.4 \pm 0.3 $~$\mu$s \citep{Rossi:1941zz}, 
was inferred from the anomalous attenuation of the cosmic ray muon flux in the Earth's atmosphere.
Subsequently, a direct determination of the muon lifetime, $ 1.5 \pm 0.3 $~$\mu$s \citep{Rasetti:1941ie}, was obtained by measurement of time intervals between stopped muons and decay electrons using cosmic rays.
This direct measurement was clear evidence that muons decay to electrons. 

When stopped in matter, negative muons were expected to form muonic  atoms and undergo nuclear capture. Such nuclear capture would be accompanied by nuclear disintegration and a large energy release. %In 1947,  Conversi {\textit et al.}\ 
\citet{PhysRev.71.209} used a magnetic lens to separate positive and negative muons and measure their lifetimes.
They found that negative muons do undergo nuclear capture,
but with a rate far lower than expected for a force carrier 
of the strong interaction.

In 1947, the pion was discovered in high-altitude cosmic ray studies.
%Powell {\textit  et al.}\ 
\citet{Lattes:1947my} 
identified pions through photographic images that showed the pion decay into a muon, followed by muon decay into an electron, the $\pi \rightarrow \mu \rightarrow e$ decay chain.
Through the observation of `star' events corresponding to nuclear disintegrations, they further concluded 
that the pion was the mediator of the strong interaction.

%In 1949, Leighton {\textit et al.}\ 
\citet{PhysRev.75.1432} measured the electron energy spectrum from muon decay with sufficient statistics to clearly establish a continuous energy distribution and definite  energy endpoint, $m_{\mu} / 2$.
They interpreted their results 
as the decay of a spin-1/2 particle into three light or massless particles such as $\mu \rightarrow e \nu \bar{\nu}$. 
The results led to the formulation of a universal weak interaction describing muon decay, muon capture and nuclear $\beta$ decay \citep{PhysRev.72.246,PhysRev.93.354}. These developments also motivated Michel's work on the possible Lorentz structure of the $\mu \rightarrow e \nu \bar{\nu}$ decay \citep{Michel1950}.

%In 1956, Lee and Yang 
\citet{Lee:1956qn} questioned whether parity was conserved in weak interaction processes and discussed several possibilities for testing its conservation.
Spanning only a few weeks in early 1957, the
violation of parity was first reported in nuclear $\beta$-decay by %Wu {\textit et al.}\ 
\citet{Wu1957},
and then confirmed in $\pi \rightarrow \mu \rightarrow e$ decay by %Garwin {\textit et al.} 
\citet{Garwin:1957hc} and %Freidman {\textit et al.}\ 
\citet{Friedman:1957mz}.
The large muon polarization observed in pion decay,
and large-decay electron asymmetry observed in muon decay, became key evidence
for the $V$-$A$ character of the weak interaction \citep{MarshakSudarshan1957,Feynman:1958ty}.

The Garwin experiment also yielded the first measurement of the muon g-factor, $g_{\mu}$.
The measurement took advantage of the muon polarization in pion decay and the electron asymmetry in muon decay. In a magnetic field, the muon polarization and resulting electron asymmetry precess at rates proportional to $g_{\mu}$.
Their result, $g_{\mu} = +2.00\pm0.10$,
was a clear verification of the muon's identity as a point-like, spin-1/2 fermion.
Within a few years, the first direct measurement of the muon anomalous magnetic moment, $a_\mu = (g_\mu-2)/2$, arising from quantum vacuum fluctuations, was performed at CERN \citep{Charpak1961}, yielding  $a_\mu = ( 1.145 \pm 0.022 ) \times 10^{-3}$.

An emerging puzzle was the non-observation of the $\mu \rightarrow e \gamma$ decay \citep{PhysRev.73.257}. 
Early weak interaction theories  with force-carrying  bosons predicted $\mu \rightarrow e \gamma$ rates that exceeded experimental limits \citep{Feinberg1958}. In such theories, the  $\mu \rightarrow e \gamma$ decay arises at one-loop order via diagrams involving virtual W bosons and neutrinos. The decay is forbidden, however, if electron and muon neutrinos have distinct flavors and lepton flavor is conserved \citep{Pontecorvo1958}.
The 1962 discovery of the muon neutrino \citep{Danby1962} was strong evidence for lepton flavor and its conservation.

It was expected that negative muons form hydrogen-like muonic atoms \citep{PhysRev.71.320,PhysRev.72.399}.
In such $\mu^-$Z atoms, the atomic radii are smaller and energy spacings are larger by a factor of $m_{\mu} / m_e  \sim 207$ relative to ordinary atoms. The direct detection of muonic x-rays from atomic transitions in muonic atoms was first achieved in the early 1950s \citep{FitchRainwater1953}. 
Such spectroscopy became a valuable tool for measurements of charged radii of protons and nuclei.

The existence of muonium, $\mu^+$$e^-$,
was less obvious. Muonium was unambiguously identified %by Hughes {\textit et al.}\  
\citep{PhysRevLett.5.63,PhysRevLett.6.19}
through detection
of its characteristic Larmor 
precession frequency in an external magnetic field.
Muonium -- a purely leptonic bound state --
became particularly important for precision measurements of fundamental constants and stringent tests of quantum electrodynamics.

\begin{table}[htbp]
\centering
\small
\begin{tabularx}{\textwidth}{p{2.1cm}|p{5.2cm}|X}
\hline
 & & \\
~~~Date & ~~~~~~~~~~~~~~~Milestone & ~~~~~~~~~~~~~~~~Significance \\
 & & \\
\hline
 & & \\
1920s & Detection of penetrating cosmic rays  &  Puzzling nature of penetrating cosmic rays \\
1936--37 & Muon identified as penetrating cosmic rays \citep{Anderson:1936zz,Street:1937me} & Discovery of new particle of mass $m_e < m < m_p$ \\
1940 & Cloud chamber image of $\mu \rightarrow e$ transformation \citep{Williams:1940} & Evidence of decay of muon into electron \\
1939-41 & Indirect / direct easurements of muon lifetime \citep{Rossi:1941zz,Rasetti:1941ie} &  $\tau_\mu \sim 2 \mu s$ lifetime incompatible with strong force mediator \\
1946 & Measurement of muon's weak nuclear interaction \citep{PhysRev.71.209} & Established absence of strong interactions of muons \\
1939-41 & Emulsion photographs of $\pi \to \mu \to e$ decay chain \citep{Lattes:1947my} & Discovery of pion and origin of cosmic ray muons   \\
1949 &  Measurement of energy spectrum of decay electrons  \citep{PhysRev.75.1432} & Established 3-body decay and evidence for $\mu \rightarrow e \nu \bar{\nu}$ hypothesis \\
1950s & Observation of muonic x-rays from high-Z materials \citep{FitchRainwater1953} & Demonstrated potential of muonic atom   for nuclear size measurements \\
1950s & Non-observation of $\mu \to e \gamma$ decay mode \citep{PhysRev.73.257} & Evidence of suppression of lepton flavor-changing interactions \\
1957 & Observation of parity violation in $\it \to \mu \to e$ decay chain \citep{Garwin:1957hc,Friedman:1957mz} & Understanding of $V$-$A$ form of weak currents \\
1957 & Measurement of muon g-factor using parity violation \citep{Garwin:1957hc} & Established $g_\mu \approx 2$, consistent with point-like, Dirac fermion \\

1960 & Detection of muonium via Lamor precession \citep{Hughes1966} & Discovery of purely leptonic  $\mu^+e^-$ bound state  \\
1961 & Direct measurement of anomalous part of muon g-factor \citep{Charpak1961} & Beginning of muon g-2 experimental program \\
1962 & Discovery of muon neutrino $\nu_\mu$ \citep{Danby1962} & Demonstration of distinct lepton families and lepton flavor conservation \\
 & & \\
\hline
\end{tabularx}
\caption{Experimental milestones in muon physics.}
\label{tab:muon_milestones}
\end{table}

\subsection{Facilities for muon experiments}
\label{ssc:beams}

Precision muon experiments depend on intense muon beams together with advanced charged particle and radiation detectors, sophisticated engineering and metrology techniques, and powerful electronics  and computing systems.
This section introduces the muon beams that are used in such experiments.

Early muon experiments relied on sea-level cosmic rays as sources of muons. 
Cosmic-ray muons are produced when energetic primary protons strike nuclei in the Earth's upper atmosphere, generating pions via hadronic interactions.
The subsequent decay of short-lived pions to longer-lived muons is the dominant source of the penetrating component of the Earth's cosmic rays.

Precision muon experiments require high-rate, high-quality, tunable muon beams. At these facilities, a high-current, $E \gtrsim 0.5$~GeV proton beam impinges on a radiation-hard, thermally robust production target. The collisions produce pions through hadronic interactions. The pions then decay into muons that are magnetically transported to the experiment.

Two broad classes of particle accelerators are used to generate muon beams.
At the Paul Scherrer Institute and the TRIUMF Meson Facility, continuous, high-intensity, 520-590~MeV proton beams are delivered by medium-energy cyclotrons.  At the Fermi National Accelerator Laboratory and the J-PARC Muon Science Facility,  time-structured, multi-GeV proton beams are provided by rapid-cycling synchrotrons. A key distinction  between these facilities is the continuous-wave operation of cyclotrons versus the time-structured operation of synchrotrons. This reflects the fixed magnetic fields and RF frequencies of proton cyclotrons and the ramped magnetic fields and RF frequencies of proton synchrotrons.

Muon beams are classified as surface, cloud, and decay-in-flight beams.
Surface muon beams are available for positive muons only and originate from at-rest pion decay at the surface of the production target. 
The decays produce fully polarized positive muons with a well-defined momentum, 29.2 MeV/c. The localized, monoenergetic source of surface $\mu^+$ beams enables the efficient delivery and precise focusing of the beam to the target.

Cloud muon beams are available for positive and negative muons and arise from in-flight pion decays in the vicinity of the production target. Cloud beams  have lower polarization and broader momentum distributions than surface beams.

Decay muon beams are also available for positive and negative muons. Decay-in-flight muon beams are produced by injecting pions of selected momentum into an extended decay region and extracting muons of chosen momentum from that region. In forward decay beamlines, muons emitted in the same direction as the parent pions are accepted, yielding higher intensities at the expense of reduced polarization and higher backgrounds. In backward decay beamlines, muons emitted opposite to the pion direction are selected, resulting in lower intensities but improved polarization and beam purity. The contrasting properties reflect the underlying kinematics of pion decay and the phase-space acceptance of the beamline.

Recent developments include ultra-low-energy $\mu^+$ and muonium beams \citep{Otani2018}.
Such beams are produced by stopping surface muons in materials that either re-emit a fraction 
of moderated $\mu^+$ into vacuum or form muonium that subsequently diffuses into vacuum.
The moderation and diffusion processes result in substantially lower beam intensities.
PSI and KEK continue to operate and further develop such beams for precision muon experiments.

\section{Muon lifetime}
\label{sec:muonlifetime}

The muon lifetime $\tau_{\mu}$ is the most precisely measured quantity in weak interaction physics. It determines the Fermi constant $G_F$ and fixes the overall scale of weak processes. 

Since muon decay is purely leptonic,
its theoretical description is essentially free of complications arising from hadron and nuclear structure.
Moreover, the 
QED corrections 
arising from radiative decays  
and radiative loops 
are finite and accurately calculable. These features underlie the exceptionally clean relation between the muon lifetime  and the Fermi constant.

The muon lifetime is also experimentally
attractive. Its microsecond timescale, 
the availability of intense muon beams,  and the capability of high-fidelity electron detection all support the precision  measurements of $\tau_\mu$.

\subsection{Role of the Fermi constant}

Fermi’s theory of weak interactions was originally introduced 
to explain $\beta$-decay using a quantum field theory analogous to quantum electrodynamics. 
Within the theory, $\beta$-decay is a local current-current interaction between a hadronic current involving the neutron and proton fields and a leptonic current involving the electron and neutrino fields. The constant $G_F$ fixes its overall interaction strength.

Over the following decades, the current-current interaction was extended to include the $\mu \rightarrow e \nu \bar{\nu}$ decay, and the weak currents were modified to incorporate their V$-$A structure. 
These developments established a precise relation between the muon lifetime and the Fermi constant. At tree level (Fig.\ \ref{GF1}),
in the $m_e \rightarrow 0$ limit,
the muon decay rate
$\Gamma_{\mu} =  { \tau_{\mu}^{-1}}$ is ({\it e.g.}, \citet{PDG2024}) 

\begin{equation}
\label{leading order tau_mu}
\Gamma_{\mu} = {{G_F^2 m_{\mu}^5} \over {192 \pi^3}} .
\end{equation}

The next advance was the calculation of first-order 
QED corrections to Eqn.\ \ref{leading order tau_mu}. These corrections arise from real photon emission in radiative decay and virtual-photon loops in non-radiative decay.
The first complete calculation of these corrections
was made by %Kinoshita and Sirlin in 1959 
\citet{KinoshitaSirlin1959}. They demonstrated 
that infrared divergences from real and virtual photons cancel in the decay rate $\Gamma_{\mu}$. They also showed that the leading-order weak interaction and first-order QED corrections factorize in 
$\Gamma_{\mu}$. The first-order 
QED correction to Eqn.\ \ref{leading order tau_mu} amounts to $0.42\%$.

A further advance was the late-1990s calculation of the complete second-order QED corrections by %van Ritbergen and Stuart 
\citet{vanRitbergenStuart1999,vanRitbergenStuart2000}. They treated the Fermi current-current interaction as a low-energy effective field theory (EFT), and computed all second-order corrections from real-photon emission and virtual-photon loops. 
They explicitly demonstrated
the second-order cancellation of infrared divergences
and factorization of the long-distance QED corrections from the short-distance weak interaction. The perturbative relation between $\Gamma_{\mu}$ and $G_F$ can then be written as
\begin{equation}
\label{higher order tau_mu}
\Gamma_{\mu}  = {{G_F^2 m_{\mu}^5} \over {192 \pi^3}} 
~ ( 1 + \Sigma_i \Delta q^{(i)} )
\end{equation}
where $\Delta q^{(0)}$ describes the small 
$m_e > 0$ phase space corrections and $\Delta q^{(i > 0)}$ the order-by-order QED corrections. The authors estimated a residual theoretical uncertainty of about 0.14~ppm in the second-order $G_F$-$\tau_\mu$ relation. The order-by-order corrections to the muon decay rate are summarized in Table \ref{tab:ratecorrections}.

\begin{table}[htbp]
\centering
\caption{Order-by-order corrections to the inclusive muon decay rate
$\Gamma_\mu$.}
\label{tab:ratecorrections}
\begin{tabular}{c|c|c}
\hline
 & & \\
Correction & Order &  $\Gamma_\mu$ contribution \\
 & & \\
\hline
& & \\
Finite $m_e$ phase space correction \citep{vanRitbergenStuart2000,Pak2008}
& $\Delta q^{(0)}$
& $187$ ppm \\

First-order QED correction \citep{KinoshitaSirlin1959}
& $\mathcal{O}(\alpha)$
& $0.42\%$ \\

Second-order QED correction \citep{vanRitbergenStuart1999,vanRitbergenStuart2000}
& $\mathcal{O}(\alpha^2)$
& $36.2$ ppm \\

Hadronic vacuum polarization \citep{vanRitbergenStuart1998Hadronic}
& $\mathcal{O}(\alpha^2_{\textrm had})$
 & $0.2$ ppm \\

 & & \\
\hline
\end{tabular}
\label{tab:muon_decay_qed_corrections}
\end{table}
These advances transformed Fermi's $\beta$-decay theory into a modern EFT applicable to $\beta$-decay, muon decay and other low-energy weak processes. 
The EFT cleanly separates the order-by-order QED corrections from the effective weak interaction and provides an unambiguous definition of the Fermi constant across such weak processes.

The Standard Model provides a more fundamental description of the effective current-current interaction and the Fermi constant. 
The electroweak sector of the Standard Model is defined by three parameters:
the $g$ and $g\prime$ gauge couplings
of the $SU(2)_L$ and $U(1)$ gauge symmetries,
and the Higgs vacuum expectation value $v$ associated with  
electroweak symmetry breaking.

In electroweak theory \citep{PDG2024}, 
charged-current weak processes such as muon decay are mediated by W-boson exchange (Fig.\ \ref{GF1}). At tree level and low momentum transfer $q^2 \ll M_W^2$, the W-propagator reduces to a local interaction corresponding to the Fermi EFT, giving

\begin{equation}
\label{GF1}
G_F = \frac{{\sqrt{2}} g^2}{8M_W^2} .
\end{equation}

Equivalently, in terms of the Higgs vacuum expectation value,  as
\begin{equation}
\label{GF2}
G_F = 1 / ( \sqrt{2} v^2 ) ,
\end{equation}
showing that the Fermi constant directly reflects the electroweak symmetry-breaking scale within the Standard Model.\footnote{In the Standard Model, the 
W-boson mass is given at tree level by 
$M_W  = g v / 2$. Substituting this relation into Eqn.\ \ref{GF1} yields Eqn.\ \ref{GF2} up to electroweak radiative corrections.}

\begin{figure}
\begin{centering}
  \includegraphics[width=0.8\columnwidth]{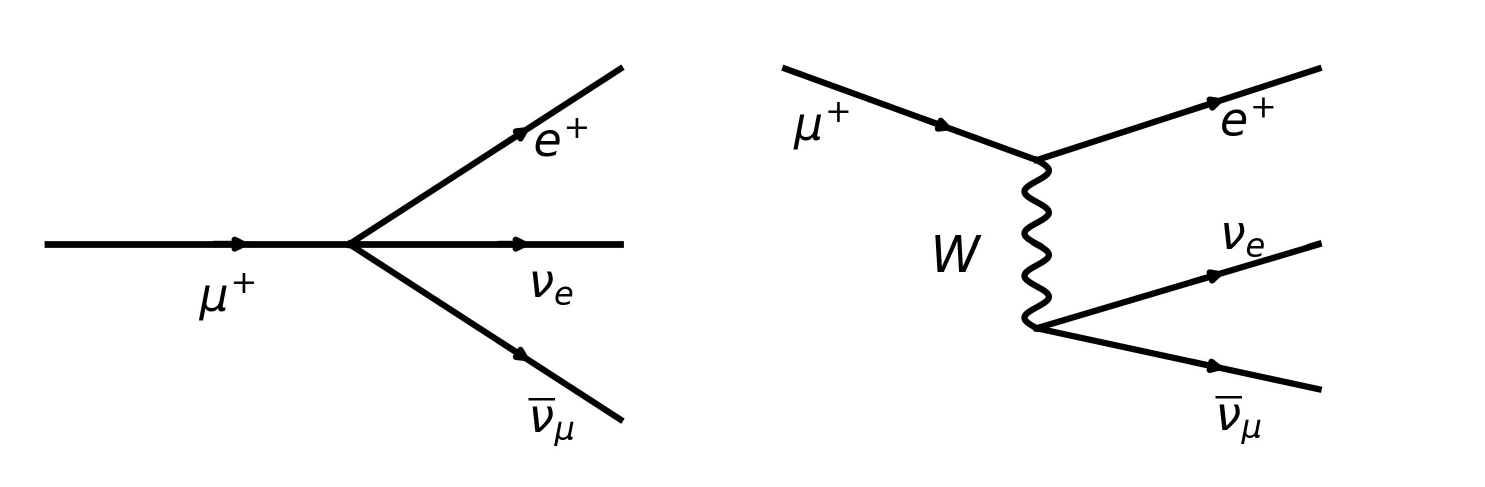}
  \caption{ \label{fg:GFfeynman} 
Left: tree-level Feynman diagram for muon decay in the weak current-current effective interaction.
Right: tree-level Feynman diagram for  muon decay in the Standard Model
showing the leading-order weak interaction
mediated by W-boson exchange between  leptonic currents.}
\end{centering}
\end{figure}

\subsection{Measurements of the muon lifetime}

Table \ref{tab:muon_lifetime} summarizes
the experimental progress in lifetime measurements
since the muon's discovery. 
Early muon lifetime measurements, spanning the 1940s to mid-1950s,  were aimed at establishing the basic properties of muons and their interactions with matter. Using cosmic rays, these early experiments were statistics limited 
and achieved few-percent precision.

\begin{table}[tp]
\centering

\begin{tabularx}{0.65\textwidth}{l|c|l}
\hline
& & \\
Year &  Lifetime result ($\mu$s) & Reference \\%& \textbf{} & Comments \\
& &  \\
\hline
& &  \\

1941  & 2.4  $\pm$ 0.3 & %Rossi {\textit et al.}\ 
\citet{Rossi:1941zz} \\ %Rossi \& Hall, Phys.\ Rev.\ 59,  223(1941) \\
1951   & 2.22  $\pm$ 0.02 & %Bell \& Hincks 
\citet{BellHincks1951} \\ %, Phys.\ Rev.\ 84,  1243(1951)  \\

1959 & 2.20 $\pm$ 0.015 & 
%Fischer {\textit et al.}\ 
\citet{Fischer1959} \\ %,Phys. Rev. Lett. 3, 349 (1959) \\

1960 &  2.211 $\pm$ 0.003  &  %Reiter {\textit et al.}\ 
\citet{Reiter1960} \\ % Phys. Rev. Lett. 5, 22 (1960) \\

1962 &   2.203  $\pm$ 0.004  & %Lundy
\citet{Lundy1962} \\%, Phys.\ Rev.\ 125,  1686(1962) \\

1972 & 2.20026  $\pm$ 0.00081  & %Williams and Williams 
\citet{Williams1972MuonLifetime} \\ 

1973 & 2.1973  $\pm$ 0.0003  & %Duclos {\textit et al.}\ 
\citet{Duclos1973} \\ % Phys.\ Lett.\ 47B 491(1973) \\
1975  & 2.19711  $\pm$ 0.00008 & %Balandin {\textit et al.}\ 
\citet{ Balandin1975} \\ % Sov.\ Phys.\ JETP 40  811(1975) \\ 

1984   &  2.19695  $\pm$ 0.00006 & %Giovanetti {\textit et al.}\ 
\citet{Giovanetti1984} \\ %, Phys.\ Rev.\ D29  343(1984) \\

1984 & 2.197078 $\pm$ 0.000073 & %Bardin {\textit et al.}\ 
\citet{Bardin1984} \\% Nucl.\ Phys.\ B 135(1984) \\

2007 & 2.197013 $\pm$ 0.000024 & %Chitwood {\textit et al.}\ 
\citet{MuLan2007} \\ % Phys.\ Rev.\ Lett.\ 99, 032001(2007)  \\

2008 & 2.197083  $\pm$ 0.000035 & %Barczyk {\textit et al.}\ 
\citet{Barczyk:2007hp} \\ %  Phys\ Lett.\ B 663, 172(2008)   \\

2013  & 2.1969803  $\pm$ 0.0000022 & %Tishchenko {\textit et al.}\ 
\citet{Tishchenko:2012ie} \\%  Phys.\ Rev.\ Lett.\ 106, 041803(2011) \\
& & \\
\hline
\end{tabularx}
\caption{Progress in the measurement of the muon lifetime from the early cosmic-ray studies to the modern meson factory experiments.}
\label{tab:muon_lifetime}
\end{table}

Precision measurements began in 1960. By then, Fermi's current-current interaction with  universal $V$-$A$ currents  motivated an accurate comparison of $\beta$-decay and muon decay rates.
This coincided with the development of low-energy muon beams from pion production at proton cyclotrons and synchrotrons, and the introduction of plastic scintillators and photomultiplier tubes for nanosecond timing.
Together, these advances 
enabled part-per-thousand measurements of $\tau_{\mu}$.

Further improvements followed in the late 1970s and early 1980s. 
At this time, the role of $G_F$ as an input into electroweak theory was well established, and  first-order QED corrections had reduced the theoretical uncertainties 
in extracting $G_F$ from $\tau_{\mu}$.
The measurements took advantage of intense muon beams at new meson factories and
advances in detector, electronics, and readout technologies.

By 1990, the world average for the muon lifetime had an 18~ppm relative uncertainty.
This average was largely determined by three experiments with individual uncertainties at the 30-40 ppm level
%Balandin {\textit et al.}\ 
\citep{Balandin1975,Giovanetti1984,Bardin1984}.
%Giovanetti {\textit et al.}\ \citep{Giovanetti1984},
%and Bardin {\textit et al.}\ \citep{Bardin1984}).
The Giovanetti {\textit et al.}\ and Balandin {\textit et al.}\ experiments employed sufficiently low beam rates
that muon stops and associated decay electrons  were recorded without beam pileup.
The avoided pileup distortions in electron time distributions, but limited the sample size to $\mathcal{O}(10^{10})$ decays the statistical uncertainty to
$\mathcal{O}(10~ppm)$ precision.

By 2000, complete second-order QED corrections
and growing interest in precision electroweak tests motivated improved measurements of $\tau_{\mu}$. 
Two experiments -- the FAST experiment \citep{Barczyk:2007hp}
and the MuLan experiment \citep{MuLan2007,Tishchenko:2012ie} -- were consequently designed to overcome the statistical limitations of single-event measurements.

FAST utilized a finely pixelated target 
to reconstruct muon-electron vertices, thereby associating decay electrons with parent muons.
In contrast, MuLan employed a time-structured muon beam 
to first accumulate a population of stopped muons 
and then measure the decay time distribution, thus eliminating the need for event-by-event association. 
Both experiments were carried out at PSI, with FAST achieving 15~ppm precision
and MuLan achieving 1~ppm precision.

\subsection{The MuLan experiment}

Part-per-million measurements of 
$\tau_\mu$ present both statistical and systematic challenges.
Achieving the statistical precision requires datasets of 10$^{12}$ $\mu \rightarrow e \nu {\bar{\nu}}$ decays, while the systematic precision requires mitigating distortions to the decay-electron time distributions arising from signal pileup, gain drifts, and muon spin rotation ($\mu$SR). 

MuLan addressed these challenges through several design features, including a high-rate time-structured muon beam, a near-hermetic finely-segmented electron detector, the waveform digitization of electron signals, and the laser-based monitoring
of detector stability. Two  different stopping targets were used to mitigate and characterize effects. Data analysis and simulation
procedures were developed to correct  and quantify the effects of pileup. To avoid analysis bias, the collaboration was blinded to the exact clock frequency.

\begin{figure}
\begin{centering}
  \includegraphics[width=0.7\columnwidth]{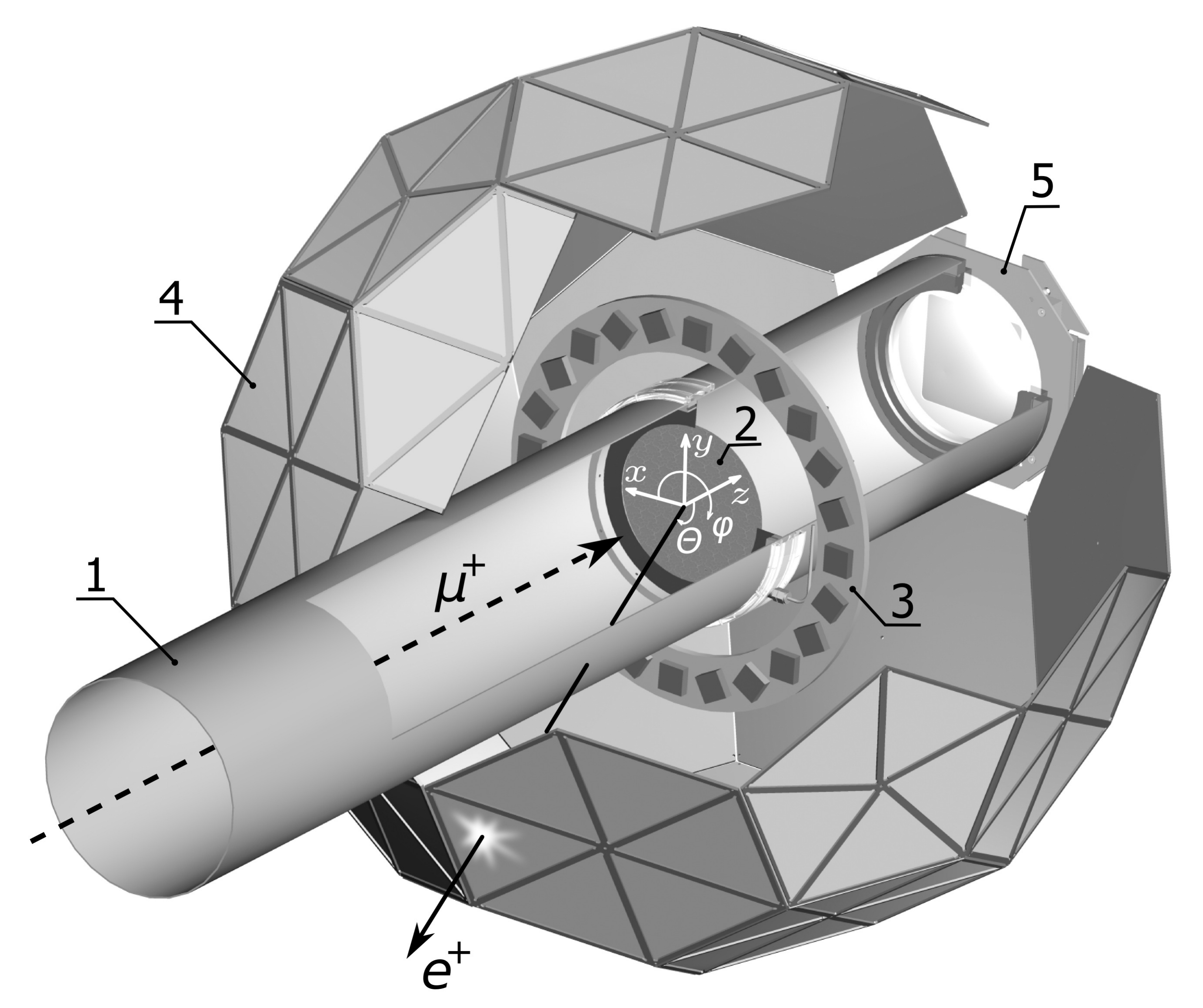}
  \caption{ \label{fg:MuLanSetup} Cutaway diagram of the MuLan experiment
indicating an incoming beam muon and outgoing decay electron and showing the
beam pipe (1), in-vacuum stopping target (2), optional permanent magnet array (3),
double-layered scintillator detector (4) and beam monitor (5). Reproduced from %Tishchenko {\textit et al.}\
\citet{Tishchenko:2012ie}.
}
\end{centering}
\end{figure}

\begin{figure}
\begin{centering}
  \includegraphics[width=0.6\columnwidth]{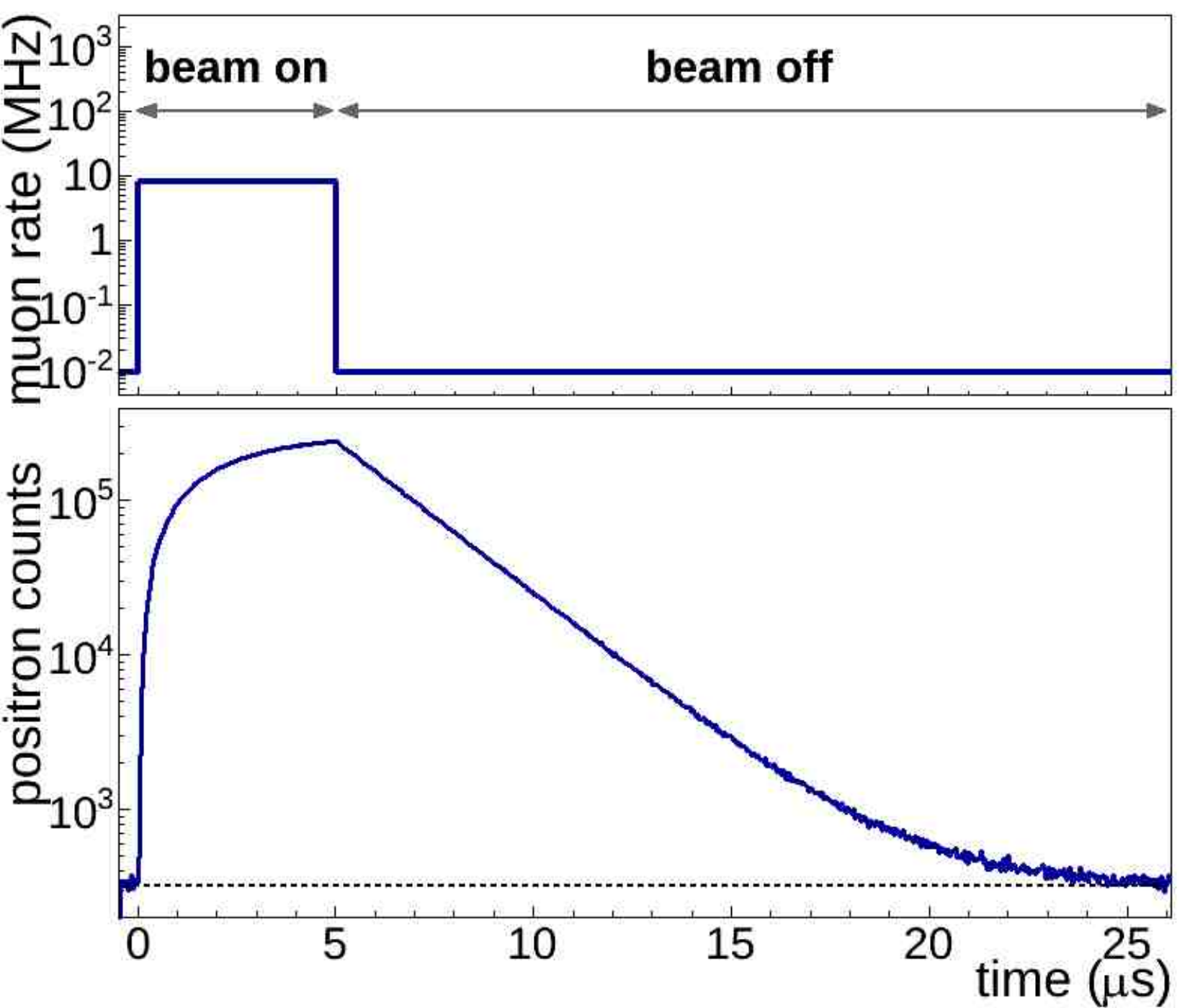}
  \caption{ \label{fg:mulanbeamonoff} 
The  time structure of the MuLan experiment comprising a 5-$\mu$s beam-on muon accumulation period and a 22-$\mu$s beam-off electron measurement period. The upper panel shows the muon arrival 
time distribution and the lower panel shows the decay electron time 
distribution. Reproduced from %Tishchenko {\textit et al.}\ 
\citep{Tishchenko:2012ie}.}
\end{centering}
\end{figure}

The MuLan 
apparatus comprised a time-structured surface muon beam, an in-vacuum stopping target, and a
segmented electron detector (see Fig.\ \ref{fg:MuLanSetup}).
The beam operated in repeating cycles consisting of 5~$\mu$s beam-on, muon accumulation periods and 22~$\mu$s beam-off, electron measurement periods (see Fig.\ \ref{fg:mulanbeamonoff}).

Surface muon beams are 
highly polarized and produce decay electrons that are asymmetrically distributed about the polarization direction.
The polarization precesses and relaxes in the local magnetic field of the target material.
This $\mu$SR leads to time-varying and geometry-dependent distortions of the electron time distribution.
 
To minimize $\mu$SR effects, the experiment employed two complementary strategies for the stopping target material and the local magnetic field.
One used a magnetized ferromagnetic foil
with an internal field of about 5000~Gauss, while the
other used a nonmagnetic quartz crystal with an external field of about 80~Gauss. 
In the ferromagnetic foil, the muons predominantly form diamagnetic $\mu^+$ ions, whereas 
in quartz they predominantly
form paramagnetic muonium.
In both cases, the resulting precession 
frequencies
yielded a thousand-fold reduction of the muon polarization through phase scrambling during muon accumulation.

To mitigate pileup effects, the experiment employed a segmented detector with waveform digitization. The detector comprised 170 double-layer plastic scintillator tiles arranged in a soccer-ball geometry, providing spatial separation of electrons. The scintillator pulses were recorded by high-speed waveform digitizers, enabling temporal separation of electrons. Together, segmentation and digitization suppressed pileup and enabled correction of residual effects.

The decay-time distribution  was constructed in several steps. Digitized waveforms were fit to pulse-shape templates to extract the times and energies of individual pulses. An adjustable software deadtime was applied to enforce a minimum resolving time between neighboring pulses. Coincidences between
inner and outer tile pairs were used to identify electrons and construct the decay-time distribution. Small data-driven corrections were applied to account  for pulse pileup and gain variations.

The time distributions 
were fit to extract $\tau_{\mu}$. 
The ferromagnetic target data showed
no evidence of $\mu$SR effects
and were fit to a single exponential decay with a time-independent background.\footnote{
The time-independent background arose from both cosmic rays
and the imperfect beam extinction during the measurement period.}
The quartz target data exhibited a small $\mu$SR modulation 
and were fit to a modified function incorporating the effects of precession and relaxation.
In both cases, fits were performed for multiple software deadtimes and extrapolated to zero deadtime. 

The two targets yielded consistent values of $\tau_\mu$, giving the combined result
$\tau_{\mu} = 2~196~980.3 \pm  2.1 (stat) \pm 0.7 (syst)$~ps,
corresponding to an overall uncertainty of 2.2~ps (1.0~ppm). %and thirty-fold improvement over earlier generations of experiments.

\subsection{Status and implications}

Early cosmic-ray measurements of the muon lifetime were significant in demonstrating that the muon was an unstable particle  and its lifetime was too long to be compatible with the force carrier of the strong interaction. Subsequent accelerator-based measurements  were central to the formulation of a universal weak interaction, as well as providing the normalization needed to reconcile strangeness-conserving and strangeness-changing weak decays through Cabibbo theory. 

Precision measurements of the muon lifetime also motivated developments in weak interaction EFT and higher-order QED corrections that enable a theoretically rigorous definition of the Fermi constant $G_F$. 
The $G_F$-$\tau_{\mu}$ relation of Eqn.\ \ref{higher order tau_mu}, together with measurements of $\tau_{\mu}$, $m_{\mu}$, 
and second-order QED corrections, yields
$G_F =
1.166\, 378\, 7(6)\times 10^{-5}~{GeV^{-2}}\ (0.5~\mathrm{ppm})$.
This provides the most precise determination of the weak interaction strength and serves as an overall  normalization for all weak processes.

Comparisons between the purely leptonic decays $\tau \rightarrow e \nu \bar{\nu}$,
$\tau \rightarrow \mu \nu \bar{\nu}$ and $\mu \rightarrow e \nu \bar{\nu}$ provide a test of lepton universality in weak interactions.
Using measurements of the $\tau$ lifetime and leptonic
branching ratios, lepton universality is confirmed at the few-per-mille level  \citep{HFLAV2021}
\begin{equation}
{g_\tau}/{g_\mu} = 1.0010 \pm 0.0014 , \qquad
{g_\tau}/{g_e}   = 1.0029 \pm 0.0014 , \qquad
{g_\mu}/{g_e}    = 1.0018 \pm 0.0014.
\end{equation}
The Fermi constant enters the  extraction of Cabibbo-Kobayashi-Maskawa (CKM) matrix elements, which describe flavor mixing in quark weak interactions.
Quark mixing 
was introduced by Cabibbo
to explain the different strengths
of strangeness-changing and strangeness-conserving
weak interactions and to restore weak universality.
CKM unitarity requires the sum of the squared couplings of each up-type quark to all down-type quarks (down, strange and bottom) to equal unity.
The up-quark row is currently tested at  the  part-per-mille level \citep{Seng2022VusVud,Cirigliano2023CKM}.

The Fermi constant $G_F$,  fine-structure constant $\alpha$, and Z-boson mass $M_Z$ define the electroweak parameters $g$, $g'$ and $v$, and enable precision comparisons between SM  predictions and measurements of the W boson mass $M_W$ and the weak mixing angle $\theta_W$ \citep{Sirlin1980,Brivio2021}.
These predictions involve loop  corrections that historically provided sensitivity to the then-unknown top-quark and Higgs-boson masses. 

\section{Muon decay}
\label{sc:decayparameters}

Precise measurements of $\mu \rightarrow e \nu \bar{\nu}$ decay observables provide a particularly clean test of the chiral structure of the weak interaction and set stringent limits on non-standard couplings such as right-handed currents. 

The experimentally accessible quantities
are the muon polarization and the energy, direction, and polarization of the decay electron. 
For unpolarized muons, the angular distribution is isotropic, and the relevant observables are the  energy spectrum and the longitudinal polarization.  
For polarized muons, the angular distribution becomes anisotropic, and the observables additionally include the energy-angle correlation and the transverse polarization.

A particularly important kinematic region is the endpoint energy $E_{max} = m_{\mu}/2$ of the electron energy spectrum. Near the endpoint, where the electron recoils against the neutrino pair, the decay exhibits maximal asymmetry and longitudinal polarization. In $\mu^+$ decay, right-handed $e^+$ are preferentially emitted along the muon spin direction, whereas in $\mu^-$ decay, left-handed $e^-$ are preferentially emitted opposite the muon spin direction.

\subsection{Form of the weak current}

In Fermi's original theory of $\beta$-decay, the weak interaction was constructed from vector currents. 
It was soon recognized, however, that the most general Lorentz-invariant current-current
interaction can be constructed from 
scalar (S), pseudoscalar (P), vector (V), axial-vector (A), and tensor (T) currents
\begin{equation}
\bar{\psi}_e \psi_\mu \;(S) ,~~ 
\bar{\psi}_e \gamma^5 \psi_\mu \;(P) ,~~
\bar{\psi}_e \gamma^\alpha \psi_\mu \;(V) ,~~
\bar{\psi}_e \gamma^\alpha\gamma^5 \psi_\mu \;(A) ,~~
\bar{\psi}_e \sigma^{\alpha\beta} \psi_\mu \;(T)
\end{equation}
which lead to distinctive decay distributions in $\mu \rightarrow e \nu \bar{\nu}$.
\citet{Michel1950} introduced a parameterization of $\mu \rightarrow e \nu \bar{\nu}$ decay in which a set of parameters serve as intermediaries between experimental observables and the underlying Lorentz structure.
For unpolarized muon decay, the electron energy spectrum depends on two Michel parameters, $\rho$ and  $\eta$.\footnote{The parameter $\eta$ enters through $m_e / E_e$ terms and only contributes at low electron energies.}

Before polarized muon beams became available,
$\mu$-decay experiments focused on  the improved measurements of the electron energy spectrum. 
The data indicated $\rho \simeq 3/4$, which strongly favored vector and/or axial-vector interactions that predict $\rho = 3/4$ (pure S or P interactions imply $\rho = 0$, while pure T interactions imply $\rho = 3/8$). However, $\rho$ alone cannot distinguish between vector and axial-vector interactions.

After polarized muon beams became available, 
the focus shifted to the
angular distribution of the decay electrons relative to the
muon polarization.
For polarized muons, the
Michel parameters $\xi$ and $\delta$
govern the electron asymmetry, with $\xi$ setting the overall scale and $\delta$ controlling its energy dependence. The observation of maximal asymmetry ($\xi \simeq 1$) provided important evidence for a maximally parity-violating interaction of either V$-$A or V$+$A structure, and excluded parity-conserving pure vector or pure axial-vector interactions. 

%In 1958, Marshak and Sudarshan 
\citet{MarshakSudarshan1957}, and %Feynman and Gell-Mann 
\citet{Feynman:1958ty}, independently  proposed the V$-$A theory of weak interactions. Measurements of electron polarization in nuclear $\beta$-decay strongly favored V$-$A over V$+$A currents.
This framework motivated a program of precision measurements of muon decay observables and Michel parameters to test the V$-$A prediction of $\rho = 3/4, \eta = 0, \xi = 1$ and  $\delta = 3/4$.

The Michel formalism was also extended to include parameters describing the electron polarization in muon decay \citep{KinoshitaSirlin1957,BouchiatMichel1957}.\footnote{Additionally, the evaluation of radiative decay modes and radiative loop corrections became necessary with the increasing precise measurement of the Michel parameters.} 
The polarization can be decomposed into a longitudinal component $P_L$,
parallel to the electron momentum,
and a transverse component $P_T$,
perpendicular to the electron momentum.
The transverse polarization 
is further resolved into a time-reversal-conserving component $P_{T1}$ lying in the decay plane, and a time-reversal-violating component $P_{T2}$ perpendicular to the decay plane, where the electron momentum and muon polarization define the decay plane.

The longitudinal polarization $P_L$  is parameterized by $\xi'$ and $\xi''$. The transverse polarization is parameterized by $\eta$ and
$ \eta''$ for $P_{T1}$, and $\alpha'/A$ and $\beta'/A$ for $P_{T2}$.
In V-A theory, these parameters take the values $\xi' = \xi'' = 1$ and $\eta =
 \eta'' = \alpha'/A = \beta'/A = 0$,
reflecting the electron's maximal handedness.  Transverse polarization arises from interference between decay amplitudes from different Lorentz structures \citep{Fetscher2021MuonDecay}, and inherits special sensitivity to exotic scalar and tensor currents and T–violating phases.

A major development during the 1970s and 1980s was the formulation of a systematic chiral framework for the weak current-current interaction \citep{PDG2024,Fetscher1986}.
Within this approach, the 
interaction is written in terms of definite electron and muon chiralities that represent the most general, Lorentz invariant, derivative-free 
interaction. The interaction Hamiltonian is 
\begin{equation}
\frac{4 G_F}{\sqrt{2}}
\sum_{\gamma = S,V,T}
\sum_{\epsilon,\mu = L,R}
g^{\gamma}_{\epsilon\mu}\,
\left(
\bar e_\epsilon \, \Gamma^\gamma \, \nu_e
\right)
\left(
\bar \nu_\mu \, \Gamma_\gamma \, \mu_\mu
\right)
\end{equation}

where $\Gamma^S,
\Gamma^V,
\Gamma^T$ denote the  scalar, vector, and tensor interaction structures, and $\epsilon, \mu$ label the left- and right-handed electron and muon chiralities. The couplings $g^{\gamma}_{\epsilon \mu}$s are complex coefficients describing all allowed combinations of interaction type and lepton chirality.
In V-A theory, the coupling $g^V_{LL} = 1$, while  all other couplings vanish.

The ten Michel parameters associated with ordinary muon decay $\mu^+ \rightarrow e^+ \nu \bar{\nu}$ ,
plus one additional parameter accessible in radiative decay $\mu^+ \rightarrow e^+ \nu \bar{\nu} \gamma$,
are insufficient to uniquely determine the ten {\textit complex} couplings  $g^{\gamma}_{\epsilon \mu}$.
To address this, Fetscher and collaborators introduced four positive semi-definite combinations $Q_{\epsilon \mu}$ of couplings $g^{\gamma}_{\epsilon \mu}$ that characterize transition probabilities for a  muon of chirality $\mu = L, R$ to an  electron of chirality $\epsilon = L, R$. 
In V-A theory, the bilinear $Q_{LL} = 1$, while the remaining three combinations vanish.

\subsection{Michel parameters and decay observables}

%In 1949, Leighton {\textit et al.}\ 
\citet{PhysRev.75.1432} measured the decay-electron energy spectrum using stopped cosmic-ray muons and magnetic momentum analysis. Although statistically limited, 
the resulting energy spectrum provided crucial evidence that muons decay via $\mu \rightarrow e \nu \bar{\nu}$.

Over the following decade, measurements of muon decay continued to focus on the electron energy spectrum and the Michel parameter $\rho$. This period saw the transition from cosmic-ray to accelerator-based experiments and thereby increased statistics (see Table II in Ref.\ \citep{Rosenson1958}). The  results indicated a value of  $\rho = 3/4$ with uncertainties of $\sim$10\%.

A major breakthrough came with the late-1950s demonstration of
parity violation in muon decay \citep{Garwin:1957hc,Friedman:1957mz}. These experiments used polarized muon beams and provided the first observation of muon spin precession in applied magnetic fields. They measured the asymmetry parameter $\alpha$ governing the decay-electron angular distribution 
about the muon polarization, and found a maximally parity-violating  asymmetry, $\alpha \simeq 1/3$, corresponding to $\xi \simeq 1$.\footnote{The asymmetry coefficient is given by $ \left| \alpha \right| = P_{\mu} \xi / 3 $ where $\xi$ is the Michel parameter of the electron angular  distribution and $P_{\mu}$ is the  muon polarization in the $\pi \rightarrow \mu \nu$ decay.}

During the 1960s, the experimental program shifted toward  precision tests of weak universality and V-A theory by extracting the Michel parameters 
from least-squares fits to energy-angle distributions. These experiments reached sub-percent precision for $\rho$ and five-percent precision for $\xi$ \citep{Plano1960,Rosenson1958,Sherwood1967,Fryberger1968,DeRenzio1969}.

Decay-electron polarization
was also recognized as a sensitive probe of the chiral structure of the weak interaction. 
The first measurements \citep{Buhler1963, Bloom1964,Duclos1964,PhysRev.162.1306} of the longitudinal polarization were performed in the 1960s 
using Bhabha scattering and in-flight annihilation for electron polarimetry. They confirmed the maximal handedness of decay electrons predicted by V$-$A theory, corresponding to $\xi' = 1$.

Following the emergence of the Standard Model,  %in the 1970s and 1980s, 
experimental efforts increasingly focused on searches for beyond-SM physics.
The theoretical cleanliness of muon decay made it particularly attractive for such searches.
Intense muon beams at meson factories enabled the first measurements of transverse polarization \citep{Burkard1985a,Corriveau1983,Burkard1985}.
These experiments produced the first constraints  on the T-violating polarization parameters $\alpha'/A$ and $\beta'/A$. They also improved the determination of the Michel parameter $\eta$, which contributes to both the T-conserving transverse polarization and the shape of the electron energy spectrum.

More recently,  
large-scale experimental programs 
have carried out precision measurements of Michel parameters.
A major initiative at the TRIUMF cyclotron was the TWIST experiment, which performed a comprehensive measurement of the electron energy-angle distribution
from polarized muon decay. This work enabled a $\mathcal{O}(10)$-fold improvement in determinations of $\rho$, $\eta$, and $P_{\mu} \xi$, substantially improving the global analyses of muon decay and tests of V$-$A theory \citep{Bayes2011,Bueno:2011fq,Bueno:2012PublisherNote,MacDonald2008}. 

At PSI, complementary experiments 
focused on electron polarization in muon decay. 
These experiments used Bhabha-scattering and in-flight-annihilation for polarimetry and were sensitive to non-V$-$A couplings and T-violating phases.
Together, they improved the determination of the longitudinal polarization parameter $\xi''$ by about an order of magnitude \citep{Prieels:2014paa} and  strengthened constraints on transverse polarization parameters by factors of about three \citep{Danneberg:2005xv}.

A summary of the progress in measurements of the Michel energy-angle parameters $\rho$, $\eta$, $\delta$ and $|P_\mu \xi|$ since the pioneering work of Leighton {\textit et al.}\ is provided in Table \ref{tab:michelcompilation}.

\begin{table}[!htbp]
\centering
\small
\begin{tabularx}{0.7\textwidth}{l|c|c|X}
\hline
& & \\
Year & Value & Uncertainty & Reference \\
& & \\
\hline
& & \\
$\rho$, & &  \\
1958 & 0.67 & 0.05 & %Rosenson
\citet{Rosenson1958}  \\
1967 & 0.7600 & 0.0090 & %Sherwood  
\citet{Sherwood1967}  \\ %Sherwood (1967), Phys.\ Rev.\ 156, 1475 (EFI) \\
1968 & 0.7620 & 0.0080 & %Fryberger 
\citet{Fryberger1968} \\ % (1968), Phys.\ Rev.\ 166, 1379 (EFI) \\
1969 & 0.7518 & 0.0026 & %DeRenzio 
\citet{DeRenzio1969} \\ % 
2005 & 0.75080 & 0.00105 & %Musser {\textit et al.}\ 
\citet{Musser2005} \\ %(2005), Phys.\ Rev.\ Lett.\ 94, 101805 (TWIST) \\
2008 & 0.75014 & 0.00048 & %MacDonald {\textit et al.}\ 
\citet{MacDonald2008} \\ % (2008), Phys.\ Rev.\ D78, 032010 (TWIST) \\
2011 & 0.74977 & 0.00026 & %Bayes {\textit et al.}\ 
\citet{Bayes2011} \\ % (2011), Phys.\ Rev.\ Lett.\ 106, 041804 (TWIST) \\
 & & & \\
 $\eta$, & & & \\
 
1960 & $-2.0$ & 0.9 & %Plano 
\citet{Plano1960} \\ %(1960), Phys.\ Rev.\ 119, 1400 (Columbia) \\
1969 & $-0.12$ & 0.21 & %DeRenzio 
\citet{DeRenzio1969} \\ %  (1969), Phys.\ Rev.\ 181, 1854 (EFI) \\
1985 & 0.011 & 0.085 & %Burkard {\textit et al.}\ 
\citet{PDG2024,Burkard1985} \\ %(1985), Phys.\ Lett.\ B150, 242 (SIN) \\
2005 & 0.071 & 0.037 & %Danneberg {\textit et al.}\ 
\citet{Danneberg:2005xv} \\% (2005), Phys.\ Rev.\ Lett.\ 94, 021802 (PSI) \\
 & & & \\
$\delta$, & & & \\
1960 & 0.78 & 0.05 & %Plano 
\citet{Plano1960} \\ % (1960), Phys.\ Rev.\ 119, 1400 (Columbia) \\
1968 & 0.752 & 0.009 & %Fryberger 
\citet{Fryberger1968} \\ %(1968), Phys.\ Rev.\ 166, 1379 (EFI) \\
1988 & 0.7486 & 0.0038 & %Balke {\textit et al.}\ 
\citet{Balke1988} \\ % (1988), Phys.\ Rev.\ D37, 587 (TRIUMF) \\
2005 & 0.74964 & 0.00130 & %Gaponenko {\textit et al.}\ 
\citet{Gaponenko2005} \\ %(2005), Phys.\ Rev.\ D71, 071101 (TWIST) \\
2008 & 0.75067 & 0.00073 & %MacDonald {\textit et al.}\ 
\citet{MacDonald2008} \\ %et al.\ (2008), Phys.\ Rev.\ D78, 032010 (TWIST) \\
2011 & 0.75049 & 0.00034 & %Bayes {\textit et al.}\ 
\citet{Bayes2011} \\ %  Bayes et al.\ (2011), Phys.\ Rev.\ Lett.\ 106, 041804 (TWIST) \\
& & & \\
$|P_\mu \xi|$, & & & \\
1959 & 0.97 & 0.05 & %Bardon {\textit et al.}\ 
\citet{Bardon1959} \\ %(1959), Phys.\ Rev.\ Lett.\ 2, 56 (Columbia) \\
1960 & 0.94 & 0.07 & %Plano 
\citet{Plano1960} \\ %Plano (1960), Phys.\ Rev.\ 119, 1400 (Columbia) \\
1987 & 1.0027 & 0.0085 & %Beltrami {\textit et al.}\ 
\citet{Beltrami1987} \\ %(1987), Phys.\ Lett.\ B194, 326 (SIN) \\
2011 & 1.00084 & $^{+0.00168}_{-0.00069}$ & %Bueno {\text et al.}\ 
\citet{Bueno:2011fq,Bueno:2012PublisherNote}\\ % (2011), TWIST \\
& & & \\
\hline
\end{tabularx}
\caption{Summary of measurements of the Michel energy-angle parameters $\rho$, $\eta$, $\delta$ and $|P_\mu \xi|$. The Michel parameter $\xi$ and the muon polarization in pion decay $P^{\pi}_{\mu}$ are experimentally inseparable. The improved determinations of $\eta$ by Burkard {\textit et al.}\ and Danneberg {\textit et al.}\ were obtained by the measurement of the electron transverse polarization $P_{T1}$. Where reported separately, statistical and systematic uncertainties are combined in quadrature; asymmetric uncertainties are retained.}
\label{tab:michelcompilation}
\end{table}

\subsection{The TWIST experiment}

The precision determination of the Michel parameters $\rho$, $\delta$ and $\xi$ is challenging because they appear experimentally as subtle variations in the decay electron energy-angle distribution. Consequently, a small energy-angle dependence of detection efficiency can bias their measurement. 

Accordingly, the TWIST experiment was designed to minimize acceptance variations and carefully control beam alignment, stopping distributions, and muon polarization. This required precision momenta reconstruction, detailed understanding of detector acceptance, and long-term detector stability.

\begin{figure}
\begin{centering}
  \includegraphics[width=0.8\columnwidth]{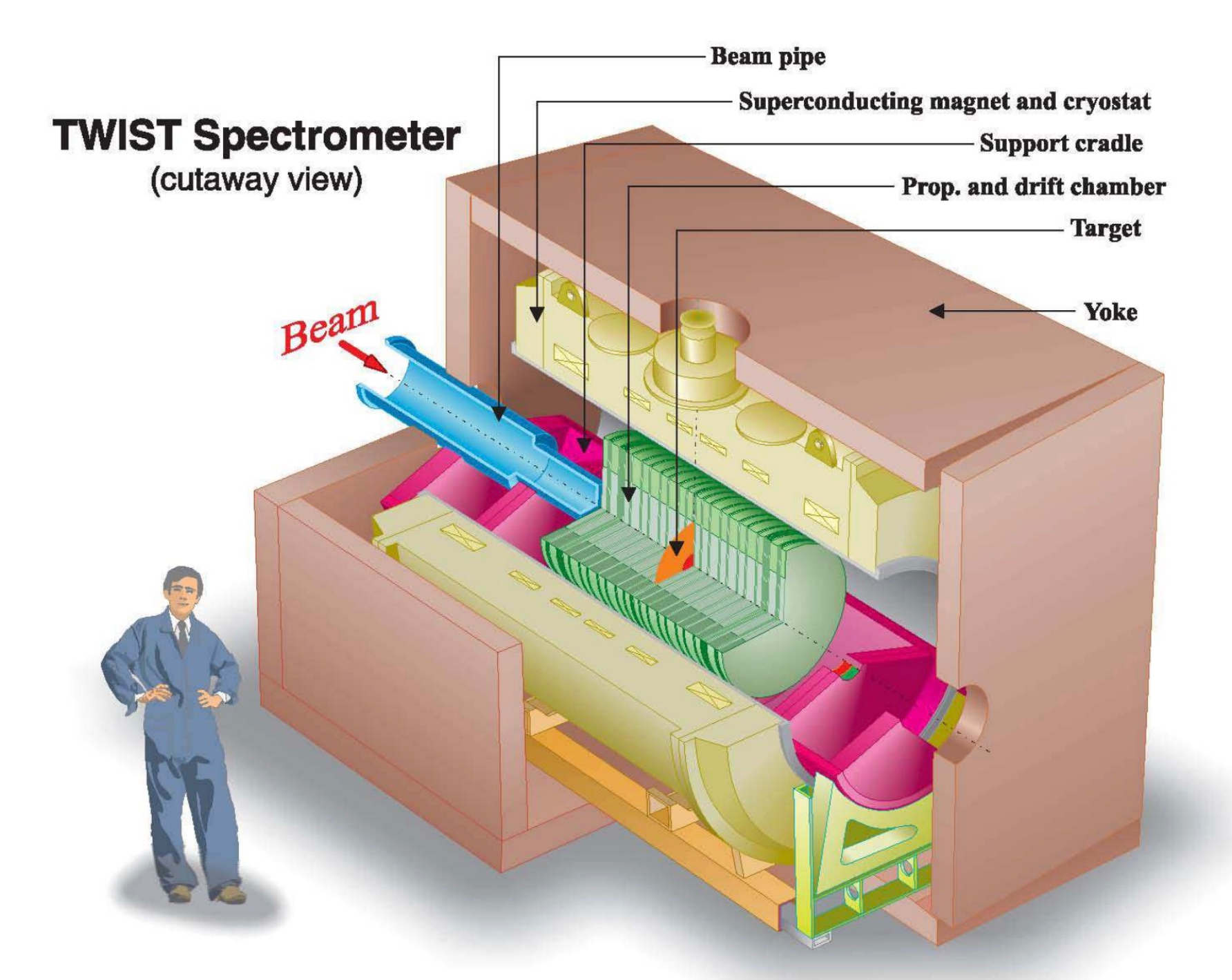}
  \caption{ \label{fg:TWISTSetup} Cutaway  view of the TWIST experiment
indicating the muon beam, 
planar drift and proportional chambers, stopping target, and cylindrical superconducting magnet. Reproduced from %Henderson {\textit et al.}\ 
\citet{Henderson2005TWIST} with permission of Elsevier.}
\end{centering}
\end{figure}

The TWIST setup (Fig.\ \ref{fg:TWISTSetup}) was based on stopping a surface muon beam
in a  thin target foil and detecting the outgoing decay electrons
in a low-mass, large-acceptance magnetic spectrometer.
The high quality beam had a sub-1\% momentum bite and a near-100\% polarization. 
It  entered the spectrometer through a thin scintillator and an adjustable gas degrader
that enabled real-time beam centering at the target. 

The spectrometer provided measurements of both the beam muons and the decay positrons. It comprised
a symmetric interleaved array of low-mass planar drift and proportional
chambers arranged along the beam axis. 
The drift chambers enabled the precise measurement
of the momentum-angle coordinates of the outgoing electron, 
while the proportional chambers provided $\mu$/e discrimination
via energy loss.

The experiment collected datasets
using Ag and Al target foils.
Both measured and simulated data
were stored in a common data format
and processed through a common analysis chain.
The two datasets
were analyzed in parallel to extract differences $\Delta \rho$, $\Delta \delta$, and $\Delta \xi$ between the  measured values and the known reference values in the simulated data.
The relative  approach 
facilitated
a blind analysis
where the absolute values
of the decay parameters
were not known
until the analysis
had been completed.

The simulation of particle interactions,
detector geometry, and readout included
the discrete nature of detector ionization, Garfield calculations of ionization drift, OPERA calculations of magnetic fields, and rate-dependent effects from overlapping tracks.
The simulated p-$\theta$ distributions of decay electrons included full radiative corrections to order O($\alpha^2$).
Systematic effects associated with muon stopping, muon depolarization, detector alignment, and detector calibration were studied through detailed comparisons between measurements and simulations.

The final results for the Michel parameters $\rho$, $\delta$, and $P^{\pi}_{\mu}$,  obtained from the
combined analysis of measured and simulated data, are listed in Table \ref{tab:decayparameters}.
The largest systematics in the measurement of the parameters $\rho$ and $\delta$ include the target-independent chamber response, resolution and calibration and the target-dependent stopping position and energy loss.   The
Michel parameter $\xi$ and the muon polarization in pion decay $P^{\pi}_{\mu}$
are experimentally inseparable. 
The uncertainty in $P^{\pi}_{\mu} \xi$ is dominated by
muon depolarization effects,  including contributions from spectrometer-beam alignment
and spectrometer fringe fields.

\subsection{Status and implications}

Muon decay observables have had a significant impact in fundamental physics.
The pioneering measurement by Leighton {\textit et al.}\ of the decay-electron energy spectrum was pivotal in establishing the three-body nature of muon decay and strongly supported the hypothesized
$\mu \rightarrow e \nu \bar{\nu}$ decay. The 
Garwin {\textit et al.}\ and Frieddman {\textit et al.}\ observations of the decay-electron asymmetry about the muon polarization direction  was essential to confirming the parity-violating nature of weak interactions and favoring either a V$-$A or V$+$A weak current. Precision measurements of this purely leptonic decay provide pristine tests of V$-$A theory and exceptional sensitivity to non-SM physics, including right-handed currents, scalar and tensor interactions, and T-violating phases of weak couplings.

Table \ref{tab:decayparameters} compiles the contemporary determinations of Michel parameters from the TWIST, %\citep{MacDonald2008,MacDonald2008,Bayes2011},
%Danneburg {\textit et al.} 
\citet{Prieels:2014paa},
and %Prieels {\textit et al.}\ 
\citet{Danneberg:2005xv} experiments.
Additionally included are the world average values of the  parameters $\xi'$ and $\bar{\eta}$ from \citet{PDG2024}.
The average for $\xi^{\prime}$ is dominated by the measurement of %Burkard {\it et al.}\ 
\citet{Burkard1985a} 
and the average
for $\bar{\eta}$ is dominated by the measurement of %Eichenberger {\it et al.}\ 
\citet{Eichenberger1984}.
Also listed are the SM values for the Michel parameters, 
{\textit i.e.}\ that correspond to $g^V_{LL} = 1$ and all other couplings zero.
In many cases the world data on decay parameters are consistent with the $V$-$A$ theory at the $10^{-3}$-$10^{-4}$ level.

A global analysis of the %Table \ref{tab:decayparameters} 
decay parameters was also performed by the TWIST collaboration \citep{Hillairet2012}.
Their results for the bilinears $Q_{e\mu}$ 
and the couplings $g^{\gamma}_{\epsilon \mu}$
are reproduced in Table \ref{tab:couplings}.
Of particular interest are 
the upper limits derived on the 
bilinear sums $Q^{\mu}_R \equiv Q_{RR} + Q_{LR}$
and $Q^{e}_R \equiv Q_{RR} + Q_{RL}$.
A non-zero value of $Q^{\mu}_R$ would 
signal a weak interaction contribution 
from a right-handed muon current, while
a non-zero value of $Q^{e}_R$ would 
signal a weak interaction contribution 
from a right-handed electron current.
The positron angular distribution parameter $\xi$ is
especially sensitive to $Q^{\mu}_R$, 
while the longitudinal polarization parameter $\xi^{\prime}$ is especially sensitive to $Q^{e}_R$.
The global analysis yields limits of $Q^{\mu}_R < 8.2 \times 10^{-4}$
and $Q^{e}_R < 0.044$.

\begin{table}[htb]
\centering \caption {Compilation of the most recent determinations
of the nine decay parameters in ordinary muon decay
and the additional $\bar{\eta}$ decay parameter in radiative muon decay.
With the exception of the ordinary decay parameter $\xi^{\prime}$
and the radiative decay parameter $\bar{\eta}$, we give the
results of the most recent precision measurements. For $\xi^{\prime}$
and $\bar{\eta}$ we give the world averages compiled in \citet{PDG2024}.
Where two uncertainties are quoted, the first is statistical and the second systematic. For $P_\mu^\pi\xi$, the systematic uncertainty is asymmetric. The Danneberg et al. values are from their general analysis.}
\begin{tabular}{c|c|c|c}
\hline
 & & & \\
  Parameter   & Extracted value & SM value & Ref.\ \\
  & & & \\
\hline
 & & & \\
$\rho$ & $0.749~77 \pm 0.000~12 \pm 0.000~23$ & 0.75 & \citet{Hillairet2012} \\
$\delta$ & $0.750~49 \pm 0.000~21 \pm 0.000~27$ & 0.75 &  \citet{Hillairet2012} \\
$P^{\pi}_{\mu} \xi$  & $1.00084 \pm 0.00029 ^{+0.00165}_{-0.00063}$ & 1.0 &  \citet{Bueno:2011fq,Bueno:2012PublisherNote} \\
$\xi^{\prime}$ & $1.00 \pm 0.04$ & 1.0 &  \citet{PDG2024} \\
$\xi^{\prime \prime}$ & $0.981 \pm 0.045 \pm 0.003$ & 1.0 &  \citet{Prieels:2014paa} \\
$\eta$ & $( 71 \pm 37 \pm 5 ) \times 10^{-3}$ & 0.0 &  \citet{Danneberg:2005xv} \\
$\eta^{\prime \prime}$ & $( 105 \pm 52 \pm 6 ) \times 10^{-3}$ & 0.0 &  \citet{Danneberg:2005xv} \\
$\alpha^{\prime}/A$ & $( -3.4 \pm 21.3 \pm 4.9 ) \times 10^{-3}$ & 0.0 & \citet{Danneberg:2005xv} \\
$\beta^{\prime}/A$ & $( -0.5 \pm 7.8 \pm 1.8 ) \times 10^{-3}$ & 0.0 & \citet{Danneberg:2005xv} \\
$\bar{\eta}$ & $0.02 \pm 0.08$ & 0.0 &  \citet{PDG2024} \\
 & & & \\
\hline
\end{tabular}
\label{tab:decayparameters}
\end{table}

\begin{table}[htb]
\centering \caption {Results for the 90\% confidence level experimental limits
on the bilinear quantities $Q_{\epsilon\mu}$ and the
coupling constants $g^{\gamma}_{\epsilon \mu}$
of the general current-current interaction
derived from the global analysis of the
muon decay parameters \citep{Hillairet2012}.
In $V$-$A$ theory the bilinear $Q_{LL}$ and 
coupling $g^V_{LL}$ are unity and
all other bilinears and coupling constants are exactly zero.
Note the limits on $|g^S_{LL}|$ and $|g^V_{LL}|$
are derived from inverse muon decay, 
$\nu_{\mu} e^- \rightarrow  \nu_e \mu^-$.}

\begin{tabular}{c|c|c|c}
\hline
 \\
 bilinears $Q_{\epsilon\mu}$ \\
\\
\hline
& & & \\
$ Q_{RR}  < 3.0\times10^{-4}$ & $ Q_{LR}  < 6.3\times10^{-4}$ & $ Q_{RL}  < 0.044$ & $ Q_{LL}  > 0.955$ \\
& & & \\
\hline
\\
couplings $g^{\gamma}_{\epsilon \mu}$  \\ 
 \\
\hline
& & & \\
$| g^S_{RR} | < 0.035$ & $| g^S_{LR} | < 0.050$ & 
$| g^S_{RL} | < 0.420$ & $| g^S_{LL} | < 0.550$ \\
$| g^V_{RR} | < 0.017$ & $| g^V_{LR} | < 0.023$ & 
$| g^V_{RL} | < 0.105$ & $| g^V_{LL} | > 0.960$ \\
$| g^T_{RR} | \equiv 0$ & $| g^T_{LR} | < 0.015$ & 
$| g^T_{RL} | < 0.105$ & $| g^T_{LL} | \equiv 0$ \\
\\
\hline
\end{tabular}
\label{tab:couplings}
\end{table}

\section{Muonium spectroscopy}

Muonium is a purely leptonic, hydrogen-like  bound state consisting of a positive muon and an electron. As such, the $\mu^+e^-$ bound state is free from complications arising from the finite size and internal structure of nuclei in ordinary atoms. It is unique as an electromagnetic bound system containing a second-generation particle and is well suited to precision spectroscopy.

The principal observables in muonium spectroscopy are the 1S ground-state hyperfine splitting, the 2S--2P Lamb shift, and the 1S--2S interval. These transitions provide sensitivity to the muon-electron charge ratio $q_e / q_\mu$, the  muon-electron mass ratio $m_\mu / m_e$, and the muon-proton magnetic moment ratio $\mu_\mu / \mu_p$. The ratio $m_\mu / m_e$ and mass $m_\mu$ are needed in the extraction of the muon's anomalous magnetic moment from muon g$-$2 experiments and the Fermi constant from muon lifetime experiments (Sec.\ \ref{sec:anomalousmoment}). Muonium spectroscopy also permits tests of bound-state quantum electrodynamics as well as sensitive searches for non-Standard-Model interactions involving muons.

\subsection{Muonium and its energy levels}

Atomic spectroscopy has played a central role in modern physics. 
The hydrogen spectrum shaped the  development of quantum mechanics, while the hydrogen Lamb shift motivated the development of quantum electrodynamics.
Fine and hyperfine structure in atomic spectra established the concepts of intrinsic spin and magnetic moments of electrons and nuclei.

However, the hydrogen nucleus has finite size and internal structure. This introduces uncertainties into theoretical calculations of atomic levels and limits comparisons to spectroscopic measurements. As proton structure became experimentally accessible, hydrogen spectroscopy increasingly focused on probing finite-size effects.

As a purely leptonic system, the $\mu^+ e^-$ atom is free of complications associated with the internal structure of its constituent particles and  represents an `ideal atom' for fundamental physics. Compared to positronium ($e^+ e^-$), it has a longer lifetime, a larger mass, and lacks an annihilation channel. The larger recoil and presence of annihilation make spectroscopic calculations for positronium more cumbersome and less accurate.

\begin{figure}
\begin{centering}
  \includegraphics[width=1.0\columnwidth]{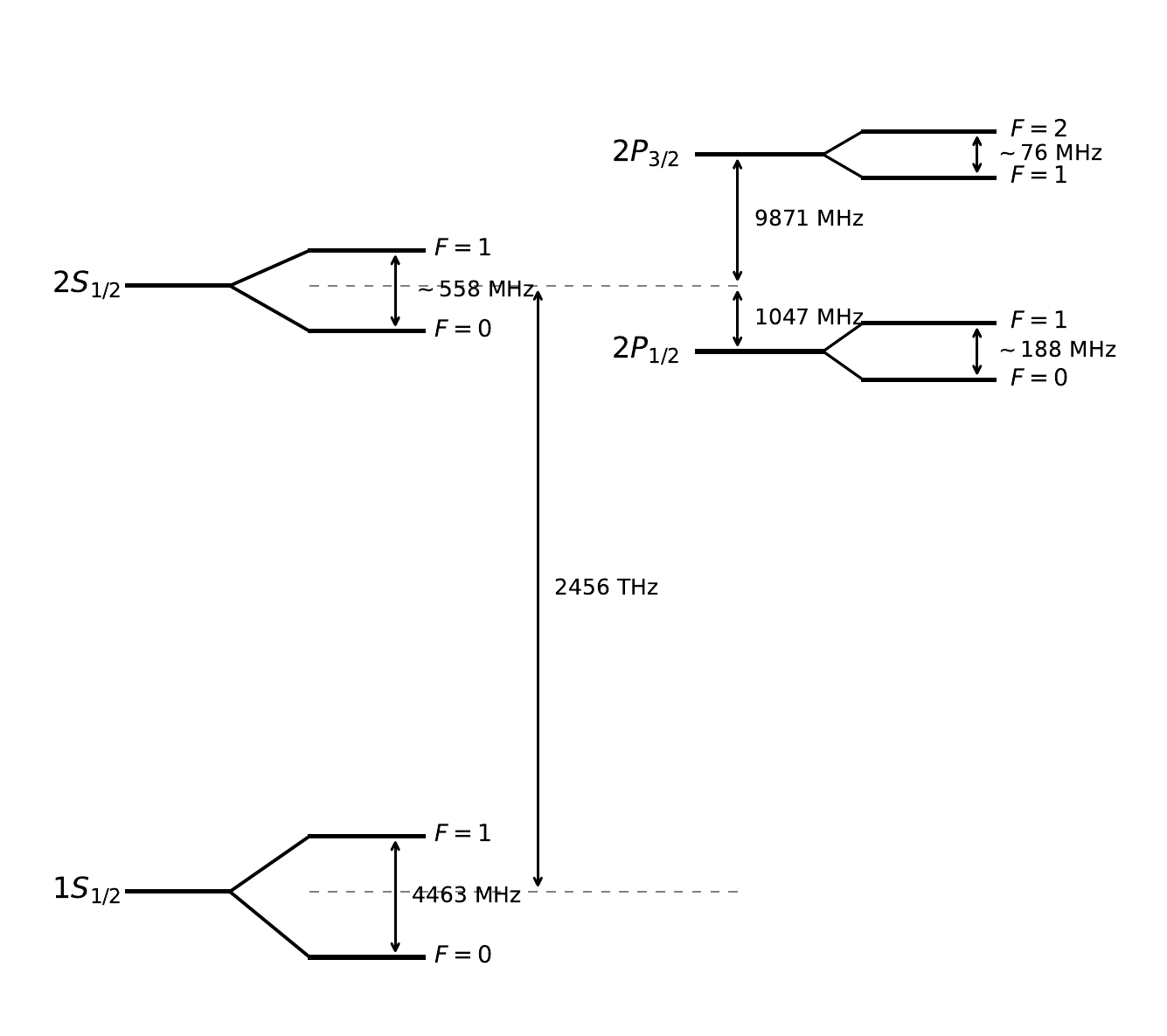}
  \caption{ \label{fig:muoniumenergylevels} Energy level diagram and transition frequencies for the low-lying $n=1, 2$ states of the muonium atom indicating the hyperfine structure, Lamb shift and 1S-2S interval.}
\end{centering}
\end{figure}

The observables most relevant to muonium spectroscopy are the 1S-2S interval, the 1S(F=1)-1S(F=0) hyperfine splitting, and the 2S$_{1/2}$-2P$_{1/2}$ Lamb shift
(see Fig.\ \ref{fig:muoniumenergylevels}). The 1S-2S interval is dominated by the long-range Coulomb binding energy of the $\mu^+ e^-$ system, while the hyperfine splitting is dominated by the short-distance spin-spin interaction of the electron and muon magnetic moments. The Lamb shift receives contributions from QED effects, including electron self-energy and vacuum polarization. 

The short-lived 2P state decays by electric-dipole photon emission with a 1.6~ns intrinsic lifetime (a natural linewidth of order 100~MHz). In contrast, the metastable 2S state decays by two-photon emission with a 0.12~s intrinsic lifetime (a natural linewidth of order 1~Hz). In muonium, however, the 2.2~$\mu$s muon lifetime imposes an effective linewidth of order 100~kHz on 1S-2S interval spectroscopy and 1S hyperfine spectroscopy .

At leading order, 
the corresponding transition frequencies 
$\nu_{1S-2S}$, $\Delta \nu_{\textrm {hfs}}$ and $ \nu_{2S-2P}$
scale as \citep{Eides2001}
\begin{equation}
\label{eqn: 1S2S}
\nu_{1S-2S} ~~\propto~~
m_r ~ \alpha^2
\end{equation}
\begin{equation}
\label{eqn: HFS}
\Delta \nu_{\textrm{hfs}} ~~\propto~~
g_e g_\mu ~ \alpha^4 ~ { m_r^3 \over m_e m_\mu}
\end{equation}
\begin{equation}
\label{eqn: Lamb}
\\nu_{2S-2P}
~~\propto~~
\alpha^5 ~ln\! {\small{1 \over \alpha^2}} ~ m_r 
\end{equation}

where $\alpha$ is the fine structure constant, $m_r = m_e m_\mu / (  m_e + m_{\mu} )$ is the muonium reduced mass, and $g_e$ and $g_\mu$ are the electron and muon g-factors. 
The presence of the product $g_e ~ g_\mu$ in Eqn.\ \ref{eqn: HFS} arises from the  proportionality of the hyperfine splitting to the product of the electron and muon  magnetic moments. The enhanced sensitivity of $\Delta \nu_{\textrm{hfs}}$ to $m_{\mu}$ %compared to $\nu_{2P-2S}$ and $\nu_{1S-2S}$  
follows from the explicit mass dependence of these magnetic moments.
An additional power of $\alpha^2$ in $\Delta \nu_{\textrm{hfs}}$  and $\nu_{2S-2P}$ arises from the short-distance character of these interactions and their dependence on the wavefunction  at the origin.

The atomic hyperfine splitting was first calculated by %Fermi in 1930 
\citet{Fermi1930}, and the hydrogen Lamb shift by %Bethe in 1947 
\citet{Bethe1947}.
Refinements to  Eqns. \ref{eqn: 1S2S}--\ref{eqn: Lamb} include relativistic, radiative and recoil corrections, as well as small contributions from weak and hadronic interactions (for example see \citet{Eides2001}). 
Modern bound-state calculations of muonium transition frequencies, 
yield QED uncertainties of about 20~kHz (2.4~ppt) for the 2455~THz 1S-2S interval \citep{Meyer2000}, 70~Hz (16~ppb) for the 4463~MHz hyperfine splitting \citep{Eides2019}, and 2~kHz (2~ppm) for the 1047~MHz Lamb shift \citep{PhysRevD.100.015010}. 

The leading uncertainty 
in calculating $\nu_{1S-2S}$ and $\Delta \nu_{\textrm{hfs}}$ arises from  the independent knowledge of the $m_\mu / m_{e}$ mass ratio.
At present, the measurement of $\Delta \nu_{hfs}$ is used for the  determination of the mass ratio $m_\mu / m_{e}$ \citep{Mohr:2012tt}.

\subsection{Hyperfine, Lamb shift, and 1S--2S measurements}

The formation of muonium, through electron capture by positive muons, was anticipated by analogy with formation of positronium in matter.
Following the discovery of muon polarization in pion decay and anisotropic electron emission in muon decay, it was  recognized that the muonium atom could be detected by its  Larmor precession in a  magnetic field.

Muonium was first observed in %1960 by Hughes et al.\ 
\citet{PhysRevLett.6.19,PhysRevLett.5.63}  via studies of positive muons stopped in argon gas.
They identified the characteristic precession of the muonium atoms in an external magnetic field via the resulting  time-varying angular distribution of the high-energy decay electrons.\footnote{In this
``weak field'' experiment 
using polarized muons 
and unpolarized electrons, 
the muonium ($F, m_{F}$)-states were populated in proportions
$(1,+1) = 1/2$, $(1,0) = 1/4$,
$(1,-1) = 0$, and $(0,0) = 1/4$.
Such experiments observe 
the precession of the 
$( F, M_{F}) = (1, +1)$ atoms.}

The formation and survival of muonium during $\mu^+$ slowing-down in matter involves a competition among energy-loss processes, muonium formation through electron capture, and muonium ionization through electron stripping. The heavier noble gases such as Ar and Kr have relatively low ionization potentials and were found to form muonium with high probability. The lighter noble gases such as He and Ne have relatively high ionization potentials and were found not to form muonium in appreciable quantities \citep{Stambaugh1974MuoniumNobleGases}. In solids, liquids and molecular gases, the chemical binding and reactions of muonium further complicate its formation and survival.

While this sensitivity of muonium to its environment is valuable for material research, it represents a complication for precision muonium spectroscopy. For example, collisions with surrounding atoms can shift transition frequencies in hyperfine spectroscopy and quench metastable 2S-state populations in Lamb-shift spectroscopy. 

Isolated muonium in vacuum promised a cleaner environment for muonium spectroscopy and drove the development of methods to produce muonium in vacuum. The beam-foil method used for in-vacuum 2S muonium production (Sec.\ \ref{sec:lamb shift}) was inspired by well developed beam-foil atomic spectroscopy, 
while the porous silica powder approach used for in-vacuum 1S muonium emission (Sec.\ \ref{sec:1s2s}) was inspired by similar use of silica-powders in positronium spectroscopy.

\subsubsection{Hyperfine structure}

\begin{figure}
\begin{centering}
  \includegraphics[width=1.0\columnwidth]{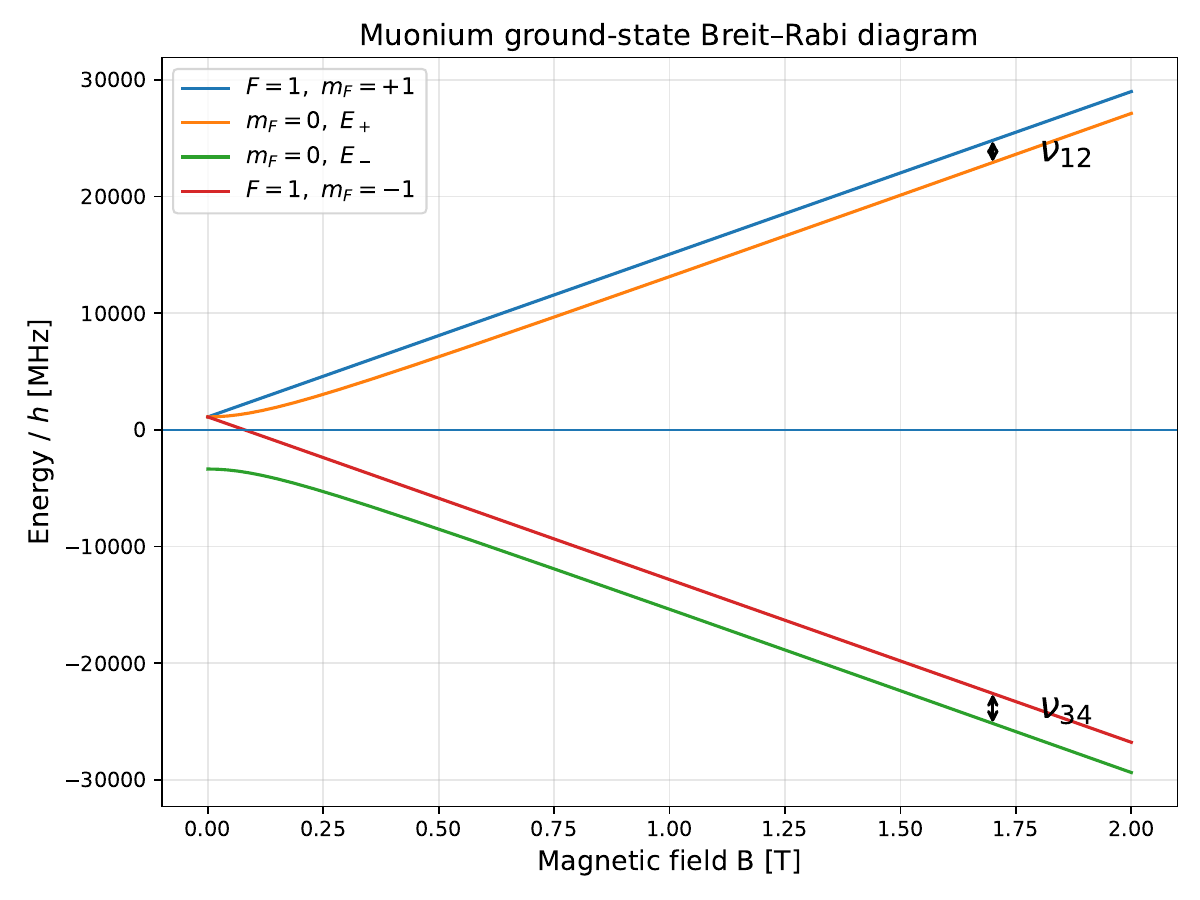}
  \caption{ \label{fg:BreitRabiDiagram} Breit-Rabi diagram of
the $F=0, m_F = 0$ and $F=1, m_F = -1, 0, +1$ energy levels of the muonium $1S$ hyperfine states
versus the applied magnetic magnetic field. Also indicated are the
two Zeeman transitions $\nu_{12}$ and $\nu_{34}$ that were measured in the double resonance hyperfine experiments.}
\end{centering}
\end{figure}

In an external magnetic field, the muonium energy levels undergo an additional Zeeman splitting. 
As shown in Fig.\ \ref{fg:BreitRabiDiagram},
the four $F= 0, 1$ ground-state hyperfine sublevels
experience energy shifts resulting from the interaction between the magnetic dipoles and the external field.
The level splitting as a function of the magnetic field strength is described by the Breit-Rabi
equations \citep{BreitRabi1931}

\begin{equation}
E(F=1, m_F=\pm1)
= \frac{h\,\Delta\nu_{\mathrm hfs}}{4} 
\pm \frac{h}{2}\left(\nu_e - \nu_\mu\right)
\label{eqn:linear states}
\end{equation}

\begin{equation}
E_{\pm}(m_F=0)
= -\frac{h\,\Delta\nu_{\mathrm hfs}}{4}
\pm \frac{h\,\Delta\nu_{\mathrm hfs}}{2}\sqrt{1 + x^2}
\label{eqn:mixed states}
\end{equation}

where $\nu_e = {g_e \mu_B B}/{h}$ is the electron Larmor frequency, $\nu_\mu = {g_\mu \mu_\mu B}/{h}$ is the muon Larmor frequency, and  $x(B) = ( \nu_e +\nu_\mu ) / {\Delta \nu_{\mathrm hfs} }$ is the dimensionless field parameter. 
The upper (lower) sign in Eqn.\ \ref{eqn:linear states} corresponds to $m_F = +1$ ($m_F = -1$), while the upper (lower) sign in Eqn.\ \ref{eqn:mixed states} denotes the higher (lower) eigenvalue of the mixed $m_F = 0$ states. The $E_{\pm}$ states evolve from the 
 $F = 1,~m_F = 0$ and  $F = 0,~m_F = 0$ states at zero magnetic field.

The $F=1, m_F=\pm1$ levels shift linearly with magnetic field, whereas the $m_F=0$ mixed-state levels exhibit a nonlinear dependence on magnetic field. 
The weak and strong field limits correspond to $x(B) \ll 1$ and  $x(B) \gg 1$, respectively.
In the weak-field limit, the hyperfine interaction dominates over the Zeeman interaction,
and $( F, m_F )$ are good quantum numbers.
In the strong-field limit, the Zeeman interaction  dominates over the hyperfine interaction,
and the electron and muon spin projections $( m_e, m_{\mu} )$ are good quantum numbers.

The early-1960s  measurements of 
the muonium hyperfine splitting were 
conducted by stopping muons in noble gases \citep{Hughes1966,PhysRevLett.13.202,Ehrlich1969}.  They used a static magnetic field to produce the  Zeeman splitting and a perpendicular microwave magnetic field  to drive transitions between the Zeeman levels.
When the microwave frequency matched a Zeeman transition frequency, the induced spin flips changed the decay-electron angular distribution and hence the observed count rate.
It was quickly realized that an extrapolation from  finite gas pressure to the zero-pressure limit was needed to correct for frequency shifts arising from collisions between muonium and gas atoms.
The experiments achieved  per-mille-level precision in  $\Delta \nu_{\mathrm{ hfs}}$.

In such strong-field experiments, extracting $\Delta \nu_{\mathrm{ hfs}}$ from a single Zeeman-frequency measurement requires both the Breit-Rabi equation and the prior knowledge of the muon magnetic moment.
This limitation led %DeVoe {\textit et al.}\ 
\citet{DeVoe:1971ph} to develop 
the double-resonance technique, involving the measurement of
two hyperfine transition frequencies in the strong-field region, denoted $\nu_{12}$ and  $\nu_{34}$ in Fig.\ \ref{fg:BreitRabiDiagram}. The frequency sum $\nu_{12}+\nu_{34}$ determines the muonium hyperfine splitting, while the frequency difference $\nu_{12} - \nu_{34}$ determines the muon magnetic moment. 
To minimize sensitivity to magnetic-field variations, the two frequencies were measured near the ``magic field"  where $d ( \nu_{12} + \nu_{34} ) / dB = 0$.
This technique yielded a 2~ppm determination of $\Delta \nu_{\mathrm{hfs}}$ and a 4~ppm determination of $\mu_{\mu} / \mu_p$. 

Over the following decades, modern meson factories, surface muon beams, improved magnetic field metrology, and higher-order QED calculations
all contributed to renewed efforts to measure the hyperfine splitting.
%In 1984,  Mariam {\textit et al.}\ 
\citet{Mariam1982} reached a 36~ppb precision in $\Delta \nu_{\mathrm{hfs}}$ and 360~ppb precision in $\mu_{\mu} / \mu_p$.
%In 1999 Liu {\textit et al.}\ 
\citet{Liu1999} improved these measurements to
11~ppb precision in $\Delta \nu_{\mathrm{hfs}}$ and 27~ppb precision in $\mu_{\mu} / \mu_p$.
Notably, Liu {\textit et al.}\ employed a time-structured muon beam with distinct muon accumulation and microwave interrogation periods, which enabled a longer microwave interaction time and thereby a narrower microwave resonance. 

A new experiment, MuSEUM \citep{Kanda2021}, is under development at J-PARC. The experiment will use the same approach as the Liu {\textit et al.}\  experiment: a double-resonance measurement of two Zeeman frequencies in the high-field region using a time-structured muon beam. 
By exploiting the intense muon beam at J-PARC,
MuSEUM aims to increase the number of recorded decay electrons by two orders of magnitude.
\citet{Kanda2021} reported a zero-field demonstrator measurement of $\Delta_{\textit{hfs}}$, 
along with plans for a zero-field measurement with 12~ppb precision and a high-field measurement with 1.2~ppb precision.

\subsubsection{Lamb shift}
\label{sec:lamb shift}

Muonium Lamb-shift experiments involve three main steps: production of 2S muonium in vacuum using beam-foil techniques,  excitation of 2S-2P transitions using radio-frequency electric fields, and detection of surviving 2S muonium through a combination of Stark quenching and Lyman-$\alpha$ detection.
The principal challenges are the low production of metastable 2S muonium and low efficiency for Lyman-$\alpha$ detection.

The first successful measurements of the muonium Lamb shift were made during the 1980s  by %Oram {\textit et al.}\ 
\citet{Oram1984} at TRIUMF and %Woodle {\textit et al.}\ 
\citet{Woodle1990} at LAMPF.
Both experiments used the beam-foil technique, in which low-momentum muons traversed a thin foil and emerged with a small population of 2S muonium. The muonium passed through an RF electric-field region, where 2S-2P transitions could be induced, followed by a static electric-field region,
where surviving 2S populations were detected by Stark quenching and Lyman-$\alpha$ photon counting. The static field caused 2S-2P Stark mixing and thereby Lyman-$\alpha$ emission.

On-resonance, the excitation of 2S-2P transitions in the RF-field region, yielded a lower downstream Lyman-$\alpha$ rate.
Off-resonance, the absence of 2S-2P transitions in the RF-field region, yielded a higher  downstream Lyman-$\alpha$ rate.
In practice, the experiments measured the
surviving 2S population with RF power on and off at multiple frequencies spanning the $2S-2P$ resonance manifold (see Fig.\ \ref{fig:muoniumenergylevels}).\footnote{The $2S-2P$ Lamb shift has a hyperfine sub-structure corresponding to different ($F$, $m_F$) sub-state transitions.}  
The measurements achieved uncertainties  of 1-2\% in $\nu_{\mathrm{2S–2P}}$ and were statistically limited by the low 2S-muonium production efficiency and the  low Lyman-${\alpha}$ detection efficiency.  

At the Paul Scherrer Institute, %Ohayon {\textit et al}.\ 
\citet{Ohayon2022} have recently improved the Lamb shift measurement by about an order of magnitude. This advance was enabled by the PSI low-energy muon (LEM) beam line,
which substantially improved the limited production of metastable muonium in earlier foil-target experiments. The LEM facility provides a tunable 1-30 keV muon beam  by first stopping a surface muon beam in a neon moderator and then electrostatically accelerating the re-emitted epithermal muons. The experiment also employed large-area microchannel-plate detectors for higher-efficiency Lyman-$\alpha$ detection.
Ohayon {\textit et al}.\ obtained $\nu_{2S-2P} =1047.2(2.5)$~MHz, with the uncertainty dominated by the statistical error in the lineshape fit to the $\sim$100~MHz-width resonance structure.

\subsubsection{1S$-$2S interval}
\label{sec:1s2s}

The 1S–2S interval is measured through a sequence of two-photon excitation of 1S-2S transitions,  photo-ionization of 2S muonium, and  acceleration and detection of liberated $\mu^+$s. The principal challenges include the production of 1S muonium in vacuum, the requirement of two-photon Doppler-free 1S-2S excitation, and the efficient detection of the small 2S population.

The first evidence of 1S-2S excitations in muonium was obtained by %Chu {\textit et al.}\ 
\citet{Chu1988Muonium1S2S} at KEK, Japan in 1988. 
The authors directed a pulsed, 28~MeV/c, surface $\mu^+$ beam onto a SiO$_2$ target, from which thermal muonium  diffused into the nearby vacuum region. Pulsed 244~nm laser light was directed into the region adjacent to the SiO$_2$ target. A multiplass mirror arrangement increased the laser intensity and provided the counter-propagating beams necessary for two-photon Doppler-free 1S-2S excitation.
A 770~ns delay between the beam arrival and the laser pulse allowed for the muonium diffusion from the target into the vacuum.
Ionization of the 2S muonium and liberation of the bound $\mu^+$, was accomplished by absorption of a third 244~nm photon. Charged particles from the target region were accelerated by a 4~kV potential and guided to a microchannel-plate detector. The ionized-$\mu^+$ signal of 2S muonium was identified by its characteristic time of flight between the target and detector. The authors reported six candidate events corresponding to 1S-2S  excitation, with an estimated background of 0.4 events.  

Two subsequent measurements by %Maas {\textit et al.}\ 
\citet{Maas1994Muonium1S2S} and %Meyer {\textit et al.}\ 
\citet{Meyer2000} developed the Chu {\textit et al.}\ method into precision 1S-2S spectroscopy. Both experiments used the same basic approach: production of thermal muonium in vacuum from a SiO$_2$  target, Doppler-free two-photon excitation with counter-propagating 244 nm laser light, photoionization of the excited 2S muonium, and detection of the liberated $\mu^+$.
The improvements came from higher statistics, improved laser frequency control, and better characterization of systematic effects. The final RAL measurement by Meyer et al. reported
$\nu_{1S-2S} = 2455528941.0(9.8)$ MHz, corresponding to a precision of about 5 ppb. The dominant systematic uncertainties arose from the frequency chirp of the pulsed laser and residual Doppler effects from imperfect cancellation in the counter-propagating geometry.

The Mu-MASS collaboration \citep{Cortinovis2023} aims to significantly improve the precision measurement of the muonium 
1S–2S interval. Using the LEM beamline at PSI and advances in continuous-wave UV laser technology, the experiment is aiming to measure $\nu_{1S-2S}$ to  10~kHz precision (a three-orders-of-magnitude improvement over the Meyer {\textit et al.}\ result). It will address the limitations of the laser chirp and the residual Doppler broadening using a narrow-linewidth continuous-wave UV laser and a reduced muonium phase space. 

\begin{table}[ht]
\centering
\small
\begin{tabularx}{0.8\textwidth}{l|c|c|X}
\hline
& & & \\
Transition & Year & Result (MHz) & Reference \\
& & & \\
\hline
& & & \\
HFS & 1964 &  4463.15(6) & %Cleland {\textit et al.} 
\citet{PhysRevLett.13.202} \\  
HFS & 1969 & 4463.317(21) & %Ehrlich {\textit et al.}\ 
\citet{Ehrlich1969}  \\ % Ar, Kr, high field, 
HFS & 1970 & 4463.302~2(89) & %DeVoe {\textit et al.}\ 
\citet{DeVoe:1971ph} \\ 
HFS & 1971 & 4463.301~2(23) & %Favart {\textit et al.}\ 
\citet{PhysRevLett.27.1336} \\ % Ar/Kr, zero field, Ramsey resonance (two successive coherent microwave pulses lengths tau, separation T)
HFS & 1973 & 4463.304~0(18)  & %Kobrak {\textit et al.}\ 
\citet{Kobrak1973} \\ % SREL, low pressure Ar, Favart method
HFS &  1975 &4463.302~2(14) & %Casperson {\it et al.} 
\citet{Casperson1975} \\ % LAMPF, low pressure Ar,Kr, zero  field, line-narrowing
HFS & 1982 & 4463.302~88(16) & %Mariam {\textit et al.}\  
\citet{Mariam1982} \\ %Phys Rev Lett 49, 993 (1982) \\
HFS & 1999 & 4463.302~765(53)  &  %Liu {\textit et al.}\ 
\citet{Liu1999} \\%Phys.\ Rev.\ Lett.\ 82, 711 (1999) \\
HFS & 2021 & 4463.302(4)  & %Kanda {\textit et al.} 
\citet{Kanda2021} \\
 & & & \\
Lamb shift & 1984 &  $1070^{+12}_{-15}$& %Oram {\textit et al.}\ 
\citet{Oram1984} \\ %, Phys. Rev. Lett. 52, 910(1984) \\
Lamb shift & 1990 & $1042^{+21}_{-23}$ & %Woodle {\textit et al.} 
\citet{Woodle1990} \\ % Phys. Rev. A 41, 93(1990)  \\
Lamb shift & 2022 & 1047.2(2.5) & %Ohayon {\textit et al.}\ 
\citet{Ohayon2022} \\ %  Phys. Rev. Lett. 128, 011802 \\
& & & \\
1S--2S & 1988 & 2 455 528 912(184) &  %Chu {\textit et al.}\ 
\citet{Chu1988Muonium1S2S,Danzmann:1989}  \\
1S--2S & 1994 &  2 455 529 002(33)(46)  & %Maas {\textit et al.}\ 
\citet{Maas1994Muonium1S2S} \\
1S--2S & 2000 & 2 455 528 941.0(9.8) & %Meyer {\textit et al.}\ 
\citet{Meyer2000} \\ %Phys.\ Rev.\ Lett.\ 84, 1136 (2000) \\
& & & \\
\hline
\end{tabularx}

\caption{Precision measurements of the hyperfine splitting, Lamb shift, and 1S-2S interval frequencies  in muonium spectroscopy. For the Chu {\textit et al.}\ 1S-2S transition frequency, we list the value quoted in Maas {\textit et al.}\ after the application of a correction for the hyperfine splitting \citet{Danzmann:1989}.}
\label{tab:muoniumspectroscopy}
\end{table}

\subsection{The LAMPF double-resonance experiment}

%Liu {\textit et al.}\ 
\citet{Liu1999} measured the muonium hyperfine splitting 
using the double-resonance technique. In this method, the hyperfine frequency $\Delta \nu_{\mathrm{hfs}}$ is obtained from the
sum of the two high-field spin-flip transition frequencies 
$\nu_{12}$ and $\nu_{34}$
(Fig.\ \ref{fg:BreitRabiDiagram}).
One challenge was accumulating the necessary decay-electron statistics over microwave resonance scans. Additional challenges were the precise determination of the magnetic field -- including its spatial uniformity, temporal stability, and absolute calibration -- as well as 
measurement and extrapolation of pressure-dependent shifts of the microwave resonances in the gas target.

The experiment used a high-rate, 
nearly 100\% polarized, surface muon beam
derived from the 120~Hz 
pulsed proton beam at LAMPF. 
An electrostatic kicker was used 
to produce cycles consisting of 4~$\mu$s muon accumulation periods 
followed by 10~$\mu$s  microwave interrogation periods.
\begin{figure}
\begin{centering} \includegraphics[width=0.8\columnwidth]{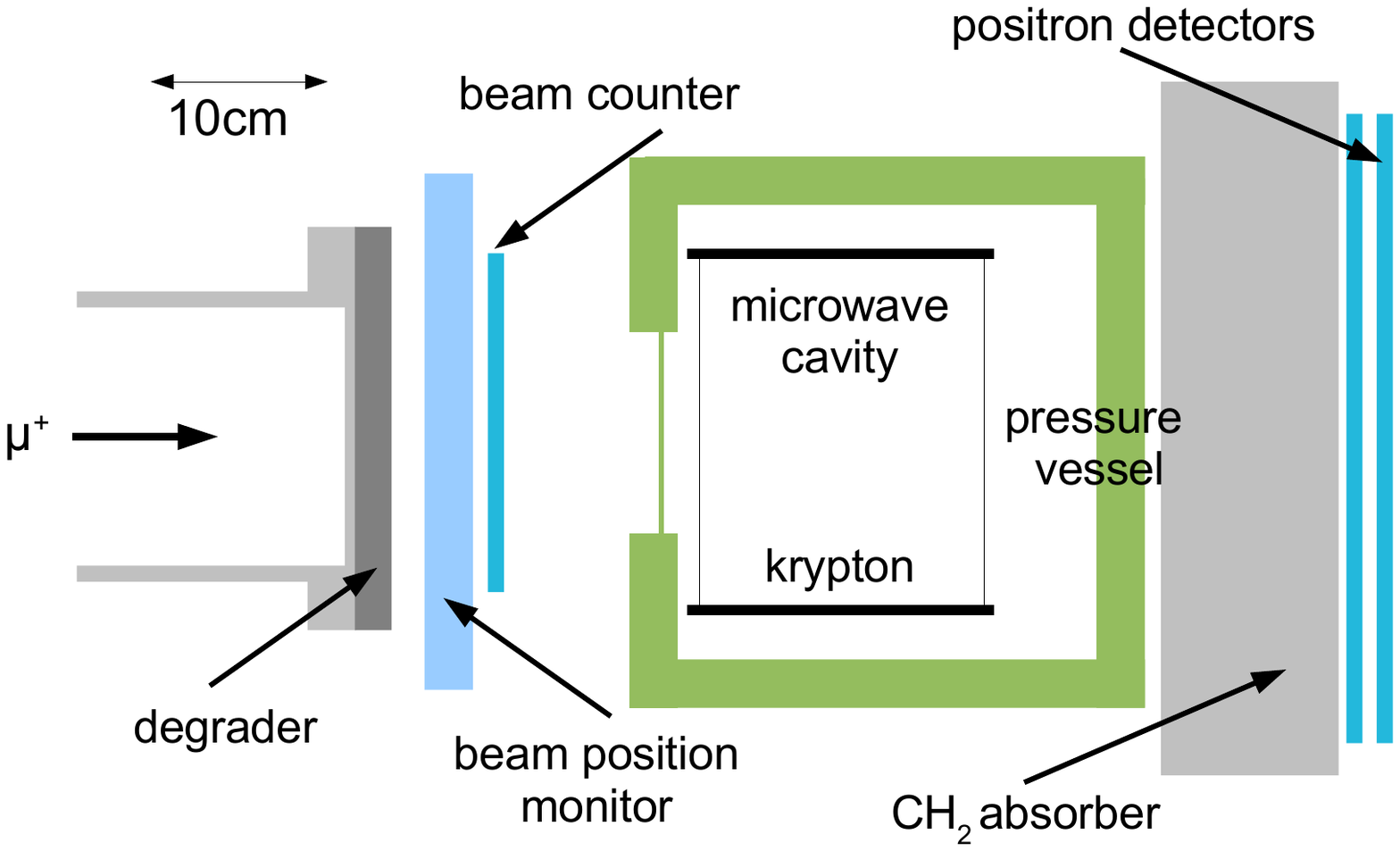}
  \caption{ \label{fg:HFSexpt} Schematic diagram of the Liu {\textit et al.}\  HFS experiment using a time-structured muon beam and double-resonance technique. Indicated are the muon beam, beam counters, krypton target, microwave cavity
and downstream absorber and positron detector. Reproduced from \citet{GorringeHertzog2015}. }
\end{centering}
\end{figure}

Fig.\ \ref{fg:HFSexpt} shows the experimental apparatus, including the incident muon beam, krypton gas target, microwave cavity, and positron detector package. The arrangement was located in a large-bore, high-uniformity electromagnet with a 1.7~T field.  
The polarized muons were stopped krypton gas and formed polarized muonium in the $( m_e$, $m_{\mu} )$ = $( \tfrac{1}{2}, -\tfrac{1}{2})$  and $( -\tfrac{1}{2}, -\tfrac{1}{2})$  substates through electron capture.
The microwave cavity was designed to resonate at both the  $\nu_{12} \simeq 1897.5$~MHz  and
$\nu_{34} \simeq 2565.8$~MHz frequencies.
Proton NMR magnetometry, employing fixed and movable probes, provided precise measurement of the static magnetic field.
A polyethylene absorber and double-layered scintillator telescope were used to detect the energetic, downstream, decay electrons.

When tuned to either $\nu_{12}$ or $\nu_{34}$,
the microwave field induces muon spin-flip transitions and changes the angular distribution and measured rate of decay electrons.
The accumulation-interrogation cycle  allowed the measurement of decay electrons from muonium atoms that experience extended interactions with the microwave field.
This timing scheme narrowed the 
resonance lineshape 
by about a factor of three.

The decay electron rate was measured with microwave power on and off during both microwave frequency scans at fixed magnetic field and magnetic-field scans at fixed microwave frequency. Near resonance, the spin-flip transitions  caused a pronounced increase in decay electron rates.

The fitted lineshapes using models incorporated the muon stopping distribution,
the static field distribution, 
the microwave power distribution, 
and the positron detection efficiency.
Data were collected at two gas pressures and extrapolated to zero pressure  to account for pressure-dependent shifts of the resonance frequencies arising from the muonium-krypton collisions.

The measured resonance curves determined the two high-field spin-flip frequencies with
precisions of 17-18~ppb, with the statistical errors
on the resonance curves dominating the overall uncertainty.
Using the Breit-Rabi Eqns.\ \ref{eqn:mixed states} and the measurements values of $\nu_{12}$ and $\nu_{34}$,  a hyperfine frequency of $\Delta \nu_{\mathrm{hfs}}$ $=$ $4~463~302~765(53)$~Hz (12~ppb) and a
magnetic moment ratio of $\mu_{\mu} / \mu_p$ $=$ $3.183~345~13(39)$ (120~ppb)
were extracted.

\subsection{Status and implications}

Muonium spectroscopy has yielded measurements of the hyperfine splitting, 1S$–$2S interval, and Lamb shift with precisions at the 10-ppb, few-ppb, and part-per-mille levels, respectively (Table \ref{tab:muoniumspectroscopy}).  Ongoing efforts to improve the precision include an order-of-magnitude improvement in $\Delta \nu_{\mathrm{hfs}}$ (MuSEUM at J-PARC, \citet{Kanda2021}), and a three-order-of-magnitude improvement in $ \nu_{\mathrm{1S-2S}}$ (Mu-MASS at PSI, \citet{Cortinovis2023}). Additionally, the %Liu {\textit et al.}\ 
\citet{Liu1999} double-resonance experiment yielded a direct measurement of the muon-proton magnetic moment ratio with 120~ppb precision.

The different transitions have distinctive sensitivities to bound-state QED, fundamental constants, and non-Standard-Model interactions. The hyperfine splitting is particularly sensitive to the muon g-factor $g_\mu$ and the electron-muon mass ratio $m_\mu / m_e$ through the muon-electron magnetic interaction. The 1S$-$2S transition offers a $g_\mu$-independent sensitivity to the mass ratio $m_\mu / m_e$. The Lamb shift directly probes the quantum fluctuations arising from electron self-energy and vacuum polarization effects.

\begin{table}[h]
\centering
\caption{Values of the muon-electron mass ratio $m_\mu / m_e$ extracted from measurements of the $\mu_\mu / \mu_p$  magnetic moment ratio, 1S hyperfine splitting frequency $\Delta \nu_{\mathrm{hfs}}$, and 1S-2S interval frequency in muonium spectroscopy.}
\label{table:mmu over me}
\begin{tabularx}{0.6\textwidth}{l|c|l}

\hline
 & & \\ 
Observable & Mass ratio $m_\mu/m_e$  & Reference \\
 & & \\ 
\hline
 & & \\ 
$\mu_\mu\!/\!\mu_p$ 
& $206.768~277(24)$ 
& %Liu {\textit et al.} 
\citet{Liu1999} \\ % Phys. Rev. Lett. 82, 711 (1999) \\

$\Delta \nu_{\mathrm{hfs}}$ 
& $206.768~281(4)$
& %Liu {\textit et al.}\ 
\citet{Liu1999,Eides2019} \\ %Rev. Mod. Phys. 93, 025010 (2021) \\

$\nu_{1S-2S}$ 
& $206.768~38(17)$ 
& %Meyer {\textit et al.}\ 
\citet{Meyer2000} \\ %Phys. Rev. Lett. 84, 1136 (2000) \\
 & & \\ 
\hline
\end{tabularx}
\end{table}

Table \ref{table:mmu over me} lists the most recent values of the $m_\mu / m_e$ ratio obtained from muonium spectroscopy. The derivations of $m_\mu / m_e$ from $\Delta \nu_{\mathrm{hfs}}$ and $\nu_{1S-2S}$ require theoretical  calculations of the corresponding bound-state QED  energy levels. The extractions of  $m_\mu / m_e$ from $\mu_\mu$\!/\!$\mu_p$  and $\Delta \nu_{\mathrm{hfs}}$  additionally require the experimental determination of the muon g-factor $g_\mu = 2 ( 1 + a_\mu)$, where  $a_\mu$ is obtained from muon g-2 experiments (Sec.\ \ref{sec:anomalousmoment}).
The resulting uncertainties in $m_\mu / m_e$ from $\Delta \nu_{\mathrm{hfs}}$, $\nu_{1S-2S}$, and $\mu_\mu$\! / \!$\mu_p$ are about 20~ppb, 820~ppb, and 120~ppb, respectively.

Precise knowledge of $m_\mu / m_e$ 
is requirewd for extracting the Fermi constant $G_F$ from the muon lifetime measurements (Sec.\ \ref{sec:muonlifetime}) and the muon anomalous magnetic moment $a_\mu$ from the muon g$-$2 experiments (Sec.\ \ref{sec:anomalousmoment}).
The use of $m_\mu / m_e$ in determining $a_\mu$, together with use of $g_\mu = 2(1 + a_\mu)$ in extracting $m_\mu / m_e$ from hyperfine measurements, may appear circular. In practice, however, no inconsistency arises because $a_\mu$ contributes to the  hyperfine splitting only at a part-per-mille level. Consequently, the extractions of $a_\mu$ and $m_\mu / m_e$ are effectively decoupled at current experimental precision.

The planned MuSEUM measurement \citep{Kanda2021} of $\Delta \nu_{\mathrm{hfs}}$ at J-PARC, and Mu-MASS measurement \citep{Cortinovis2023} of $\nu_{\mathrm{1S-2S}}$ at PSI, offer opportunities for order-of-magnitude improvements in determining $m_\mu / m_e$. Moreover, the combination of precision measurements of $\Delta \nu_{\mathrm{hfs}}$ and $\nu_{\mathrm{1S-2S}}$ could provide a theory-dependent extraction of $a_\mu$, complementary to direct measurements from muon g$-$2 experiments.

The muonium 1S$-$2S transition also provides a sensitive test of muon-electron charge equality. At leading order the 1S$-$2S frequency is proportional to $( q_\mu q_e)^2$. Using an independent determination of $m_\mu / m_e$, the comparison between measured and calculated 1S$-$2S frequencies constrains the charge ratio, $q_\mu / q_e$. 
From their few-ppb measurement of the 1S$-$2S frequency, %Meyer {\textit et al.}\ 
\citet{Meyer2000} derived a 1~ppb limit on deviations from muon-electron charge equality.

\section{Muon anomalous magnetic moment}
\label{sec:anomalousmoment}

The muon anomalous magnetic moment $a_\mu$ arises from quantum fluctuations involving quarks, leptons and their electromagnetic, weak, and strong interactions. Its precise experimental measurement and theoretical calculation yield a stringent test of the Standard Model.  Its broad sensitivity to virtual effects from physics beyond the Standard Model complements more targeted probes, including electric dipole moments and charged-lepton flavor violation. 

Muon storage-ring measurements of $a_\mu$  have a long history, beginning in the 1960s at CERN, continuing in the 1990s at BNL, and most recently at Fermilab \citep{FarleyPicasso1979,Miller2007}. The CERN experiments pioneered key experimental techniques, including a co-magnetometer approach, in which $a_\mu$ is determined from the ratio of muon and proton precession frequencies, and magic momentum operation, in which $a_\mu$ is measured in a uniform magnetic field with electrostatic focusing.

The greatest theoretical challenge 
in determining $a_\mu$ is the contribution arising from hadronic vacuum fluctuations. For many decades, it was estimated using a data-driven approach based on  $e^+ e^- \rightarrow$ hadrons cross section measurements, while more recently it has been calculated using ab initio lattice-QCD methods.

\subsection{Theory of the magnetic anomaly}

The concept of intrinsic spin emerged from attempts to explain several puzzling features of atomic structure and spectral lines. In 1924, Pauli introduced an additional two-valued quantum number of non-classical origin to reproduce the observed atomic shell structure. %\footnote{Together, the two-valued quantum number and Pauli exclusion principle was able to reproduce the noble-gas closed shells with $Z = 2, 8, 18, $ {\textit etc}.}  
Soon afterwards, Goudsmit and Uhlenbeck interpreted this quantum number as the two allowed projections of an electron's intrinsic angular momentum.

For a charged particle, intrinsic spin implies an associated magnetic dipole moment. The spin $\vec{s}$ and magnetic dipole moment $\vec{\mu}$ are related by 
\begin{equation}
\vec{\mu} = g {q \over 2m } \vec{s} 
\end{equation}
where $q$ and $m$ are the particle's charge and mass, and  $g$ is a dimensionless proportionality constant known as the 
g factor. The electron's magnetic moment provided a natural explanation for the Stern-Gerlach splitting
of atomic beams
and the anomalous Zeeman splitting of spectral lines. 

Measurements indicated, however, that  $g \simeq 2$, in contrast to the classical expectation $g = 1$ for a rotating charge distribution.
In 1928, Dirac introduced a 
relativistic wave equation for the electron that incorporated spin and reproduced the fine structure of atomic hydrogen. 
It also, predicted, unexpectedly, 
$g = 2$ for a free, point-like spin-1/2 particle.

Dirac’s theory represented a major step toward a quantum field theory of electromagnetism, in which both electrons and photons are described as quantized fields. Such a theory gives rise to new effects through processes in which electrons emit and reabsorb virtual photons (self-energy), and in which vacuum fluctuations create virtual $e^+$$e^-$ pairs (vacuum polarization). 
Progress in quantum electrodynamics stalled, however when early attempts to calculate these effects yielded divergent results.

In 1947, two experimental discoveries revitalized theoretical work. First, %Lamb and Retherford 
\citet{LambRetherford1947} measured a splitting between the hydrogen 
2S$_{1/2}$ and 2P$_{1/2}$
levels, which are degenerate in Dirac theory. Second, %Kusch and Foley 
\citet{KuschFoley1948} found from Zeeman measurements of alkali atoms that the electron g-factor was about 0.1\% larger than the Dirac value. 
These unexpected results motivated 
the development of renormalization procedures for  
handling divergences. 
%Schwinger's 1948 calculation 
The \citet{Schwinger1948} calculation of the
electron's magnetic anomaly $a_e = (g-2)/2 \simeq \alpha / 2\pi = 0.0012$
represented a landmark success of renormalized QED.

By the mid 1960s, the first- and second-order QED corrections to the electron and muon magnetic anomalies had been computed \citep{Sommerfield1957,Elend1966}.
The leading $(\alpha /  \pi) $ term arises from one-loop virtual-photon diagrams and contributes identically to electrons and muons. 
The  $( \alpha /  \pi )^2$ corrections include additional virtual processes 
with mass-dependent effects, leading to small differences between electron and muon anomalies.

It was recognized that contributions to magnetic anomalies from virtual heavy particles scale as $( m_\ell / M )^2$, where $m_\ell$ and $M$ are the masses of the lepton and heavy particle, respectively \citep{Fujikawa1972,JackiwWeinberg1972}. Consequently, such effects are enhanced by factors of $( m_\mu / m_e )^2 ~\simeq~  $43,000 for muons compared to electrons.
This enhancement makes measurements of $a_\mu$ a powerful probe of heavy virtual particles.

At the time, the known heavy particles included pions and kaons, protons and neutrons, and several strange baryons.
Based on the $( m_\ell / M )^2$ scaling, the pion was expected to dominate the hadronic contributions to $a_\mu$ through vacuum fluctuations into virtual $\pi^+ \pi^-$ pairs. However, the strong interaction was poorly understood, and computing the $\pi^+ \pi^-$ vacuum polarization from first principles appeared intractable.

This difficulty was overcome
through two developments.
Experimentally, the first measurements of $e^+ e^- \rightarrow $ hadronic cross sections became available from electron-positron colliders.
Theoretically, it was recognized that the hadronic vacuum polarization contribution could be related to these data \citep{BouchiatMichel1961,GourdinDeRafael1969}.
This enabled a data-driven determination of the leading-order hadronic vacuum polarization, $a^{had,LO}_\mu$, through a dispersion integral over the measured $e^+e^- \rightarrow$ hadron cross sections. The largest contribution to $a^{had,LO}_\mu$ arises from the 2$\pi$-channel in the  
$\rho$-resonance region. 

From pioneering work at Frascati and Novosibirsk, which identified
the $\rho$, $\omega$ and $\phi$ resonances contributions, to modern high-luminosity $e^+e^-$ colliders, which have performed high-statistics energy-scans and initial-state radiation measurements, the low-energy hadronic cross sections have been mapped in key channels 
to the sub-percent level.   However, tensions among modern $e^+e^-$ datasets  presently limit the application of this approach to determining $a^{had,LO}_\mu$ \citep{Aoyama2020,Aliberti2025}.\footnote{Specifically, significant tensions exist between the high-statistics  BaBar \citep{BaBar2012},
KLOE \citep{KLOE2013}, BESIII \citep{BESIII2016}, SND \citep{SND2020update}, and CMD-3 \citep{CMD32024} $\pi \pi$-channel datasets.}

\begin{figure}[h!]
  \includegraphics[width=\columnwidth]{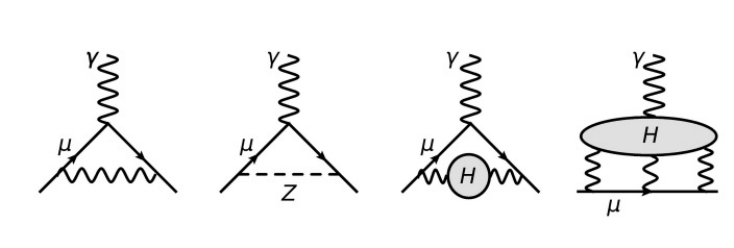}
  \caption{Feynman diagrams of representative SM contributions to the muon anomalous magnetic moment. From left to right: the first-order QED contribution, first-order $Z^o$ weak contribution,  leading-order hadronic (H) vacuum polarization contribution, and higher-order  hadronic light-by-light contribution.
}
\label{fig:g-2feynmandiagrams}
\end{figure}

Interest in $a_\mu$ broadened in the 1960s and 1970s as the Standard Model took shape. The weak interaction, mediated by heavy W and Z bosons, gives rise to electroweak contributions to $a_\mu$, while the strong interaction, described by quantum chromodynamics (QCD), lead to hadronic contributions including  $q \bar{q}$-pair vacuum polarization (see Fig.\ \ref{fig:g-2feynmandiagrams}).
The motivation for measuring $a_\mu$ thus shifted from a test of the QED prediction to a comprehensive test of the Standard Model prediction
\begin{equation}
a^{SM}
= a^{QED}_\mu
+ a^{EW}_\mu
+ a^{had}_\mu 
\end{equation}
incorporating electromagnetic, weak, and hadronic effects.

A complementary, ab initio method for calculating  $a^{had,LO}_\mu$ is provided by lattice QCD. Within this approach, quark and gluon fields are represented on a discrete space–time lattice.
The finite lattice spacing provides an ultraviolet cutoff that regularizes the ultraviolet divergences and enables nonperturbative calculations through large-scale numerical simulations.
Advances in lattice QCD algorithms and increasing computational power have driven rapid progress in ab initio determinations of $a^{had,LO}_\mu$ \citep{Blum2018,Borsanyi2021}.

The evolution of approaches to hadronic contributions -- from data-driven methods to ab initio calculations -- is reflected in the work of the Muon g-2 Theory Initiative \citep{Aoyama2020,Aliberti2025}.
The Initiative was formed to provide a community-based SM prediction for $a^{SM}_\mu$. Its first report was published in 2020 and an updated evaluation was released in 2025. These papers were produced during a period of increasing tension among $e^+e^-$ datasets and steadily improving precision in lattice-QCD calculations. The 2020 estimate of $a_\mu^{SM}$ was based on the data-driven approach, while the 2025 update used a lattice-QCD determination of the leading-order hadronic vacuum polarization.

\subsection{Principles of g-2 measurements}

Parity violation in weak interactions provides the  foundation for the measurement
of the muon g-factor $g_\mu$  and its anomaly $a_\mu$. It gives rise to highlyy-polarized muons from $\pi \rightarrow \mu \nu$  decay and directionally correlated decay electrons from $\mu \rightarrow e \nu {\bar{\nu}}$ decay.  These features 
enable the production and analysis of ensembles of polarized muons.

The muon g-factor was first measured as a by-product of the %Garwin {\textit et al.}\ 
\citet{Garwin:1957hc} parity-violation experiment. A beam of polarized muons was stopped in a transverse magnetic field, and the muon spin precession inferred from the decay-electron angular distribution. From the precession frequency, the authors obtained $g_\mu = 2.0 \pm 0.2$, consistent with the muon's identification as a point-like, spin-1/2 Dirac particle.

Over the next years, further spin-precession measurements both refined the method and increased the statistics \citep{Cassels1957,Garwin1960}.
These efforts yielded a precision  $g =  2.00244 \pm 0.00016$, sufficient to test the lepton-independent ($\alpha /  \pi$) correction to the muon g-factor at the 10\% level. The results furnished further evidence that the muon behaves as a heavy electron.

An important advance was the 1959 formulation 
of the Bargmann–Michel–Telegdi equation \citep{Bargmann1959} describing  the spin motion of a relativistic spin-1/2 particle in an electromagnetic field.
In a uniform perpendicular magnetic field, the spin precession frequency $\omega_s$ and cyclotron frequency $\omega_c$ differ by the anomalous precession frequency 
\begin{equation}
\vec{\omega_a} = \vec{\omega_s} - \vec{\omega_c} = 
~-~ \left[ ( a + {1  \over \gamma} ) {q B \over m} \right] ~-~  \left[ - {q \vec{B} \over \gamma m} \right] =
- a {q \vec{B} \over m} 
\end{equation}
which is proportional to the magnetic anomaly  $a$.\footnote{Note the spin precession frequency is the net difference between the Lamor frequency and the Thomas frequency $\omega_s = \omega_L - \omega_T$. The Thomas frequency accounts for the relativistic rotation of the particle reference frame in the laboratory reference frame.} Importantly, this relation between the anomalous precession frequency and the anomalous magnetic moment is independent of the particle momentum, although $\omega_c$ and $\omega_s$ themselves depend on momentum.

The possibility of directly measuring the muon anomaly $a_\mu$ through the proportionality between $\omega_a$ and $a_\mu$ offers a nearly thousand-fold greater sensitivity to quantum fluctuations compared to  measurements of $g_\mu$. Throughout the 1960s and 1970s, a series of three muon g$-$2  experiments were conducted at CERN to determine $a_\mu$ by measuring the anomalous precession frequency $\omega_a$.   Together these experiments represented a sequence of innovations that improved the precision on $a_\mu$ from 0.43\% to 7~ppm, an improvement of four orders of magnitude.

The first muon g-2 experiment \citep{Charpak1961} was conducted using $\sim$$100$ MeV/c muons from the CERN synchrocyclotron. 
The experiment used a 6.0~m-long dipole magnet with a transverse magnetic-field gradient. Longitudinally polarized muons were injected through a degrader at one end of the magnet and extracted into a polarization analyzer at the other end. 
Within the magnet,
the muons followed a trochoidal path arising from cyclotron motion in the dipole field combined with the field-gradient drift.
The muons were held for $t \sim 2$-$8$~$\mu$s and completed approximately $10^3$ turns. During this period, the spin advanced relative to the muon momentum by an angle $\Delta \theta (t) = a_\mu~ (q B / m)~ t = \omega_a t$. 
The accumulated spin advance $\Delta \theta (t)$ was determined from the measured sinusoidal modulation of the decay-electron asymmetry in the polarization analyzer.

The experiment provided the first  measurement of the anomalous precession frequency $\omega_a$ and enabled the determination of $a_\mu$ to 0.43\%, thereby providing sensitivity to ($\alpha / \pi)^2$ 
QED corrections. 
Since then, the extraction of $\omega_a$ from a  `wiggle plot' of  the $\omega_a$-modulated decay-electron signal has become the hallmark of g-2 experiments. The measurement was statistically limited by the low beam rate from the CERN synchrocyclotron and the relatively short lab-frame lifetime of the $\sim$$100$ MeV/c muons.

By the early 1960s, the CERN proton synchrotron offered both increased muon beam intensities and longer lab-frame  lifetimes due to time dilation. This motivated the development of a dedicated storage ring for a second muon g$-$2 experiment \citep{Bailey1968}.
The second experiment stored 1.27 GeV/c muons in a 1.7~T, 5.0~m-diameter storage ring with a radial magnetic-field gradient providing vertical focusing.
Pions were produced by protons striking an internal production target, and muons were captured 
from forward pion decays into the ring orbit. 
The ring operated 
on repeating cycles of short muon accumulation periods and longer decay-electron measurement periods. The accumulation periods were set by
the time structure of the proton synchrotron, while the measurement periods were set by the $27$~$\mu$s time-dilated lifetime of the 1.27 GeV/c muons.

The anomalous frequency $\omega_a$ was directly observed via decay electrons from 
in-flight muon decay. This is possible because high-energy decay electrons are preferentially emitted along the muon spin direction. When the muon spin and momentum are parallel, the decay electrons are boosted to higher lab energies; when they are antiparallel, the decay electrons are boosted to lower lab energies. Thus, the anomalous precession is observable as a time-dependent modulation of the decay-electron energy spectrum in the lab frame.

The decay electrons were continuously measured during muon storage by energy-sensitive detectors around the inner circumference of the ring. 
By selecting high-energy electrons, 
decays that occur when spin and momentum are aligned are preferentially sampled.
The resulting time distribution of energetic decay electrons is described by
\begin{equation}
N(t) \propto e^{-t/ \gamma \tau} [ 1 - A \cos ( \omega_a t + \phi )],
\label{eqn:5parfit}
\end{equation}
where 
$\gamma \tau$ is the time-dilated muon lifetime, and A, $\omega_a$, and $\phi$ are the asymmetry, frequency, and phase of the anomalous precession signal. The decay-electron energy threshold is set to optimize the statistical power in measuring $\omega_a$.

The experiment marked the first storage-ring measurement of the anomalous precession frequency. The result, $a_\mu = (116616 \pm 31) \times 10^{-8}$, provided an important test of higher-order, mass-dependent QED corrections.
Its agreement with theory strengthened the evidence that  muons are point-like, spin-1/2 Dirac particles with identical electromagnetic interactions to those of  electrons.

The experiment highlighted several challenges for future measurements. 
First, muon storage via pion production and decay within the storage ring resulted in poor capture efficiency, low muon polarization, and large backgrounds. 
Second, vertical focusing using magnetic field gradients complicated the precision determination of the  magnetic field experienced by the stored muons (and therefore the  extraction of $a_\mu$). This required meticulous magnetic field  mapping and a detailed understanding of the muon orbits, a limiting factor for higher-precision measurements.

In 1970, a new storage-ring experiment \citep{Bailey1979} was undertaken
to achieve the sensitivity required to probe hadronic vacuum polarization.
The design involved injecting pion bunches through a pulsed inflector magnet into the ring's storage region.\footnote{The  pulsed inflector is a field-canceling magnet that enables the pion bunches to enter the storage magnet without deflection} The pions then decayed to muons within the storage volume, with muons from forward decays captured onto stable orbits. Injecting pions with a narrow momentum bite from an external beam,  
rather than collecting a broad pion momentum distribution  
from an internal target, significantly improved both capture efficiency and longitudinal polarization of stored muons. 

The new design also replaced magnetic gradient focusing with electrostatic focusing. This allowed the use of a uniform magnetic field and its precise measurement using proton NMR. This transformed the experiment into a co-magnetometer measurement of the anomalous precession frequency of muons and the Larmor precession frequency of protons.

In the presence of magnetic and electric fields, 
the anomalous precession frequency  is
\begin{equation}
\vec{\omega_a} =  - {q \over m}  ~ \left[ ~ a \vec{B} ~-~  a \left( {\gamma \over 1 + \gamma} \right) ~ ( \vec{\beta} \cdot {\vec{B}} ) \vec{\beta} ~-~
\left( a - ( {1 \over 1 - \gamma^2} \right)  ~ \vec{\beta} \times \vec{E} ~ \right] .
\label{eqn:BMTforamu}
\end{equation}
The  electric-field contribution to the anomalous frequency $\omega_a$ vanishes for a Lorentz factor $\gamma = (1 + 1 / a_\mu )^{1/2} = 29.3$.
This magic $\gamma$ dictated the magic momentum 3.09~GeV/c
for  muon storage.
The electrostatic focusing was provided by 
eight electrostatic quadrupoles with 36$^o$ azimuthal coverage, spaced around the ring's circumference with gaps at the injection region and its antipodal location.

The combined result from measurements
with positive and negative muons was
$a_\mu = 1 165 924 (8.5) \times 10^{-9}$. The $7$~ppm precision was sufficient to probe the hadronic vacuum polarization contribution 
and third order QED corrections.  The electroweak contribution from virtual W and Z bosons was 
below experimental sensitivity.

In the 1990s, the muon $g$-$2$ program moved to Brookhaven National Laboratory (BNL).
The BNL experiment built on CERN innovations including the uniform-field co-magnetometer strategy and its operation at the magic momentum.
It replaced pion injection with direct muon injection in order to improve storage efficiency and reduce injection-related backgrounds. 

The BNL experiment \citep{Bennett2006} developed a 
superconducting inflector and fast kicker to inject muons 
onto stable orbits.
The inflector provided a field-free channel through the magnet yoke into the storage region.
It used
a compact double-cosine-theta coil to cancel the ring's 1.45~T field inside the entrance channel and minimize  perturbations to the uniform field in the storage ring.
The kicker provided a magnetic kick to injected particles on the first turn of the storage ring and placd muons
onto stable orbits.
The magnetic pulse had a 100~ns duration and fast rise and fall times 
in order to act only on the first turn.

Completed in 2001, 
the experiment reached a 0.54~ppm precision, below the  1.3~ppm contribution from virtual W and Z bosons.
It indicated a 2.5~$\sigma$ discrepancy with the Standard Model prediction based on data-driven determinations of hadronic vacuum polarization.
Subsequent evaluations of $a^{SM}$ increased the discrepancy to more than 3~$\sigma$ and stimulated discussions of possible new physics as well as tensions among data-driven inputs.  This discrepancy motivated the recent Fermilab muon g$-$2 experiment.

\subsection{The Fermilab g-2 experiment}

The Fermilab muon $g$-$2$ experiment \citep{Muong-2:2025xyk} is the latest-generation storage-ring measurement.
While continuing the BNL approach, it improved the precision on $a_\mu$  by a factor of four through a  
twenty-fold increase in positron statistics. It also reduced the systematics uncertainties associated with measurements of the anomalous precession frequency and the proton Larmor frequency by factors of approximately an order of magnitude.

The experiment employed
the relocated BNL storage ring,
with upgraded $\omega_a$, $\omega_p$, and beam measurement systems, together with improved magnetic-field uniformity.
The storage-ring fill rate was increased from 2-3~Hz at BNL to 12~Hz at Fermilab.
This enabled a large increase in decay-electron statistics while maintaining a reasonable instantaneous decay-electron rate.

A new detector, electronics, and data acquisition system was developed
for the Fermilab $\omega_a$ measurement. 
Each electromagnetic calorimeter comprised a 9$\times$5 array of rectangular lead fluoride crystals.
The calorimeters permitted spatial discrimination of pileup through their segmentation  and  temporal discrimination of pileup through their fast-timing characteristics. 
The readout system recorded contiguous digitized waveforms of detector signals throughout the 0.7 ms measurement period after each storage-ring fill.

An updated 
proton NMR system was developed for the Fermilab $\omega_p$ measurement.
The magnetic field was tracked continuously
during beam-on by 400 NMR probes located immediately outside the vacuum vessel. Every few days, a beam-off measurement using a movable trolley equipped with NMR probes recorded a 9000-point, three-dimensional map of the field inside the vacuum vessel. The data acquisition system recorded digitized waveforms of NMR signals for both systems.

A new in-vacuum positron tracking system was constructed to monitor the stored beam. From measured positron trajectories, it  reconstructed the stored beam distribution as a function of time after injection. These data enable muon-weighted averaging of the magnetic-field maps to determine the mean field experienced by the muons.

The value of $a_\mu$ is obtained from the ratio of the measured anomalous precession frequency $\omega_a$,
the measured Larmor precession frequency $\tilde{\omega}_p$, and well-known constants via
\begin{equation}
a_\mu = {\omega_a \over \tilde{\omega}_p} {\mu_p \over \mu_N } {m_\mu \over m_e} 
\end{equation}
where ${\mu_p / \mu_N }$ is known to 5 ppb and ${m_\mu / m_e}$ is known to 22 ppb. The tilde on $\tilde{\omega_p}$ denotes the convolution of the three-dimensional $\omega_p$-distribution with the muon spatial distribution. In practice, both measured frequencies require small corrections to account for effects including the electric field, the kicker magnet, and other systematic contributions.

\begin{figure}

  \includegraphics[width=\columnwidth]{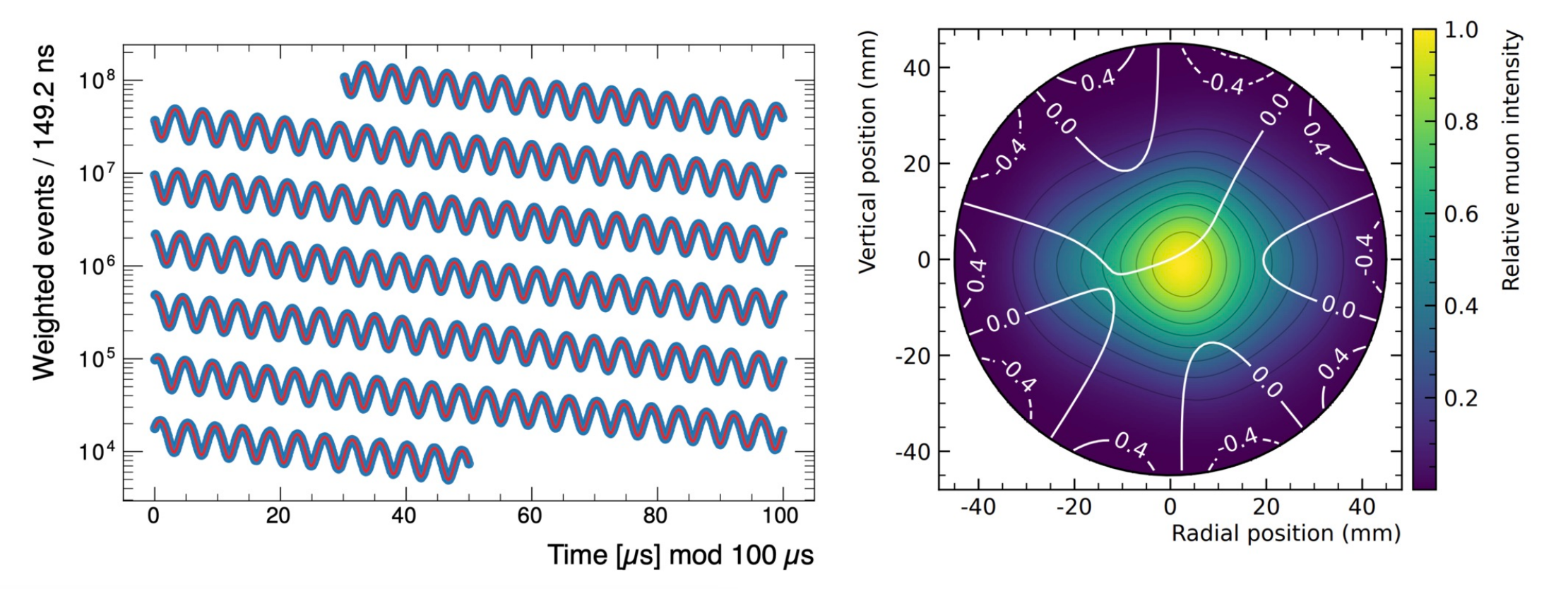}

  \caption{\label{fig:fnalg-2}
    Left: example wiggle fit  to the measured time distribution decay electrons in the Fermilab g-2 experiment. Right: example time-averaged muon distribution (color scale, relative intensity units) and azimuthally-averaged magnetic field plot (white contours, ppm units) in the Fermilab g-2 experiment.}
\end{figure}

The frequency $\omega_a$ is derived from the electromagnetic calorimeter data (see Fig.\ \ref{fig:fnalg-2}).
The signal traces are first fit to template pulse-shapes to extract the times and energies of crystal hits. The crystal hits 
are then clustered into positron candidates, and the time distribution of the positrons constructed. 
Data-driven corrections are applied to the positron time distributions to account for cyclotron rotation of stored muons, unresolved pileup of decay electrons, and gain changes over the duration of the fill.

The 5-parameter function in Eqn.\ \ref{eqn:5parfit} is insufficient to fit the measured  electron time distributions. The residuals from fits using the 5-parameter function indicate 
the effects of the 
beam oscillations about the equilibrium orbit and the beam losses from the storage ring. 
The function was therefore supplemented with additional terms to account for such distortions. The modulation arising from the beam oscillations at the fundamental horizontal and vertical betatron frequencies is complicated and appears at aliased, higher-order, and mixed frequencies.
Phenomenological modeling or data-driven descriptions of the modulation envelopes were also required.
A number of independently blinded analyses were performed utilizing different strategies to ensure robustness in the extraction of the frequency $\omega_a$.

The frequency  $\tilde{\omega_p}$ is derived from a multi-step procedure  involving calibration NMR probes, fixed NMR probes, trolley NMR probes, and tracker measurements of the muon distribution (see Fig.\ \ref{fig:fnalg-2}).
The 392 fixed probes provide a temporal measurement of the magnetic field during the data taking at locations just above and below the storage region.
The 17 trolley probes provide
a discrete, three-dimensional spatial measurement of the magnetic field in the storage region every few days during data taking.
The two tracker detectors provide a continuous temporal measurement of the two-dimensional muon profile at two azimuthal locations in the storage ring during data taking. 

Combining the NMR and tracker data 
involves fitting the trolley-determined spatial maps to a multipole expansion of the storage ring magnetic field, interpolating the resulting magnetic-field moments across the data taking periods using the fixed-probe temporal measurements, and extrapolating tracker measurements of the muon beam distribution around the storage ring using beam simulations. The magnetic footprint of the movable trolley on the fixed NMR probes is also accounted for in determining $\tilde{\omega_p}$. 

The measured frequencies $\omega_a$ and $\tilde{\omega_p}$ require several corrections.
For  $\omega_a$, one class of corrections arises from true $\omega_a$-frequency shifts associated with the $\vec{\beta} \times \vec{E}$ and $\vec{\beta}\cdot\vec{B}$ terms
in Eqn.\ \ref{eqn:BMTforamu}. The $\vec{\beta} \times \vec{E}$ contribution reflects the  momentum spread of the stored beam, while the $\vec{\beta}\cdot\vec{B}$ term arises from its vertical betatron motion. 
A second class of corrections arises from apparent shifts in the fitted $\omega_a$-frequency, due to time-dependent variations in the precession phase during the measurement period. These phase shifts originate from the coupling between the calorimeter acceptance and time-dependent change in the beam position and momentum. The total $\omega_a$ correction was $+$515~ppb for Runs 4-6.

For $\tilde{\omega_p}$, the dominant corrections arise from transient magnetic fields synchronous with muon injection. During data taking, the fixed NMR probes sample the field asynchronously with respect to muon injection and therefore do not  fully account for these magnetic transients. The transients arise from the pulsing of the kicker at injection and the energizing of the electrostatic quadrupoles prior to injection. The resulting time-varying currents induce eddy currents in the neighboring conductors, generating transient magnetic fields within the storage volume during the measurement period.  The total $\tilde{\omega_p}$ correction was $-$58~ppb for Runs 4-6.

The data taking spanned six running periods over six years (2018–2023). The analyses were conducted in three run groups with Run 1 published in 2021, Runs 2-3 in 2023, and Runs 4-6 in 2025. The experiment yielded  21 times the decay-electron statistics of the BNL experiment, and a four-fold reduction in systematic uncertainties.The largest uncertainty was the $\omega_a$ statistical uncertainty, and amounted to 434~ppb, 201~ppb, and 114~ppb for Runs 1, 2-3, and  4-6, respectively (the $\omega_p$ statistical uncertainty is negligible).  The total systematic uncertainties for the three datasets were 157~ppb, 70~ppb, and 76~ppb, respectively.

The largest systematic uncertainties in the  $\omega_a$ measurement arose from the modeling of beam dynamics in the wiggle fits and the correction for gain shifts during the measurement period. These uncertainties were each approximately 20~ppb 
for Runs 4-6. The largest systematic uncertainty in the $\tilde{\omega_p}$ measurement arose from the magnetic-field mapping and was approximately 30~ppb 
for Runs 4-6. 
In addition, the $\omega_a$ electric field correction   had a systematic uncertainty of about 30~ppb, 
while the $\tilde{\omega_p}$ kicker transient correction had a systematic uncertainty of about 20~ppb,  
for Runs 4-6.

\subsection{Status and implications}

\begin{figure}
\begin{centering}
  \includegraphics[width=.95\columnwidth]{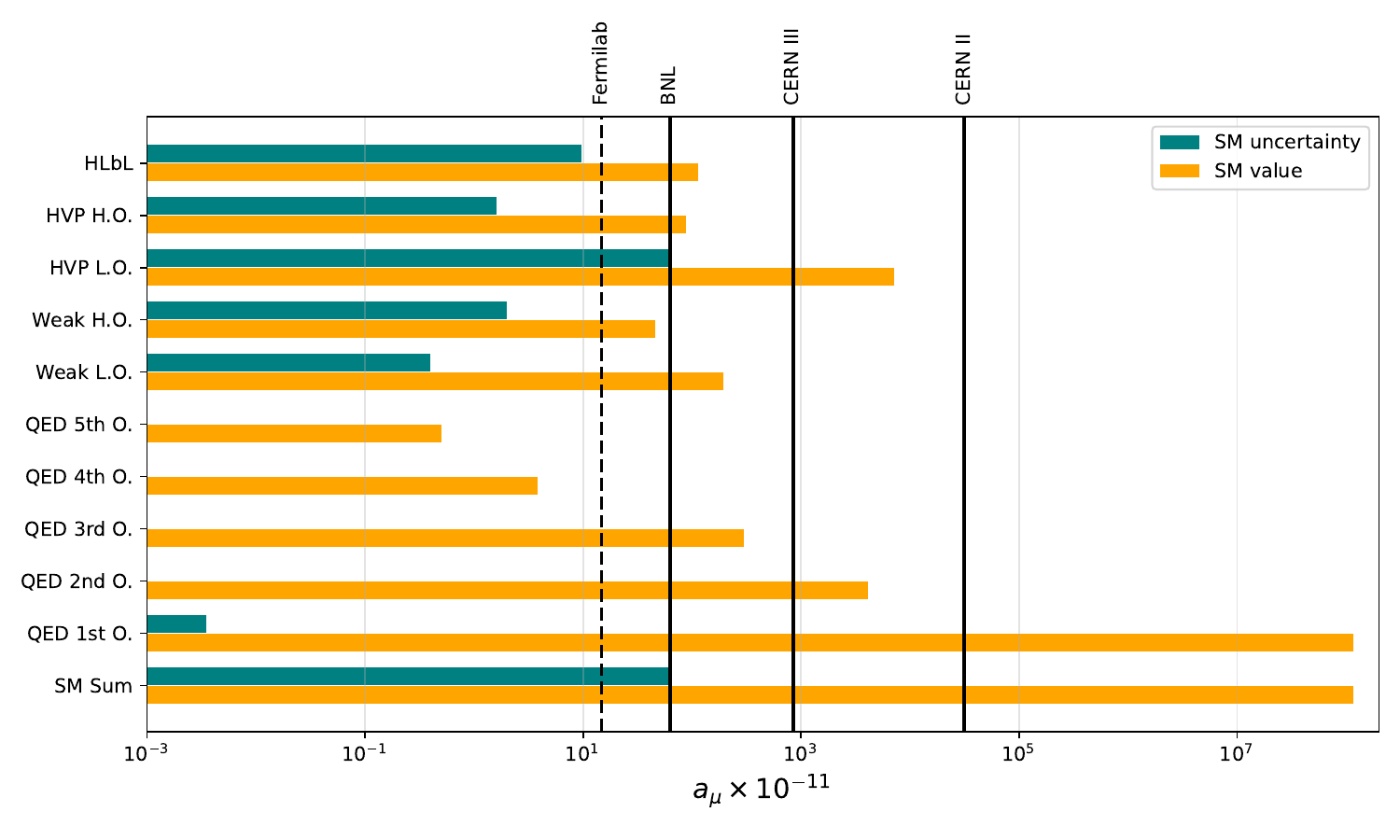}
  \caption{\label{fg:amu-contributions}
    Compilation of QED, weak and hadronic contributions to the muon anomalous magnetic moment \citep{Aliberti2025,Aoyama2020}. The horizontal bars indicate the magnitudes (orange) and uncertaintes (teal) of the contributions.  Terms include QED through 5$^{th}$ order in $\alpha$, leading- and higher-order weak, leading- and higher-order hadronic vacuum polarization, and hadronic light-by-light scattering. The solid vertical black lines indicate the precision achieved by the CERN, BNL and Fermilab g$-$2 experiments. Data from \citet{Aoyama2020, Aliberti2025}. }
\end{centering}
\end{figure}

The Fermilab g$-$2 result is the culmination of more than half a century of experimental effort, pioneered at CERN, refined at Brookhaven, and realized to highest-precision at Fermilab. Across these generations of experiments the essential methods were developed: polarized muon beam injection, proton NMR co-magnetometry, magic-momentum operation, and the direct measurement of the anomalous frequency.  The final results of the CERN, BNL, and Fermilab programs are compiled in Table \ref{tab:g-2results} and Fig.\ \ref{fg:amu-contributions}.

\begin{table}[ht]
\centering
\small
\begin{tabularx}{0.7\textwidth}{l|c|c|c|X}
\hline
 & & & & \\
Experiment & Year & Anomalous moment, $a_\mu$ & Uncertainty & Reference \\
 & & & & \\
\hline
 & & & & \\
CERN-I & 1962 & 0.001~162(5)  & 0.43\%  & %Charpak {\textit et al.}\ 
\citet{Charpak1962} \\ %, Phys. Lett. 1, 16 (1962) \\
CERN-II  & 1968 & 0.001~166~16(31) & 265~ppm  & %Bailey {\textit et al.}\ 
\citet{Bailey1968} \\ % Phys. Lett. 28B, 287 (1968)  \\
CERN-III & 1979 &  0.001~165~924~0(85) & 7.3~ppm &  %Bailey {\textit et al.}\ 
\citet{Bailey1979} \\ % Nucl.\ Phys.\ B150, 1 (1979) \\
% mu+ result 0.001~165~911(11)
& & & & \\
BNL E821 & 2006 & 0.001~165~920~80(63) & 540~ppb & 
%Bennett {\textit et al.}\ 
\citet{Bennett2006} \\ %Phys.\ Rev.\ D 73, 072003 (2006) \\

 & & & & \\
FNAL E989 & 2025 & 0.001~165~920~705(148)  & 127~ppb & %Aguillard {\textit et al.}\
\citet{Muong-2:2025xyk} \\ % Muon g-2 Collaboration, Phys.\ Rev.\ Lett.\ (2025) \\
 & & & & \\
\bottomrule
\end{tabularx}
\caption{Final results for the muon anomalous magnetic moment $a_\mu = (g-2)/2$ from CERN, BNL, and Fermilab g$-$2 experiments. Uncertainties are the total statistical and systematic uncertainties combined in quadrature.}
\label{tab:g-2results}
\end{table}

The CERN I and II measurements reached sensitivities to QED corrections of order $(\alpha / \pi)^2$ and  $(\alpha / \pi)^3$ QED correction, respectively.  They verified QED for muons beyond the 
Schwinger term, and demonstrated with increasing precision that muons behave as heavy electrons. 
By CERN III,  focus had shifted from verifying QED to testing the full SM prediction, including hadronic  contributions to vacuum polarization and weak contributions from virtual W and Z bosons. 
CERN III provided the first quantitative confirmation of the
hadronic vacuum polarization contribution. 

The following BNL experiment reached a precision of 540 ppb, well below the scale of electroweak  contributions, while the recent Fermilab experiment achieved a precision of 127 ppb,  far below the scale of hadronic contributions. Such sensitivity underscores the role of $a_\mu$ as a comprehensive test of the electromagnetic, weak, and strong sectors of the 
Standard Model and its quantum-field-theory framework. The BNL and Fermilab results are in agreement  and yield a world average \citep{Muong-2:2025xyk}  
\begin{equation}
a_\mu = 116~592~071.5 (14.5) \times 10^{-11} (124~ {\mathrm{ppb}}).
\label{eqn:amuworldaverage}
\end{equation}

The QED, electroweak, and hadronic contributions to $a_\mu$ are depicted
in Fig.\ \ref{fg:amu-contributions}.
The QED contribution accounts for approximately 99.994\%  of $a_\mu^{SM}$ and  has been calculated through order $(\alpha / \pi)^5$ in perturbation theory. Its uncertainty, at the level of about 1~ppb, is dominated by the knowledge of the fine-structure constant $\alpha$, with smaller uncertainties arising from the determinations of the lepton masses and the numerical computation of the order $(\alpha / \pi)^4$ and $(\alpha / \pi)^5$ terms.

The electroweak contribution arises from diagrams involving the W, Z, and Higgs bosons. These effects are suppressed by the large masses of the weak bosons and contribute about 1.3~ppm to $a_\mu^{SM}$. The largest electroweak terms are one-loop diagrams involving either a W or a Z. The approximately 3~ppb theoretical uncertainty in $a_\mu^{EW}$ arises from hadronic contributions to electroweak two-loop diagrams.

By far the largest uncertainty in calculating $a_\mu^{SM}$ arises from the hadronic contribution $a^{had}_\mu$. This contribution is
dominated by the leading-order  hadronic vacuum polarization term  $a^{had,LO}_\mu$, with higher-order hadronic vacuum polarization and hadronic light-by-light scattering contributing at the percent level.  For decades, the prediction of $a_\mu^{had}$ has relied primarily on a data-driven approach using the measured $e^+ e^- \rightarrow$ hadrons cross sections. 

The largest contribution to $a^{had,LO}_\mu$ arises from the $e^+ e^- \rightarrow \pi^+ \pi^-$ channel. High-luminosity measurements from BaBar \citep{BaBar2012} KLOE \citep{KLOE2013}, BESIII \citep{BESIII2016}, SND \citep{SND2020update}, and CMD-3 \citep{CMD32024} quote sub-percent precision, yet show few-percent, energy-dependent differences in the $e^+ e^- \rightarrow \pi^+ \pi^-$  cross sections. These discrepancies highlight the challenge of achieving sub-percent control of the absolute normalization, radiative corrections, and detection efficiency.  

The largest outlier among high-luminosity $e^+ e^- \rightarrow \pi^+ \pi^-$ datasets is CMD-3 \citep{CMD32024}. Before the CMD-3 result, the 2020 g-2 Theory Initiative report quoted a SM prediction of $a_\mu^{SM}$ based on a data-driven estimate of the hadronic vacuum polarization term $a_\mu^{had,LO}$. After the CMD-3 result, the 2025 g-2 Theory Initiative update did not provide a $a_\mu^{SM}$ estimate using a data-driven $a_\mu^{had,LO}$ owing  to the increased inconsistencies among the $e^+e^- \rightarrow \pi^+ \pi^-$  datasets. 

More recently, lattice QCD has emerged as an alternative, ab initio approach to calculating hadronic vacuum polarization. 
Over the past decade, lattice determinations have evolved from 5-10\%-level estimates, uncompetitive with the data-driven approach, to sub-percent results. 
A milestone was the 0.8\% result of the BMW collaboration \citep{Borsanyi2021}, which showed the potential for lattice QCD to achieve precisions comparable with experiment.

At present, lattice calculations are advancing rapidly, with multiple collaborations pursuing determinations of $a_\mu^{had}$ and the reduction of its systematic
uncertainties. Reflecting this progress, the 2025  g-2 Theory Initiative update uses a lattice-QCD determination of hadronic vacuum polarization. The 2025 lattice-QCD estimate, $7132(61)\times10^{-11}$, exceeds the 2020 data-driven estimate, $6931(40)\times10^{-11}$, and yields
a 2025 Standard Model prediction 
\citep{Aliberti2025}
%\citep{Aliberti2025,Aoyama:2012wk,Volkov:2019phy,Volkov:2024yzc,Aoyama:2024aly,Parker:2018vye,Morel:2020dww,Fan:2022eto,Czarnecki:2002nt,Gnendiger:2013pva,Ludtke:2024ase,Hoferichter:2025yih,RBC:2018dos,Giusti:2019xct,Borsanyi:2020mff,Lehner:2020crt,Wang:2022lkq,Aubin:2022hgm,Ce:2022kxy,ExtendedTwistedMass:2022jpw,RBC:2023pvn,Kuberski:2024bcj,Boccaletti:2024guq,Spiegel:2024dec,RBC:2024fic,Djukanovic:2024cmq,ExtendedTwistedMass:2024nyi,MILC:2024ryz,Keshavarzi:2019abf,DiLuzio:2024sps,Kurz:2014wya,Colangelo:2015ama,Masjuan:2017tvw,Colangelo:2017fiz,Hoferichter:2018kwz,Eichmann:2019tjk,Bijnens:2019ghy,Leutgeb:2019gbz,Cappiello:2019hwh,Masjuan:2020jsf,Bijnens:2020xnl,Bijnens:2021jqo,Danilkin:2021icn,Stamen:2022uqh,Leutgeb:2022lqw,Hoferichter:2023tgp,Hoferichter:2024fsj,Estrada:2024cfy,Deineka:2024mzt,Eichmann:2024glq,Bijnens:2024jgh,Hoferichter:2024bae,Holz:2024diw,Cappiello:2025fyf,Colangelo:2014qya,Blum:2019ugy,Chao:2021tvp,Chao:2022xzg,Blum:2023vlm,Fodor:2024jyn}
\begin{equation}
a_\mu^{SM} = 116~592~033.(62.) \times 10^{-11} (530~ppb),
\end{equation}
that is consistent with the measured world average (Eqn.\ \ref{eqn:amuworldaverage}).

\begin{figure}[h!]
\includegraphics[width=0.95\columnwidth]{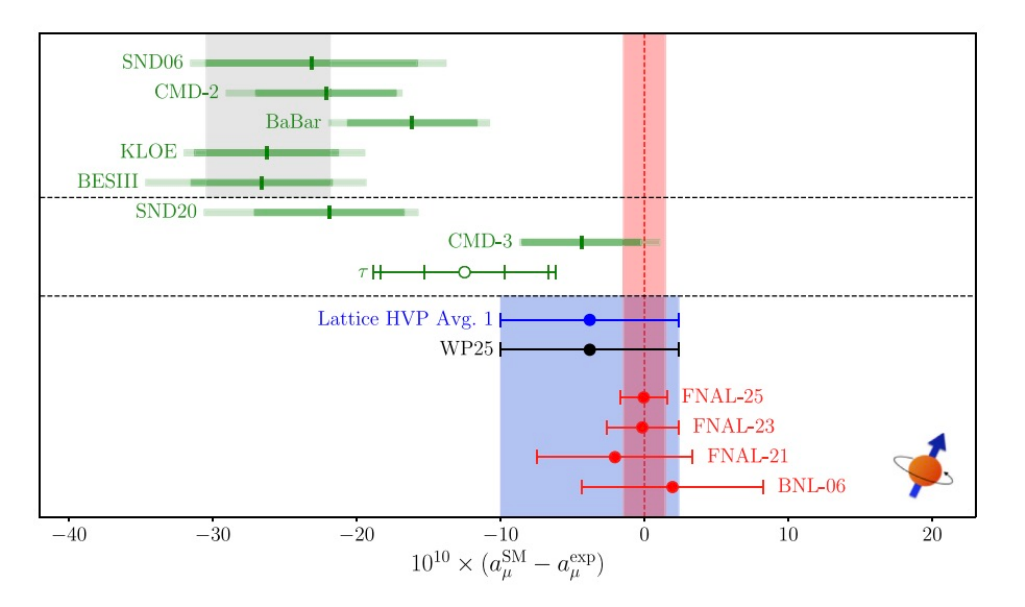}
\caption{Comparison of the measured and predicted values of the muon anomalous magnetic moment. The red, blue and green vertical bands indicate the experimental world average \citep{Muong-2:2025xyk}, 2025 lattice-QCD theory update \citep{Aliberti2025}, and 2020 data-driven theory result \citep{Aliberti2025}, respectively. The red data points are the Fermilab Run 1, Runs 2-3, Runs -4-6 results and BNL result that are used in the  experimental world average. The green data points above the upper dashed line are  data-driven SM predictions based on the individual $e^+ e^-$ datasets included in the 2020 theory result. The green data points below the upper dashed line 
are the data-driven SM predictions based on the post-2020 SND and CMD3 $e^+ e^-$ datasets and the $\tau$-decay data. The black data point is the updated 2025 SM prediction derived by averaging the lattice QCD results for hadronic vacuum polarization. Reproduced with permission from Muon g$-2$ Theory Initiative collaboration. Reproduced from \citet{Aliberti2025}. }
\label{fig:g-2results}
\end{figure}

The situation regarding the comparison between measurement and calculations for $a_\mu$ is summarized in Fig.\ \ref{fig:g-2results} reproduced from Ref.\ \citep{Aliberti2025}. The red, blue and green vertical bands indicate the experimental world average \citep{Muong-2:2025xyk}, the 2025 lattice-QCD theory update \citep{Aliberti2025}, and the earlier 2020 data-driven theory result \citep{Aliberti2025}.
The red data points indicate the  consistency between Fermilab and BNL experiments,
while the green data points indicate the tensions among the $e^+ e^-$ datasets.
The 2025 theoretical uncertainty of 530~ppb is several times larger than the experimental uncertainty.

\section{Conclusion}

The Standard Model consists of
three generations of elementary fermions with common gauge interactions but differing masses. While remarkably successful, it provides no explanation for the existence of the generations or the hierarchy of their masses. The muon is arguably the archetypal example of the SM's empirical success and its deeper incompleteness.

Muon experiments offer powerful opportunities to study elementary interactions and determine fundamental constants. The muon lifetime determines the Fermi constant $G_F$ %and the overall strength
of the weak interaction,  while decay observables establish the V$-$A structure of the weak current. Muonium spectroscopy determines the charge ratio $q_\mu / q_e$, mass ratio $m_\mu / m_e$, and  magnetic moment ratio $\mu_\mu / \mu_p$.
Storage-ring experiments provide a precise determination of the anomalous part of the muon magnetic moment $a_\mu$.

Such precision measurements also enable stringent tests of quantum electrodynamics, quantum chromodynamics, and the overall completeness of the Standard Model.
Muon decay measurements constrain 
right-handed currents, scalar and tensor interactions, and T-violating phases of weak couplings.
Muon g$-$2 measurements constrain  supersymmetry, leptoquarks, and the dynamics of the Higgs sector. Muonium spectroscopy offers sensitivity to dark-sector particles, new muonic forces, and  Lorentz and CPT violation.

The special opportunities of muon physics arise from basic physical properties. Its 2.2~$\mu$s lifetime permits muons to be stopped in matter for decay experiments, confined in storage rings for g$-2$ measurements, and bound in  muonium atoms for spectroscopic studies. Its mass of about 200~$m_e$ enhances its sensitivity to short-distance physics 
and enables its reach into physics beyond the Standard Model. Its purely leptonic decay is  essentially free of hadronic complications and is therefore theoretically pristine.

Spin is another common feature
of precision muon physics. The parity-violating decays 
\(\pi \to \mu\nu_\mu\) and  \(\mu \to e\nu_e\bar{\nu}_\mu\) enable the production and analysis of polarized muons.  These features underlie the V$-$A sensitivity of decay experiments, spin-precession signal of g$-$2  measurements, and use of microwave resonance techniques in 
muonium hyperfine spectroscopy.

Together, these features have enabled determinations of the Fermi constant $G_F$ to 0.5~ppm, the anomalous magnetic moment $a_\mu$ to 124~ppb, the  mass ratio $m_\mu / m_e$ to 22~ppb, and the charge ratio  $| q_\mu / q_e |$  to about 1~ppb.  Measurements of muon decay have sharpened tests of the $V$-$A$ structure of the weak interaction to the $10^{-4}$ level. Alongside exacting tests of QED, electroweak theory, QCD and SM completeness, the muon portfolio reflects its outsized role in fundamental physics.

Muon physics is a vibrant convergence of methods drawn from accelerator, atomic, nuclear and particle physics. Over almost a century, the unexpected particle has become a versatile tool for fundamental physics and testing the laws of nature.

\begin{ack}[Acknowledgments]

The  author would like to thank all  members of the MuLan, MuCap, and Fermilab muon g-2
collaborations for 
their perseverance and ingenuity in advancing these measurements, as well as for contributing to a cordial, creative and engaging working environment. 
The author also acknowledges partial support of this work by the U.S. National Science Foundation. 

\end{ack}

%\seealso{article title article title}

%\bibliographystyle{Harvard}
% References are embedded below from the verified BibTeX output.
%\bibliographystyle{unsrtnat}


\begin{thebibliography}{141}
\expandafter\ifx\csname natexlab\endcsname\relax\def\natexlab#1{#1}\fi
\providecommand{\url}[1]{\texttt{#1}}
\providecommand{\href}[2]{#2}
\providecommand{\path}[1]{#1}
\providecommand{\DOIprefix}{doi:}
\providecommand{\ArXivprefix}{arXiv:}
\providecommand{\URLprefix}{URL: }
\providecommand{\Pubmedprefix}{pmid:}
\providecommand{\doi}[1]{\href{http://dx.doi.org/#1}{\path{#1}}}
\providecommand{\Pubmed}[1]{\href{pmid:#1}{\path{#1}}}
\providecommand{\bibinfo}[2]{#2}
\ifx\xfnm\relax \def\xfnm[#1]{\unskip,\space#1}\fi
%Type = Article
\bibitem[{Keshavarzi et~al.(2022)Keshavarzi, Khaw, and
  Yoshioka}]{KeshavarziKhawYoshioka2022}
\bibinfo{author}{A.~Keshavarzi}, \bibinfo{author}{K.~S. Khaw},
  \bibinfo{author}{T.~Yoshioka},
\newblock \bibinfo{title}{Muon {$g-2$}: a review},
\newblock \bibinfo{journal}{Nuclear Physics B} \bibinfo{volume}{975}
  (\bibinfo{year}{2022}) \bibinfo{pages}{115675}.
  \DOIprefix\doi{10.1016/j.nuclphysb.2022.115675}.
  \href{http://arxiv.org/abs/2106.06723}{{\tt arXiv:2106.06723}}.
%Type = Article
\bibitem[{Hertzog and Hoferichter(2026)}]{HertzogHoferichter2026}
\bibinfo{author}{D.~W. Hertzog}, \bibinfo{author}{M.~Hoferichter},
\newblock \bibinfo{title}{The anomalous magnetic moment of the muon: status and
  perspectives},
\newblock \bibinfo{journal}{Annual Review of Nuclear and Particle Science}
  (\bibinfo{year}{2026}). \DOIprefix\doi{10.1146/annurev-nucl-102422-040841}.
  \href{http://arxiv.org/abs/2512.16980}{{\tt arXiv:2512.16980}},
  \bibinfo{note}{in press}.
%Type = Article
\bibitem[{Ohayon et~al.(2021)Ohayon, Burkley, and
  Crivelli}]{OhayonBurkleyCrivelli2021}
\bibinfo{author}{B.~Ohayon}, \bibinfo{author}{Z.~Burkley},
  \bibinfo{author}{P.~Crivelli},
\newblock \bibinfo{title}{Current status and prospects of muonium spectroscopy
  at {PSI}},
\newblock \bibinfo{journal}{SciPost Physics Proceedings} \bibinfo{volume}{5}
  (\bibinfo{year}{2021}) \bibinfo{pages}{029}.
  \DOIprefix\doi{10.21468/SciPostPhysProc.5.029}.
%Type = Article
\bibitem[{Chen et~al.(2026)Chen, Sun, and Tang}]{ChenSunTang2026}
\bibinfo{author}{S.~Chen}, \bibinfo{author}{M.~Sun}, \bibinfo{author}{J.~Tang},
\newblock \bibinfo{title}{Muon beams towards muonium physics: progress and
  prospects},
\newblock \bibinfo{journal}{arXiv}  (\bibinfo{year}{2026}).
  \DOIprefix\doi{10.48550/arXiv.2601.15818}.
  \href{http://arxiv.org/abs/2601.15818}{{\tt arXiv:2601.15818}},
  \bibinfo{note}{review article}.
%Type = Article
\bibitem[{Echenard and Petrov(2026)}]{EchenardPetrov2026}
\bibinfo{author}{B.~Echenard}, \bibinfo{author}{A.~A. Petrov},
\newblock \bibinfo{title}{Precision physics with muons: a decade of theoretical
  and experimental advances},
\newblock \bibinfo{journal}{Annual Review of Nuclear and Particle Science}
  \bibinfo{volume}{76} (\bibinfo{year}{2026}) \bibinfo{pages}{1--25}.
  \DOIprefix\doi{10.1146/annurev-nucl-100324-024335}.
  \href{http://arxiv.org/abs/2602.12442}{{\tt arXiv:2602.12442}},
  \bibinfo{note}{in press}.
%Type = Article
\bibitem[{Gorringe and Hertzog(2015)}]{GorringeHertzog2015}
\bibinfo{author}{T.~P. Gorringe}, \bibinfo{author}{D.~W. Hertzog},
\newblock \bibinfo{title}{Precision muon physics},
\newblock \bibinfo{journal}{Progress in Particle and Nuclear Physics}
  \bibinfo{volume}{84} (\bibinfo{year}{2015}) \bibinfo{pages}{73--123}.
  \DOIprefix\doi{10.1016/j.ppnp.2015.06.001}.
  \href{http://arxiv.org/abs/1506.01465}{{\tt arXiv:1506.01465}}.
%Type = Article
\bibitem[{Anderson and Neddermeyer(1936)}]{Anderson:1936zz}
\bibinfo{author}{C.~D. Anderson}, \bibinfo{author}{S.~H. Neddermeyer},
\newblock \bibinfo{title}{{Cloud chamber observations of cosmic rays at 4300
  meters elevation and near sea-level}},
\newblock \bibinfo{journal}{Phys.Rev.} \bibinfo{volume}{50}
  (\bibinfo{year}{1936}) \bibinfo{pages}{263--271}.
  \DOIprefix\doi{10.1103/PhysRev.50.263}.
%Type = Article
\bibitem[{Street and Stevenson(1937)}]{Street:1937me}
\bibinfo{author}{J.~Street}, \bibinfo{author}{E.~Stevenson},
\newblock \bibinfo{title}{{New evidence for the existence of a particle of mass
  intermediate between the proton and electron}},
\newblock \bibinfo{journal}{Phys.Rev.} \bibinfo{volume}{52}
  (\bibinfo{year}{1937}) \bibinfo{pages}{1003--1004}.
  \DOIprefix\doi{10.1103/PhysRev.52.1003}.
%Type = Article
\bibitem[{Williams and Roberts(1940)}]{Williams:1940}
\bibinfo{author}{E.~J. Williams}, \bibinfo{author}{G.~E. Roberts},
\newblock \bibinfo{title}{Evidence for transformation of mesotrons into
  electrons},
\newblock \bibinfo{journal}{Nature} \bibinfo{volume}{145}
  (\bibinfo{year}{1940}) \bibinfo{pages}{102--103}.
  \DOIprefix\doi{10.1038/145102a0}.
%Type = Article
\bibitem[{Rossi and Hall(1941)}]{Rossi:1941zz}
\bibinfo{author}{B.~Rossi}, \bibinfo{author}{D.~B. Hall},
\newblock \bibinfo{title}{{Variation of the rate of decay of mesotrons with
  momentum}},
\newblock \bibinfo{journal}{Phys.Rev.} \bibinfo{volume}{59}
  (\bibinfo{year}{1941}) \bibinfo{pages}{223--228}.
  \DOIprefix\doi{10.1103/PhysRev.59.223}.
%Type = Article
\bibitem[{Rasetti(1941)}]{Rasetti:1941ie}
\bibinfo{author}{F.~Rasetti},
\newblock \bibinfo{title}{{Disintegration of slow mesotrons}},
\newblock \bibinfo{journal}{Phys.Rev.} \bibinfo{volume}{60}
  (\bibinfo{year}{1941}) \bibinfo{pages}{198--204}.
  \DOIprefix\doi{10.1103/PhysRev.60.198}.
%Type = Article
\bibitem[{Conversi et~al.(1947)Conversi, Pancini, and
  Piccioni}]{PhysRev.71.209}
\bibinfo{author}{M.~Conversi}, \bibinfo{author}{E.~Pancini},
  \bibinfo{author}{O.~Piccioni},
\newblock \bibinfo{title}{On the disintegration of negative mesons},
\newblock \bibinfo{journal}{Phys. Rev.} \bibinfo{volume}{71}
  (\bibinfo{year}{1947}) \bibinfo{pages}{209--210}.
  \DOIprefix\doi{10.1103/PhysRev.71.209}.
%Type = Article
\bibitem[{Lattes et~al.(1947)Lattes, Occhialini, and Powell}]{Lattes:1947my}
\bibinfo{author}{C.~M.~G. Lattes}, \bibinfo{author}{G.~P.~S. Occhialini},
  \bibinfo{author}{C.~F. Powell},
\newblock \bibinfo{title}{{Observations on the tracks of slow mesons in
  photographic emulsions. 2}},
\newblock \bibinfo{journal}{Nature} \bibinfo{volume}{160}
  (\bibinfo{year}{1947}) \bibinfo{pages}{486--492}.
  \DOIprefix\doi{10.1038/160486a0}.
%Type = Article
\bibitem[{Leighton et~al.(1949)Leighton, Anderson, and
  Seriff}]{PhysRev.75.1432}
\bibinfo{author}{R.~B. Leighton}, \bibinfo{author}{C.~D. Anderson},
  \bibinfo{author}{A.~J. Seriff},
\newblock \bibinfo{title}{The energy spectrum of the decay particles and the
  mass and spin of the mesotron},
\newblock \bibinfo{journal}{Phys. Rev.} \bibinfo{volume}{75}
  (\bibinfo{year}{1949}) \bibinfo{pages}{1432--1437}.
  \DOIprefix\doi{10.1103/PhysRev.75.1432}.
%Type = Article
\bibitem[{Pontecorvo(1947)}]{PhysRev.72.246}
\bibinfo{author}{B.~Pontecorvo},
\newblock \bibinfo{title}{Nuclear capture of mesons and the meson decay},
\newblock \bibinfo{journal}{Phys. Rev.} \bibinfo{volume}{72}
  (\bibinfo{year}{1947}) \bibinfo{pages}{246--247}.
  \DOIprefix\doi{10.1103/PhysRev.72.246}.
%Type = Article
\bibitem[{Michel and Wightman(1954)}]{PhysRev.93.354}
\bibinfo{author}{L.~Michel}, \bibinfo{author}{A.~Wightman},
\newblock \bibinfo{title}{Muon decay, beta radioactivity, and universal fermi
  interaction},
\newblock \bibinfo{journal}{Phys. Rev.} \bibinfo{volume}{93}
  (\bibinfo{year}{1954}) \bibinfo{pages}{354--355}.
  \DOIprefix\doi{10.1103/PhysRev.93.354}.
%Type = Article
\bibitem[{Michel(1950)}]{Michel1950}
\bibinfo{author}{L.~Michel},
\newblock \bibinfo{title}{Interaction between four half-spin particles and the
  decay of the $\mu$ meson},
\newblock \bibinfo{journal}{Proceedings of the Physical Society. Section A}
  \bibinfo{volume}{63} (\bibinfo{year}{1950}) \bibinfo{pages}{514--531}.
  \DOIprefix\doi{10.1088/0370-1298/63/5/311}.
%Type = Article
\bibitem[{Lee and Yang(1956)}]{Lee:1956qn}
\bibinfo{author}{T.~Lee}, \bibinfo{author}{C.-N. Yang},
\newblock \bibinfo{title}{{Question of parity conservation in weak
  interactions}},
\newblock \bibinfo{journal}{Phys.Rev.} \bibinfo{volume}{104}
  (\bibinfo{year}{1956}) \bibinfo{pages}{254--258}.
  \DOIprefix\doi{10.1103/PhysRev.104.254}.
%Type = Article
\bibitem[{Wu et~al.(1957)Wu, Ambler, Hayward, Hoppes, and Hudson}]{Wu1957}
\bibinfo{author}{C.~S. Wu}, \bibinfo{author}{E.~Ambler}, \bibinfo{author}{R.~W.
  Hayward}, \bibinfo{author}{D.~D. Hoppes}, \bibinfo{author}{R.~P. Hudson},
\newblock \bibinfo{title}{Experimental test of parity conservation in beta
  decay},
\newblock \bibinfo{journal}{Physical Review} \bibinfo{volume}{105}
  (\bibinfo{year}{1957}) \bibinfo{pages}{1413--1415}.
  \DOIprefix\doi{10.1103/PhysRev.105.1413}.
%Type = Article
\bibitem[{Garwin et~al.(1957)Garwin, Lederman, and Weinrich}]{Garwin:1957hc}
\bibinfo{author}{R.~Garwin}, \bibinfo{author}{L.~Lederman},
  \bibinfo{author}{M.~Weinrich},
\newblock \bibinfo{title}{{Observations of the failure of conservation of
  parity and charge conjugation in meson decays: the magnetic moment of the
  free muon}},
\newblock \bibinfo{journal}{Phys.Rev.} \bibinfo{volume}{105}
  (\bibinfo{year}{1957}) \bibinfo{pages}{1415--1417}.
  \DOIprefix\doi{10.1103/PhysRev.105.1415}.
%Type = Article
\bibitem[{Friedman and Telegdi(1957)}]{Friedman:1957mz}
\bibinfo{author}{J.~Friedman}, \bibinfo{author}{V.~Telegdi},
\newblock \bibinfo{title}{{Nuclear emulsion evidence for parity nonconservation
  in the decay chain $\pi^+ \to \mu^+ \to e^+$}},
\newblock \bibinfo{journal}{Phys.Rev.} \bibinfo{volume}{106}
  (\bibinfo{year}{1957}) \bibinfo{pages}{1290--1293}.
  \DOIprefix\doi{10.1103/PhysRev.106.1290}.
%Type = Inproceedings
\bibitem[{Sudarshan and Marshak(1958)}]{MarshakSudarshan1957}
\bibinfo{author}{E.~C.~G. Sudarshan}, \bibinfo{author}{R.~E. Marshak},
\newblock \bibinfo{title}{The nature of the four-fermion interaction},
\newblock in: \bibinfo{booktitle}{Proceedings of the Padua-Venice conference on
  mesons and recently discovered particles}, \bibinfo{year}{1958}, pp.
  \bibinfo{pages}{V--14--V--24}. \bibinfo{note}{Conference held in
  Padua--Venice, September 1957; reprinted in The Development of Weak
  Interaction Theory, edited by P. K. Kabir, Gordon and Breach, 1963, pp.
  119--128}.
%Type = Article
\bibitem[{Feynman and Gell-Mann(1958)}]{Feynman:1958ty}
\bibinfo{author}{R.~Feynman}, \bibinfo{author}{M.~Gell-Mann},
\newblock \bibinfo{title}{{Theory of Fermi interaction}},
\newblock \bibinfo{journal}{Phys.Rev.} \bibinfo{volume}{109}
  (\bibinfo{year}{1958}) \bibinfo{pages}{193--198}.
  \DOIprefix\doi{10.1103/PhysRev.109.193}.
%Type = Article
\bibitem[{Charpak et~al.(1961)Charpak, Farley, Garwin, Muller, Sens, Telegdi,
  and Zichichi}]{Charpak1961}
\bibinfo{author}{G.~Charpak}, \bibinfo{author}{F.~J.~M. Farley},
  \bibinfo{author}{R.~L. Garwin}, \bibinfo{author}{T.~Muller},
  \bibinfo{author}{J.~C. Sens}, \bibinfo{author}{V.~L. Telegdi},
  \bibinfo{author}{A.~Zichichi},
\newblock \bibinfo{title}{Measurement of the anomalous magnetic moment of the
  muon},
\newblock \bibinfo{journal}{Phys. Rev. Lett.} \bibinfo{volume}{6}
  (\bibinfo{year}{1961}) \bibinfo{pages}{128--132}.
  \DOIprefix\doi{10.1103/PhysRevLett.6.128}.
%Type = Article
\bibitem[{Hincks and Pontecorvo(1948)}]{PhysRev.73.257}
\bibinfo{author}{E.~P. Hincks}, \bibinfo{author}{B.~Pontecorvo},
\newblock \bibinfo{title}{Search for gamma-radiation in the 2.2-microsecond
  meson decay process},
\newblock \bibinfo{journal}{Phys. Rev.} \bibinfo{volume}{73}
  (\bibinfo{year}{1948}) \bibinfo{pages}{257--258}.
  \DOIprefix\doi{10.1103/PhysRev.73.257}.
%Type = Article
\bibitem[{Feinberg(1958)}]{Feinberg1958}
\bibinfo{author}{G.~Feinberg},
\newblock \bibinfo{title}{Muon decay and conservation of lepton number},
\newblock \bibinfo{journal}{Physical Review} \bibinfo{volume}{110}
  (\bibinfo{year}{1958}) \bibinfo{pages}{1482--1483}.
  \DOIprefix\doi{10.1103/PhysRev.110.1482}.
%Type = Article
\bibitem[{Pontecorvo(1958)}]{Pontecorvo1958}
\bibinfo{author}{B.~Pontecorvo},
\newblock \bibinfo{title}{Inverse beta processes and nonconservation of lepton
  charge},
\newblock \bibinfo{journal}{Soviet Physics JETP} \bibinfo{volume}{7}
  (\bibinfo{year}{1958}) \bibinfo{pages}{172--173}. \bibinfo{note}{Original
  Russian publication: Zh. Eksp. Teor. Fiz. 34, 247 (1957)}.
%Type = Article
\bibitem[{Danby et~al.(1962)}]{Danby1962}
\bibinfo{author}{G.~Danby}, et~al.,
\newblock \bibinfo{title}{Observation of high-energy neutrino reactions and the
  existence of two kinds of neutrinos},
\newblock \bibinfo{journal}{Physical Review Letters} \bibinfo{volume}{9}
  (\bibinfo{year}{1962}) \bibinfo{pages}{36--44}.
  \DOIprefix\doi{10.1103/PhysRevLett.9.36}.
%Type = Article
\bibitem[{Wheeler(1947)}]{PhysRev.71.320}
\bibinfo{author}{J.~A. Wheeler},
\newblock \bibinfo{title}{Mechanism of capture of slow mesons},
\newblock \bibinfo{journal}{Phys. Rev.} \bibinfo{volume}{71}
  (\bibinfo{year}{1947}) \bibinfo{pages}{320--321}.
  \DOIprefix\doi{10.1103/PhysRev.71.320}.
%Type = Article
\bibitem[{Fermi and Teller(1947)}]{PhysRev.72.399}
\bibinfo{author}{E.~Fermi}, \bibinfo{author}{E.~Teller},
\newblock \bibinfo{title}{The capture of negative mesotrons in matter},
\newblock \bibinfo{journal}{Phys. Rev.} \bibinfo{volume}{72}
  (\bibinfo{year}{1947}) \bibinfo{pages}{399--408}.
  \DOIprefix\doi{10.1103/PhysRev.72.399}.
%Type = Article
\bibitem[{Fitch and Rainwater(1953)}]{FitchRainwater1953}
\bibinfo{author}{V.~L. Fitch}, \bibinfo{author}{J.~Rainwater},
\newblock \bibinfo{title}{Studies of x-rays from mu-mesonic atoms},
\newblock \bibinfo{journal}{Physical Review} \bibinfo{volume}{92}
  (\bibinfo{year}{1953}) \bibinfo{pages}{789--800}.
  \DOIprefix\doi{10.1103/PhysRev.92.789}.
%Type = Article
\bibitem[{Hughes et~al.(1960)Hughes, McColm, Ziock, and
  Prepost}]{PhysRevLett.5.63}
\bibinfo{author}{V.~W. Hughes}, \bibinfo{author}{D.~W. McColm},
  \bibinfo{author}{K.~Ziock}, \bibinfo{author}{R.~Prepost},
\newblock \bibinfo{title}{Formation of muonium and observation of its larmor
  precession},
\newblock \bibinfo{journal}{Phys. Rev. Lett.} \bibinfo{volume}{5}
  (\bibinfo{year}{1960}) \bibinfo{pages}{63--65}.
  \DOIprefix\doi{10.1103/PhysRevLett.5.63}.
%Type = Article
\bibitem[{Prepost et~al.(1961)Prepost, Hughes, and Ziock}]{PhysRevLett.6.19}
\bibinfo{author}{R.~Prepost}, \bibinfo{author}{V.~W. Hughes},
  \bibinfo{author}{K.~Ziock},
\newblock \bibinfo{title}{Observation of the hyperfine structure splitting of
  muonium by use of a static magnetic field},
\newblock \bibinfo{journal}{Phys. Rev. Lett.} \bibinfo{volume}{6}
  (\bibinfo{year}{1961}) \bibinfo{pages}{19--21}.
  \DOIprefix\doi{10.1103/PhysRevLett.6.19}.
%Type = Article
\bibitem[{Hughes(1966)}]{Hughes1966}
\bibinfo{author}{V.~W. Hughes},
\newblock \bibinfo{title}{Muonium},
\newblock \bibinfo{journal}{Annu. Rev. Nucl. Sci.} \bibinfo{volume}{16}
  (\bibinfo{year}{1966}) \bibinfo{pages}{445--470}.
  \DOIprefix\doi{10.1146/annurev.ns.16.120166.002305}.
%Type = Article
\bibitem[{Otani et~al.(2018)}]{Otani2018}
\bibinfo{author}{M.~Otani}, et~al.,
\newblock \bibinfo{title}{Re-acceleration of ultra cold muon in j-parc muon
  facility},
\newblock \bibinfo{journal}{Proceedings of IPAC2018}  (\bibinfo{year}{2018})
  \bibinfo{pages}{4547--4550}. \DOIprefix\doi{10.18429/JACoW-IPAC2018-FRXGBF1}.
%Type = Article
\bibitem[{Navas et~al.(2024)}]{PDG2024}
\bibinfo{author}{S.~Navas}, et~al.,
\newblock \bibinfo{title}{Review of particle physics},
\newblock \bibinfo{journal}{Physical Review D} \bibinfo{volume}{110}
  (\bibinfo{year}{2024}) \bibinfo{pages}{030001}.
  \DOIprefix\doi{10.1103/PhysRevD.110.030001}, \bibinfo{note}{particle Data
  Group}.
%Type = Article
\bibitem[{Kinoshita and Sirlin(1959)}]{KinoshitaSirlin1959}
\bibinfo{author}{T.~Kinoshita}, \bibinfo{author}{A.~Sirlin},
\newblock \bibinfo{title}{Radiative corrections to fermi interactions},
\newblock \bibinfo{journal}{Physical Review} \bibinfo{volume}{113}
  (\bibinfo{year}{1959}) \bibinfo{pages}{1652--1660}.
  \DOIprefix\doi{10.1103/PhysRev.113.1652}.
%Type = Article
\bibitem[{van Ritbergen and Stuart(1999)}]{vanRitbergenStuart1999}
\bibinfo{author}{T.~van Ritbergen}, \bibinfo{author}{R.~G. Stuart},
\newblock \bibinfo{title}{Complete 2-loop quantum electrodynamic contributions
  to the muon lifetime in the fermi model},
\newblock \bibinfo{journal}{Physical Review Letters} \bibinfo{volume}{82}
  (\bibinfo{year}{1999}) \bibinfo{pages}{488--491}.
  \DOIprefix\doi{10.1103/PhysRevLett.82.488}.
%Type = Article
\bibitem[{van Ritbergen and Stuart(2000)}]{vanRitbergenStuart2000}
\bibinfo{author}{T.~van Ritbergen}, \bibinfo{author}{R.~G. Stuart},
\newblock \bibinfo{title}{On the precise determination of the fermi coupling
  constant from the muon lifetime},
\newblock \bibinfo{journal}{Nuclear Physics B} \bibinfo{volume}{564}
  (\bibinfo{year}{2000}) \bibinfo{pages}{343--390}.
  \DOIprefix\doi{10.1016/S0550-3213(99)00572-6}.
%Type = Article
\bibitem[{Pak and Czarnecki(2008)}]{Pak2008}
\bibinfo{author}{A.~Pak}, \bibinfo{author}{A.~Czarnecki},
\newblock \bibinfo{title}{Mass effects in muon and semileptonic $b \to c$
  decays},
\newblock \bibinfo{journal}{Physical Review Letters} \bibinfo{volume}{100}
  (\bibinfo{year}{2008}) \bibinfo{pages}{241807}.
  \DOIprefix\doi{10.1103/PhysRevLett.100.241807}.
%Type = Article
\bibitem[{van Ritbergen and Stuart(1998)}]{vanRitbergenStuart1998Hadronic}
\bibinfo{author}{T.~van Ritbergen}, \bibinfo{author}{R.~G. Stuart},
\newblock \bibinfo{title}{Hadronic contributions to the muon lifetime},
\newblock \bibinfo{journal}{Physics Letters B} \bibinfo{volume}{437}
  (\bibinfo{year}{1998}) \bibinfo{pages}{201--208}.
  \DOIprefix\doi{10.1016/S0370-2693(98)00895-8}.
%Type = Article
\bibitem[{Bell and Hincks(1951)}]{BellHincks1951}
\bibinfo{author}{W.~E. Bell}, \bibinfo{author}{E.~P. Hincks},
\newblock \bibinfo{title}{The lifetime of the $\mu^+$ meson},
\newblock \bibinfo{journal}{Physical Review} \bibinfo{volume}{84}
  (\bibinfo{year}{1951}) \bibinfo{pages}{1243--1244}.
  \DOIprefix\doi{10.1103/PhysRev.84.1243}.
%Type = Article
\bibitem[{Fischer et~al.(1959)Fischer, Leontic, Lundby, Meunier, and
  Stroot}]{Fischer1959}
\bibinfo{author}{J.~Fischer}, \bibinfo{author}{B.~Leontic},
  \bibinfo{author}{A.~Lundby}, \bibinfo{author}{R.~Meunier},
  \bibinfo{author}{J.~P. Stroot},
\newblock \bibinfo{title}{Measurement of $\mu^+$ lifetime},
\newblock \bibinfo{journal}{Physical Review Letters} \bibinfo{volume}{3}
  (\bibinfo{year}{1959}) \bibinfo{pages}{349--350}.
  \DOIprefix\doi{10.1103/PhysRevLett.3.349}.
%Type = Article
\bibitem[{Reiter et~al.(1960)Reiter, Romanowski, Sutton, and
  Chidley}]{Reiter1960}
\bibinfo{author}{R.~A. Reiter}, \bibinfo{author}{T.~A. Romanowski},
  \bibinfo{author}{R.~B. Sutton}, \bibinfo{author}{B.~G. Chidley},
\newblock \bibinfo{title}{Precise measurements of the mean lives of $\mu^+$ and
  $\mu^-$ mesons in carbon},
\newblock \bibinfo{journal}{Physical Review Letters} \bibinfo{volume}{5}
  (\bibinfo{year}{1960}) \bibinfo{pages}{22--23}.
  \DOIprefix\doi{10.1103/PhysRevLett.5.22}.
%Type = Article
\bibitem[{Lundy(1962)}]{Lundy1962}
\bibinfo{author}{R.~A. Lundy},
\newblock \bibinfo{title}{Precision measurement of the $\mu^+$ lifetime},
\newblock \bibinfo{journal}{Physical Review} \bibinfo{volume}{125}
  (\bibinfo{year}{1962}) \bibinfo{pages}{1686--1690}.
  \DOIprefix\doi{10.1103/PhysRev.125.1686}.
%Type = Article
\bibitem[{Williams and Williams(1972)}]{Williams1972MuonLifetime}
\bibinfo{author}{R.~W. Williams}, \bibinfo{author}{D.~L. Williams},
\newblock \bibinfo{title}{Lifetime of the positive muon and a new value for
  $g_f$},
\newblock \bibinfo{journal}{Physical Review D} \bibinfo{volume}{6}
  (\bibinfo{year}{1972}) \bibinfo{pages}{737--742}.
  \DOIprefix\doi{10.1103/PhysRevD.6.737}.
%Type = Article
\bibitem[{Duclos et~al.(1973)Duclos, Magnon, and Picard}]{Duclos1973}
\bibinfo{author}{J.~Duclos}, \bibinfo{author}{A.~Magnon},
  \bibinfo{author}{J.~Picard},
\newblock \bibinfo{title}{A new measurement of the muon lifetime},
\newblock \bibinfo{journal}{Physics Letters B} \bibinfo{volume}{47}
  (\bibinfo{year}{1973}) \bibinfo{pages}{491--493}.
  \DOIprefix\doi{10.1016/0370-2693(73)90023-3}.
%Type = Article
\bibitem[{Balandin et~al.(1975)Balandin, Grebenyuk, Zinov, Konin, and
  Ponomarev}]{Balandin1975}
\bibinfo{author}{M.~P. Balandin}, \bibinfo{author}{V.~M. Grebenyuk},
  \bibinfo{author}{V.~G. Zinov}, \bibinfo{author}{A.~D. Konin},
  \bibinfo{author}{A.~N. Ponomarev},
\newblock \bibinfo{title}{Measurement of the lifetime of the positive muon},
\newblock \bibinfo{journal}{Sov. Phys. JETP} \bibinfo{volume}{40}
  (\bibinfo{year}{1975}) \bibinfo{pages}{811--814}. \bibinfo{note}{Zh. Eksp.
  Teor. Fiz. 67, 1631 (1974)}.
%Type = Article
\bibitem[{Giovanetti et~al.(1984)Giovanetti, Dey, Eckhause, Hart, Hartmann
  et~al.}]{Giovanetti1984}
\bibinfo{author}{K.~L. Giovanetti}, \bibinfo{author}{W.~Dey},
  \bibinfo{author}{M.~Eckhause}, \bibinfo{author}{R.~D. Hart},
  \bibinfo{author}{R.~Hartmann}, et~al.,
\newblock \bibinfo{title}{Mean life of the positive muon},
\newblock \bibinfo{journal}{Physical Review D} \bibinfo{volume}{29}
  (\bibinfo{year}{1984}) \bibinfo{pages}{343--348}.
  \DOIprefix\doi{10.1103/PhysRevD.29.343}.
%Type = Article
\bibitem[{Bardin et~al.(1984)Bardin, Duclos, Magnon, Martino, Zavattini
  et~al.}]{Bardin1984}
\bibinfo{author}{G.~Bardin}, \bibinfo{author}{J.~Duclos},
  \bibinfo{author}{A.~Magnon}, \bibinfo{author}{J.~Martino},
  \bibinfo{author}{E.~Zavattini}, et~al.,
\newblock \bibinfo{title}{A new measurement of the positive muon lifetime},
\newblock \bibinfo{journal}{Physics Letters B} \bibinfo{volume}{137}
  (\bibinfo{year}{1984}) \bibinfo{pages}{135--140}.
  \DOIprefix\doi{10.1016/0370-2693(84)91121-3}.
%Type = Article
\bibitem[{Chitwood et~al.(2007)}]{MuLan2007}
\bibinfo{author}{D.~B. Chitwood}, et~al.,
\newblock \bibinfo{title}{Improved measurement of the positive muon lifetime
  and determination of the fermi constant},
\newblock \bibinfo{journal}{Physical Review Letters} \bibinfo{volume}{99}
  (\bibinfo{year}{2007}) \bibinfo{pages}{032001}.
  \DOIprefix\doi{10.1103/PhysRevLett.99.032001}.
%Type = Article
\bibitem[{Barczyk et~al.(2008)}]{Barczyk:2007hp}
\bibinfo{author}{A.~Barczyk}, et~al. (\bibinfo{collaboration}{FAST
  Collaboration}),
\newblock \bibinfo{title}{{Measurement of the Fermi constant by FAST}},
\newblock \bibinfo{journal}{Phys.Lett.} \bibinfo{volume}{B663}
  (\bibinfo{year}{2008}) \bibinfo{pages}{172--180}.
  \DOIprefix\doi{10.1016/j.physletb.2008.04.006}.
%Type = Article
\bibitem[{Tishchenko et~al.(2013)}]{Tishchenko:2012ie}
\bibinfo{author}{V.~Tishchenko}, et~al. (\bibinfo{collaboration}{MuLan
  Collaboration}),
\newblock \bibinfo{title}{{Detailed report of the MuLan measurement of the
  positive muon lifetime and determination of the Fermi constant}},
\newblock \bibinfo{journal}{Phys.Rev.} \bibinfo{volume}{D87}
  (\bibinfo{year}{2013}) \bibinfo{pages}{052003}.
  \DOIprefix\doi{10.1103/PhysRevD.87.052003}.
%Type = Article
\bibitem[{Amhis et~al.(2021)}]{HFLAV2021}
\bibinfo{author}{Y.~Amhis}, et~al.,
\newblock \bibinfo{title}{Averages of b-hadron, c-hadron, and $\tau$-lepton
  properties},
\newblock \bibinfo{journal}{Eur. Phys. J. C} \bibinfo{volume}{81}
  (\bibinfo{year}{2021}) \bibinfo{pages}{226}.
  \DOIprefix\doi{10.1140/epjc/s10052-020-8156-7}.
%Type = Article
\bibitem[{Seng et~al.(2022)Seng, Galviz, Marciano, and
  Meißner}]{Seng2022VusVud}
\bibinfo{author}{C.-Y. Seng}, \bibinfo{author}{D.~Galviz},
  \bibinfo{author}{W.~J. Marciano}, \bibinfo{author}{U.-G. Meißner},
\newblock \bibinfo{title}{Update on $|v_{us}|$ and $|v_{us}/v_{ud}|$ from
  semileptonic kaon and pion decays},
\newblock \bibinfo{journal}{Phys. Rev. D} \bibinfo{volume}{105}
  (\bibinfo{year}{2022}) \bibinfo{pages}{013005}.
  \DOIprefix\doi{10.1103/PhysRevD.105.013005}.
%Type = Article
\bibitem[{Cirigliano et~al.(2023)Cirigliano, Crivellin, Hoferichter, and
  Moulson}]{Cirigliano2023CKM}
\bibinfo{author}{V.~Cirigliano}, \bibinfo{author}{A.~Crivellin},
  \bibinfo{author}{M.~Hoferichter}, \bibinfo{author}{M.~Moulson},
\newblock \bibinfo{title}{Scrutinizing ckm unitarity with a new perspective on
  the cabibbo angle anomaly},
\newblock \bibinfo{journal}{Physics Letters B} \bibinfo{volume}{838}
  (\bibinfo{year}{2023}) \bibinfo{pages}{137748}.
  \DOIprefix\doi{10.1016/j.physletb.2023.137748}.
%Type = Article
\bibitem[{Sirlin(1980)}]{Sirlin1980}
\bibinfo{author}{A.~Sirlin},
\newblock \bibinfo{title}{Radiative corrections in the su(2)l × u(1) theory: a
  simple renormalization framework},
\newblock \bibinfo{journal}{Physical Review D} \bibinfo{volume}{22}
  (\bibinfo{year}{1980}) \bibinfo{pages}{971--981}.
  \DOIprefix\doi{10.1103/PhysRevD.22.971}.
%Type = Article
\bibitem[{Brivio et~al.(2021)Brivio, Dawson, de~Blas, Durieux, Savard, Denner,
  Freitas, Hays, Pecjak, and Vicini}]{Brivio2021}
\bibinfo{author}{I.~Brivio}, \bibinfo{author}{S.~Dawson},
  \bibinfo{author}{J.~de~Blas}, \bibinfo{author}{G.~Durieux},
  \bibinfo{author}{P.~Savard}, \bibinfo{author}{A.~Denner},
  \bibinfo{author}{A.~Freitas}, \bibinfo{author}{C.~Hays},
  \bibinfo{author}{B.~Pecjak}, \bibinfo{author}{A.~Vicini},
\newblock \bibinfo{title}{Electroweak input parameters}
  (\bibinfo{year}{2021}). \DOIprefix\doi{10.48550/arXiv.2111.12515}.
  \href{http://arxiv.org/abs/2111.12515}{{\tt arXiv:2111.12515}}.
%Type = Article
\bibitem[{Kinoshita and Sirlin(1957)}]{KinoshitaSirlin1957}
\bibinfo{author}{T.~Kinoshita}, \bibinfo{author}{A.~Sirlin},
\newblock \bibinfo{title}{Polarization of electrons in muon decay with general
  parity-nonconserving interactions},
\newblock \bibinfo{journal}{Physical Review} \bibinfo{volume}{108}
  (\bibinfo{year}{1957}) \bibinfo{pages}{844--851}.
  \DOIprefix\doi{10.1103/PhysRev.108.844}.
%Type = Article
\bibitem[{Bouchiat and Michel(1957)}]{BouchiatMichel1957}
\bibinfo{author}{C.~Bouchiat}, \bibinfo{author}{L.~Michel},
\newblock \bibinfo{title}{Theory of $\mu$-meson decay with the hypothesis of
  nonconservation of parity},
\newblock \bibinfo{journal}{Physical Review} \bibinfo{volume}{106}
  (\bibinfo{year}{1957}) \bibinfo{pages}{170--172}.
  \DOIprefix\doi{10.1103/PhysRev.106.170}.
%Type = Article
\bibitem[{Fetscher(2021)}]{Fetscher2021MuonDecay}
\bibinfo{author}{W.~Fetscher},
\newblock \bibinfo{title}{Muon decay},
\newblock \bibinfo{journal}{SciPost Physics Proceedings} \bibinfo{volume}{5}
  (\bibinfo{year}{2021}) \bibinfo{pages}{006}.
  \DOIprefix\doi{10.21468/SciPostPhysProc.5.006}.
%Type = Article
\bibitem[{Fetscher et~al.(1986)Fetscher, Gerber, and Johnson}]{Fetscher1986}
\bibinfo{author}{W.~Fetscher}, \bibinfo{author}{H.-J. Gerber},
  \bibinfo{author}{K.~F. Johnson},
\newblock \bibinfo{title}{Muon decay: complete determination of the interaction
  and comparison with the standard model},
\newblock \bibinfo{journal}{Physics Letters B} \bibinfo{volume}{173}
  (\bibinfo{year}{1986}) \bibinfo{pages}{102--106}.
  \DOIprefix\doi{10.1016/0370-2693(86)91239-6}.
%Type = Article
\bibitem[{Rosenson(1958)}]{Rosenson1958}
\bibinfo{author}{L.~Rosenson},
\newblock \bibinfo{title}{Momentum spectrum of positrons from the decay of
  $\mu^+$ mesons},
\newblock \bibinfo{journal}{Physical Review} \bibinfo{volume}{109}
  (\bibinfo{year}{1958}) \bibinfo{pages}{958--972}. \URLprefix
  \url{https://doi.org/10.1103/PhysRev.109.958}.
  \DOIprefix\doi{10.1103/PhysRev.109.958}.
%Type = Article
\bibitem[{Plano(1960)}]{Plano1960}
\bibinfo{author}{R.~J. Plano},
\newblock \bibinfo{title}{Momentum and asymmetry spectrum of $\mu$-meson
  decay},
\newblock \bibinfo{journal}{Physical Review} \bibinfo{volume}{119}
  (\bibinfo{year}{1960}) \bibinfo{pages}{1400--1408}.
  \DOIprefix\doi{10.1103/PhysRev.119.1400}.
%Type = Article
\bibitem[{Sherwood(1967)}]{Sherwood1967}
\bibinfo{author}{B.~A. Sherwood},
\newblock \bibinfo{title}{Measurement of the muon-decay spectrum with a wire
  spark-chamber spectrometer},
\newblock \bibinfo{journal}{Physical Review} \bibinfo{volume}{156}
  (\bibinfo{year}{1967}) \bibinfo{pages}{1475--1489}.
  \DOIprefix\doi{10.1103/PhysRev.156.1475}.
%Type = Article
\bibitem[{Fryberger(1968)}]{Fryberger1968}
\bibinfo{author}{D.~Fryberger},
\newblock \bibinfo{title}{Measurement of the muon decay spectrum with a wire
  spark-chamber spectrometer},
\newblock \bibinfo{journal}{Physical Review} \bibinfo{volume}{166}
  (\bibinfo{year}{1968}) \bibinfo{pages}{1379--1391}.
  \DOIprefix\doi{10.1103/PhysRev.166.1379}.
%Type = Article
\bibitem[{Derenzo(1969)}]{DeRenzio1969}
\bibinfo{author}{S.~E. Derenzo},
\newblock \bibinfo{title}{Measurement of the low-energy end of the $\mu^+$
  decay spectrum},
\newblock \bibinfo{journal}{Physical Review} \bibinfo{volume}{181}
  (\bibinfo{year}{1969}) \bibinfo{pages}{1854--1866}.
  \DOIprefix\doi{10.1103/PhysRev.181.1854}.
%Type = Article
\bibitem[{B\"uhler et~al.(1963)B\"uhler, Cabibbo, Fidecaro, Massam, M\"uller,
  Schneegans, and Zichichi}]{Buhler1963}
\bibinfo{author}{A.~B\"uhler}, \bibinfo{author}{N.~Cabibbo},
  \bibinfo{author}{M.~Fidecaro}, \bibinfo{author}{T.~Massam},
  \bibinfo{author}{T.~M\"uller}, \bibinfo{author}{M.~Schneegans},
  \bibinfo{author}{A.~Zichichi},
\newblock \bibinfo{title}{A measurement of the $e^+$ polarisation in muon
  decay: the $e^+$ annihilation method},
\newblock \bibinfo{journal}{Phys. Lett.} \bibinfo{volume}{7}
  (\bibinfo{year}{1963}) \bibinfo{pages}{368--371}.
  \DOIprefix\doi{10.1016/0031-9163(63)90074-X}.
%Type = Article
\bibitem[{Bloom et~al.(1964)Bloom, Dick, Feuvrais, Henry, Macq, and
  Spighel}]{Bloom1964}
\bibinfo{author}{S.~Bloom}, \bibinfo{author}{L.~A. Dick},
  \bibinfo{author}{L.~Feuvrais}, \bibinfo{author}{G.~R. Henry},
  \bibinfo{author}{P.~C. Macq}, \bibinfo{author}{M.~Spighel},
\newblock \bibinfo{title}{Measurement, by bremsstrahlung transmission, of the
  polarisation of the {$e^+$} from {$\mu^+ \to e^+ + \nu_e + \bar{\nu}_\mu$}},
\newblock \bibinfo{journal}{Phys. Lett.} \bibinfo{volume}{8}
  (\bibinfo{year}{1964}) \bibinfo{pages}{87--89}.
  \DOIprefix\doi{10.1016/0031-9163(64)90810-8}.
%Type = Article
\bibitem[{Duclos et~al.(1964)Duclos, Heintze, de~R\'ujula, and
  Soergel}]{Duclos1964}
\bibinfo{author}{J.~Duclos}, \bibinfo{author}{J.~Heintze},
  \bibinfo{author}{A.~de~R\'ujula}, \bibinfo{author}{V.~Soergel},
\newblock \bibinfo{title}{Measurement of the $e^+$ polarisation in muon decay
  by means of {Bhabha} scattering},
\newblock \bibinfo{journal}{Phys. Lett.} \bibinfo{volume}{9}
  (\bibinfo{year}{1964}) \bibinfo{pages}{62--65}.
  \DOIprefix\doi{10.1016/0031-9163(64)90215-X}.
%Type = Article
\bibitem[{Schwartz(1967)}]{PhysRev.162.1306}
\bibinfo{author}{D.~M. Schwartz},
\newblock \bibinfo{title}{Helicity of electrons from
  ${\ensuremath{\mu}}^{\ensuremath{-}}$ decay},
\newblock \bibinfo{journal}{Phys. Rev.} \bibinfo{volume}{162}
  (\bibinfo{year}{1967}) \bibinfo{pages}{1306--1319}.
  \DOIprefix\doi{10.1103/PhysRev.162.1306}.
%Type = Article
\bibitem[{Burkard et~al.(1985)Burkard, Corriveau, Egger, Fetscher, Gerber,
  Johnson, Kaspar, Mahler, Salzmann, and Scheck}]{Burkard1985a}
\bibinfo{author}{H.~Burkard}, \bibinfo{author}{F.~Corriveau},
  \bibinfo{author}{J.~Egger}, \bibinfo{author}{W.~Fetscher},
  \bibinfo{author}{H.-J. Gerber}, \bibinfo{author}{K.~F. Johnson},
  \bibinfo{author}{H.~Kaspar}, \bibinfo{author}{H.~J. Mahler},
  \bibinfo{author}{M.~Salzmann}, \bibinfo{author}{F.~Scheck},
\newblock \bibinfo{title}{Muon decay: measurement of the positron polarization
  and implications for the spectrum shape parameter $\eta$, {V-A} and
  {T}-invariance},
\newblock \bibinfo{journal}{Physics Letters B} \bibinfo{volume}{150}
  (\bibinfo{year}{1985}) \bibinfo{pages}{242--246}.
  \DOIprefix\doi{10.1016/0370-2693(85)90178-9}.
%Type = Article
\bibitem[{Corriveau et~al.(1983)Corriveau, Egger, Fetscher, Gerber, Johnson,
  Kaspar, Mahler, Salzmann, and Scheck}]{Corriveau1983}
\bibinfo{author}{F.~Corriveau}, \bibinfo{author}{J.~Egger},
  \bibinfo{author}{W.~Fetscher}, \bibinfo{author}{H.-J. Gerber},
  \bibinfo{author}{K.~F. Johnson}, \bibinfo{author}{H.~Kaspar},
  \bibinfo{author}{H.~J. Mahler}, \bibinfo{author}{M.~Salzmann},
  \bibinfo{author}{F.~Scheck},
\newblock \bibinfo{title}{Does the positron from muon decay have transverse
  polarization?},
\newblock \bibinfo{journal}{Phys. Lett. B} \bibinfo{volume}{129}
  (\bibinfo{year}{1983}) \bibinfo{pages}{260--264}.
  \DOIprefix\doi{10.1016/0370-2693(83)90857-2}.
%Type = Article
\bibitem[{Burkard et~al.(1985)Burkard, Corriveau, Egger, Fetscher, Gerber,
  Johnson, Kaspar, Mahler, Salzmann, and Scheck}]{Burkard1985}
\bibinfo{author}{H.~Burkard}, \bibinfo{author}{F.~Corriveau},
  \bibinfo{author}{J.~Egger}, \bibinfo{author}{W.~Fetscher},
  \bibinfo{author}{H.-J. Gerber}, \bibinfo{author}{K.~F. Johnson},
  \bibinfo{author}{H.~Kaspar}, \bibinfo{author}{H.~J. Mahler},
  \bibinfo{author}{M.~Salzmann}, \bibinfo{author}{F.~Scheck},
\newblock \bibinfo{title}{Muon decay: measurement of the transverse positron
  polarization and general analysis},
\newblock \bibinfo{journal}{Phys. Lett. B} \bibinfo{volume}{160}
  (\bibinfo{year}{1985}) \bibinfo{pages}{343--348}.
  \DOIprefix\doi{10.1016/0370-2693(85)91340-1}.
%Type = Article
\bibitem[{Bayes et~al.(2011)}]{Bayes2011}
\bibinfo{author}{R.~Bayes}, et~al.,
\newblock \bibinfo{title}{Experimental constraints on left-right symmetric
  models from muon decay},
\newblock \bibinfo{journal}{Physical Review Letters} \bibinfo{volume}{106}
  (\bibinfo{year}{2011}) \bibinfo{pages}{041804}.
  \DOIprefix\doi{10.1103/PhysRevLett.106.041804}.
%Type = Article
\bibitem[{Bueno et~al.(2011)}]{Bueno:2011fq}
\bibinfo{author}{J.~Bueno}, et~al. (\bibinfo{collaboration}{TWIST
  Collaboration}),
\newblock \bibinfo{title}{{Precise measurement of parity violation in polarized
  muon decay}},
\newblock \bibinfo{journal}{Phys.Rev.} \bibinfo{volume}{D84}
  (\bibinfo{year}{2011}) \bibinfo{pages}{032005}.
  \DOIprefix\doi{10.1103/PhysRevD.84.032005}.
  \href{http://arxiv.org/abs/1104.3632}{{\tt arXiv:1104.3632}}.
%Type = Article
\bibitem[{Bueno et~al.(2012)}]{Bueno:2012PublisherNote}
\bibinfo{author}{J.~Bueno}, et~al. (\bibinfo{collaboration}{TWIST
  Collaboration}),
\newblock \bibinfo{title}{{Publisher's note: precise measurement of parity
  violation in polarized muon decay [phys. rev. d 84, 032005 (2011)]}},
\newblock \bibinfo{journal}{Phys.Rev.} \bibinfo{volume}{D85}
  (\bibinfo{year}{2012}) \bibinfo{pages}{039908}.
  \DOIprefix\doi{10.1103/PhysRevD.85.039908}.
%Type = Article
\bibitem[{MacDonald et~al.(2008)}]{MacDonald2008}
\bibinfo{author}{R.~P. MacDonald}, et~al.,
\newblock \bibinfo{title}{Precision measurement of the muon decay parameters
  $\rho$ and $\delta$},
\newblock \bibinfo{journal}{Physical Review D} \bibinfo{volume}{78}
  (\bibinfo{year}{2008}) \bibinfo{pages}{032010}.
  \DOIprefix\doi{10.1103/PhysRevD.78.032010}.
%Type = Article
\bibitem[{Prieels et~al.(2014)Prieels, Naviliat-Cuncic, Knowles, Van~Hove,
  Morelle et~al.}]{Prieels:2014paa}
\bibinfo{author}{R.~Prieels}, \bibinfo{author}{O.~Naviliat-Cuncic},
  \bibinfo{author}{P.~Knowles}, \bibinfo{author}{P.~Van~Hove},
  \bibinfo{author}{X.~Morelle}, et~al.,
\newblock \bibinfo{title}{{Measurement of the parameter $\xi''$ in polarized
  muon decay and implications on exotic couplings of the leptonic weak
  interaction}},
\newblock \bibinfo{journal}{Phys.Rev.} \bibinfo{volume}{D90}
  (\bibinfo{year}{2014}) \bibinfo{pages}{112003}.
  \DOIprefix\doi{10.1103/PhysRevD.90.112003}.
  \href{http://arxiv.org/abs/1408.1472}{{\tt arXiv:1408.1472}}.
%Type = Article
\bibitem[{Danneberg et~al.(2005)Danneberg, Fetscher, K{\"o}hler, Lang,
  Schweizer et~al.}]{Danneberg:2005xv}
\bibinfo{author}{N.~Danneberg}, \bibinfo{author}{W.~Fetscher},
  \bibinfo{author}{K.~K{\"o}hler}, \bibinfo{author}{J.~Lang},
  \bibinfo{author}{T.~Schweizer}, et~al.,
\newblock \bibinfo{title}{{Muon decay: measurement of the transverse
  polarization of the decay positrons and its implications for the Fermi
  coupling constant and time reversal invariance}},
\newblock \bibinfo{journal}{Phys.Rev.Lett.} \bibinfo{volume}{94}
  (\bibinfo{year}{2005}) \bibinfo{pages}{021802}.
  \DOIprefix\doi{10.1103/PhysRevLett.94.021802}.
%Type = Article
\bibitem[{Musser et~al.(2005)}]{Musser2005}
\bibinfo{author}{J.~R. Musser}, et~al.,
\newblock \bibinfo{title}{Measurement of the michel parameter $\rho$ in muon
  decay},
\newblock \bibinfo{journal}{Physical Review Letters} \bibinfo{volume}{94}
  (\bibinfo{year}{2005}) \bibinfo{pages}{101805}.
  \DOIprefix\doi{10.1103/PhysRevLett.94.101805}.
%Type = Article
\bibitem[{Balke et~al.(1988)}]{Balke1988}
\bibinfo{author}{B.~Balke}, et~al.,
\newblock \bibinfo{title}{Precise measurement of the asymmetry parameter
  $\delta$ in muon decay},
\newblock \bibinfo{journal}{Phys. Rev. D} \bibinfo{volume}{37}
  (\bibinfo{year}{1988}) \bibinfo{pages}{587--617}.
  \DOIprefix\doi{10.1103/PhysRevD.37.587}.
%Type = Article
\bibitem[{Gaponenko et~al.(2005)}]{Gaponenko2005}
\bibinfo{author}{A.~Gaponenko}, et~al.,
\newblock \bibinfo{title}{Measurement of the muon decay parameter $\delta$},
\newblock \bibinfo{journal}{Phys. Rev. D} \bibinfo{volume}{71}
  (\bibinfo{year}{2005}) \bibinfo{pages}{071101}.
  \DOIprefix\doi{10.1103/PhysRevD.71.071101}.
%Type = Article
\bibitem[{Bardon et~al.(1959)}]{Bardon1959}
\bibinfo{author}{M.~Bardon}, et~al.,
\newblock \bibinfo{title}{Asymmetry parameter in muon decay},
\newblock \bibinfo{journal}{Phys. Rev. Lett.} \bibinfo{volume}{2}
  (\bibinfo{year}{1959}) \bibinfo{pages}{56--57}.
  \DOIprefix\doi{10.1103/PhysRevLett.2.56}.
%Type = Article
\bibitem[{Beltrami et~al.(1987)}]{Beltrami1987}
\bibinfo{author}{I.~Beltrami}, et~al.,
\newblock \bibinfo{title}{Muon decay: measurement of the integral asymmetry
  parameter},
\newblock \bibinfo{journal}{Physics Letters B} \bibinfo{volume}{194}
  (\bibinfo{year}{1987}) \bibinfo{pages}{326--330}.
  \DOIprefix\doi{10.1016/0370-2693(87)90552-1}.
%Type = Article
\bibitem[{Henderson et~al.(2005)Henderson, Davydov, Faszer, Koetke, Miasoedov,
  Openshaw, Quraan, Schaapman, Selivanov, Sheffer, Stanislaus, and
  Torokhov}]{Henderson2005TWIST}
\bibinfo{author}{R.~S. Henderson}, \bibinfo{author}{Y.~I. Davydov},
  \bibinfo{author}{W.~Faszer}, \bibinfo{author}{D.~D. Koetke},
  \bibinfo{author}{L.~V. Miasoedov}, \bibinfo{author}{R.~Openshaw},
  \bibinfo{author}{M.~A. Quraan}, \bibinfo{author}{J.~Schaapman},
  \bibinfo{author}{V.~Selivanov}, \bibinfo{author}{G.~Sheffer},
  \bibinfo{author}{T.~D.~S. Stanislaus}, \bibinfo{author}{V.~Torokhov},
\newblock \bibinfo{title}{Precision planar drift chambers and cradle for the
  {TWIST} muon decay spectrometer},
\newblock \bibinfo{journal}{Nuclear Instruments and Methods in Physics Research
  Section A: Accelerators, Spectrometers, Detectors and Associated Equipment}
  \bibinfo{volume}{548} (\bibinfo{year}{2005}) \bibinfo{pages}{306--335}.
  \DOIprefix\doi{10.1016/j.nima.2005.02.043}.
  \href{http://arxiv.org/abs/hep-ex/0409066}{{\tt arXiv:hep-ex/0409066}}.
%Type = Article
\bibitem[{Eichenberger et~al.(1984)Eichenberger, Engfer, and van~der
  Schaaf}]{Eichenberger1984}
\bibinfo{author}{W.~Eichenberger}, \bibinfo{author}{R.~Engfer},
  \bibinfo{author}{A.~van~der Schaaf},
\newblock \bibinfo{title}{Measurement of the parameter {$\bar{\eta}$} in the
  radiative decay of the muon as a test of the {$V\!-\!A$} structure of the
  weak interaction},
\newblock \bibinfo{journal}{Nucl. Phys. A} \bibinfo{volume}{412}
  (\bibinfo{year}{1984}) \bibinfo{pages}{523--533}.
  \DOIprefix\doi{10.1016/0375-9474(84)90534-7}.
%Type = Article
\bibitem[{Hillairet et~al.(2012)}]{Hillairet2012}
\bibinfo{author}{A.~Hillairet}, et~al. (\bibinfo{collaboration}{TWIST
  Collaboration}),
\newblock \bibinfo{title}{{Precision muon decay measurements and improved
  constraints on the weak interaction}},
\newblock \bibinfo{journal}{Phys.Rev.} \bibinfo{volume}{D85}
  (\bibinfo{year}{2012}) \bibinfo{pages}{092013}.
  \DOIprefix\doi{10.1103/PhysRevD.85.092013}.
  \href{http://arxiv.org/abs/1112.3606}{{\tt arXiv:1112.3606}}.
%Type = Article
\bibitem[{Eides et~al.(2001)Eides, Grotch, and Shelyuto}]{Eides2001}
\bibinfo{author}{M.~I. Eides}, \bibinfo{author}{H.~Grotch},
  \bibinfo{author}{V.~A. Shelyuto},
\newblock \bibinfo{title}{Theory of light hydrogenlike atoms},
\newblock \bibinfo{journal}{Phys. Rep.} \bibinfo{volume}{342}
  (\bibinfo{year}{2001}) \bibinfo{pages}{63--261}.
  \DOIprefix\doi{10.1016/S0370-1573(00)00077-6}.
%Type = Article
\bibitem[{Fermi(1930)}]{Fermi1930}
\bibinfo{author}{E.~Fermi},
\newblock \bibinfo{title}{über die magnetischen momente der atomkerne},
\newblock \bibinfo{journal}{Z. Phys.} \bibinfo{volume}{60}
  (\bibinfo{year}{1930}) \bibinfo{pages}{320--333}.
%Type = Article
\bibitem[{Bethe(1947)}]{Bethe1947}
\bibinfo{author}{H.~A. Bethe},
\newblock \bibinfo{title}{The electromagnetic shift of energy levels},
\newblock \bibinfo{journal}{Phys. Rev.} \bibinfo{volume}{72}
  (\bibinfo{year}{1947}) \bibinfo{pages}{339--341}.
  \DOIprefix\doi{10.1103/PhysRev.72.339}.
%Type = Article
\bibitem[{Meyer et~al.(2000)}]{Meyer2000}
\bibinfo{author}{V.~Meyer}, et~al.,
\newblock \bibinfo{title}{Measurement of the {$1S$--$2S$} energy interval in
  muonium},
\newblock \bibinfo{journal}{Phys. Rev. Lett.} \bibinfo{volume}{84}
  (\bibinfo{year}{2000}) \bibinfo{pages}{1136--1139}.
  \DOIprefix\doi{10.1103/PhysRevLett.84.1136}.
%Type = Article
\bibitem[{Eides(2019)}]{Eides2019}
\bibinfo{author}{M.~I. Eides},
\newblock \bibinfo{title}{Hyperfine splitting in muonium: accuracy of the
  theoretical prediction},
\newblock \bibinfo{journal}{Phys. Lett. B} \bibinfo{volume}{795}
  (\bibinfo{year}{2019}) \bibinfo{pages}{113--116}.
  \DOIprefix\doi{10.1016/j.physletb.2019.06.011}.
%Type = Article
\bibitem[{Frugiuele et~al.(2019)Frugiuele, P\'erez-R\'{\i}os, and
  Peset}]{PhysRevD.100.015010}
\bibinfo{author}{C.~Frugiuele}, \bibinfo{author}{J.~P\'erez-R\'{\i}os},
  \bibinfo{author}{C.~Peset},
\newblock \bibinfo{title}{Current and future perspectives of positronium and
  muonium spectroscopy as dark sectors probe},
\newblock \bibinfo{journal}{Phys. Rev. D} \bibinfo{volume}{100}
  (\bibinfo{year}{2019}) \bibinfo{pages}{015010}.
  \DOIprefix\doi{10.1103/PhysRevD.100.015010}.
%Type = Article
\bibitem[{Mohr et~al.(2012)Mohr, Taylor, and Newell}]{Mohr:2012tt}
\bibinfo{author}{P.~J. Mohr}, \bibinfo{author}{B.~N. Taylor},
  \bibinfo{author}{D.~B. Newell},
\newblock \bibinfo{title}{{Codata recommended values of the fundamental
  physical constants: 2010}},
\newblock \bibinfo{journal}{Rev.Mod.Phys.} \bibinfo{volume}{84}
  (\bibinfo{year}{2012}) \bibinfo{pages}{1527--1605}.
  \DOIprefix\doi{10.1103/RevModPhys.84.1527}.
  \href{http://arxiv.org/abs/1203.5425}{{\tt arXiv:1203.5425}}.
%Type = Article
\bibitem[{Stambaugh et~al.(1974)Stambaugh, Casperson, Crane, Hughes, Kaspar,
  Souder, Thompson, Orth, zu~Putlitz, and
  Denison}]{Stambaugh1974MuoniumNobleGases}
\bibinfo{author}{R.~D. Stambaugh}, \bibinfo{author}{D.~E. Casperson},
  \bibinfo{author}{T.~W. Crane}, \bibinfo{author}{V.~W. Hughes},
  \bibinfo{author}{H.~F. Kaspar}, \bibinfo{author}{P.~Souder},
  \bibinfo{author}{P.~A. Thompson}, \bibinfo{author}{H.~Orth},
  \bibinfo{author}{G.~zu~Putlitz}, \bibinfo{author}{A.~B. Denison},
\newblock \bibinfo{title}{Muonium formation in noble gases and noble-gas
  mixtures},
\newblock \bibinfo{journal}{Physical Review Letters} \bibinfo{volume}{33}
  (\bibinfo{year}{1974}) \bibinfo{pages}{568--571}.
  \DOIprefix\doi{10.1103/PhysRevLett.33.568}.
%Type = Article
\bibitem[{Breit and Rabi(1931)}]{BreitRabi1931}
\bibinfo{author}{G.~Breit}, \bibinfo{author}{I.~I. Rabi},
\newblock \bibinfo{title}{Measurement of nuclear spin},
\newblock \bibinfo{journal}{Phys. Rev.} \bibinfo{volume}{38}
  (\bibinfo{year}{1931}) \bibinfo{pages}{2082--2083}.
  \DOIprefix\doi{10.1103/PhysRev.38.2082}.
%Type = Article
\bibitem[{Cleland et~al.(1964)Cleland, Bailey, Eckhause, Hughes, Mobley,
  Prepost, and Rothberg}]{PhysRevLett.13.202}
\bibinfo{author}{W.~E. Cleland}, \bibinfo{author}{J.~M. Bailey},
  \bibinfo{author}{M.~Eckhause}, \bibinfo{author}{V.~W. Hughes},
  \bibinfo{author}{R.~M. Mobley}, \bibinfo{author}{R.~Prepost},
  \bibinfo{author}{J.~E. Rothberg},
\newblock \bibinfo{title}{New value for the fine-structure constant
  $\ensuremath{\alpha}$ from muonium hyperfine structure interval},
\newblock \bibinfo{journal}{Phys. Rev. Lett.} \bibinfo{volume}{13}
  (\bibinfo{year}{1964}) \bibinfo{pages}{202--205}.
  \DOIprefix\doi{10.1103/PhysRevLett.13.202}.
%Type = Article
\bibitem[{Ehrlich et~al.(1969)Ehrlich, Hofer, Magnon, Stowell, Swanson, and
  Telegdi}]{Ehrlich1969}
\bibinfo{author}{R.~D. Ehrlich}, \bibinfo{author}{H.~Hofer},
  \bibinfo{author}{A.~Magnon}, \bibinfo{author}{D.~Stowell},
  \bibinfo{author}{R.~A. Swanson}, \bibinfo{author}{V.~L. Telegdi},
\newblock \bibinfo{title}{Determination of the muonium hyperfine splitting at
  low pressure from a field-independent zeeman transition},
\newblock \bibinfo{journal}{Phys. Rev. Lett.} \bibinfo{volume}{23}
  (\bibinfo{year}{1969}) \bibinfo{pages}{513--517}.
  \DOIprefix\doi{10.1103/PhysRevLett.23.513}.
%Type = Article
\bibitem[{De~Voe et~al.(1970)De~Voe, McIntyre, Magnon, Stowell, Swanson
  et~al.}]{DeVoe:1971ph}
\bibinfo{author}{R.~De~Voe}, \bibinfo{author}{P.~McIntyre},
  \bibinfo{author}{A.~Magnon}, \bibinfo{author}{D.~Stowell},
  \bibinfo{author}{R.~Swanson}, et~al.,
\newblock \bibinfo{title}{{Measurement of the muonium HFS splitting and of the
  muon moment by double resonance, and a new value of alpha}},
\newblock \bibinfo{journal}{Phys.Rev.Lett.} \bibinfo{volume}{25}
  (\bibinfo{year}{1970}) \bibinfo{pages}{1779--1783}.
  \DOIprefix\doi{10.1103/PhysRevLett.25.1779}.
%Type = Article
\bibitem[{Mariam et~al.(1982)Mariam, Beer, Bolton, Egan, Gardner, Hughes, Lu,
  Souder, Orth, Vetter, Moser, and zu~Putlitz}]{Mariam1982}
\bibinfo{author}{F.~G. Mariam}, \bibinfo{author}{W.~Beer},
  \bibinfo{author}{P.~R. Bolton}, \bibinfo{author}{P.~O. Egan},
  \bibinfo{author}{C.~J. Gardner}, \bibinfo{author}{V.~W. Hughes},
  \bibinfo{author}{D.~C. Lu}, \bibinfo{author}{P.~A. Souder},
  \bibinfo{author}{H.~Orth}, \bibinfo{author}{J.~Vetter},
  \bibinfo{author}{U.~Moser}, \bibinfo{author}{G.~zu~Putlitz},
\newblock \bibinfo{title}{Higher precision measurement of the hyperfine
  structure interval of muonium and of the muon magnetic moment},
\newblock \bibinfo{journal}{Phys. Rev. Lett.} \bibinfo{volume}{49}
  (\bibinfo{year}{1982}) \bibinfo{pages}{993--996}.
  \DOIprefix\doi{10.1103/PhysRevLett.49.993}.
%Type = Article
\bibitem[{Liu et~al.(1999)}]{Liu1999}
\bibinfo{author}{W.~Liu}, et~al.,
\newblock \bibinfo{title}{High precision measurements of the muonium
  ground-state hyperfine structure interval and of the muon magnetic moment},
\newblock \bibinfo{journal}{Phys. Rev. Lett.} \bibinfo{volume}{82}
  (\bibinfo{year}{1999}) \bibinfo{pages}{711--714}.
  \DOIprefix\doi{10.1103/PhysRevLett.82.711}.
%Type = Article
\bibitem[{Kanda et~al.(2021)}]{Kanda2021}
\bibinfo{author}{S.~Kanda}, et~al.,
\newblock \bibinfo{title}{New precise measurement of the muonium hyperfine
  structure with a high-intensity pulsed muon beam},
\newblock \bibinfo{journal}{Phys. Lett. B} \bibinfo{volume}{815}
  (\bibinfo{year}{2021}) \bibinfo{pages}{136154}.
  \DOIprefix\doi{10.1016/j.physletb.2021.136154}.
%Type = Article
\bibitem[{Oram et~al.(1984)}]{Oram1984}
\bibinfo{author}{C.~J. Oram}, et~al.,
\newblock \bibinfo{title}{Measurement of the lamb shift in muonium},
\newblock \bibinfo{journal}{Phys. Rev. Lett.} \bibinfo{volume}{52}
  (\bibinfo{year}{1984}) \bibinfo{pages}{910--913}.
  \DOIprefix\doi{10.1103/PhysRevLett.52.910}.
%Type = Article
\bibitem[{Woodle et~al.(1990)}]{Woodle1990}
\bibinfo{author}{K.~A. Woodle}, et~al.,
\newblock \bibinfo{title}{Measurement of the lamb shift in the n = 2 state of
  muonium},
\newblock \bibinfo{journal}{Phys. Rev. A} \bibinfo{volume}{41}
  (\bibinfo{year}{1990}) \bibinfo{pages}{93--105}.
  \DOIprefix\doi{10.1103/PhysRevA.41.93}.
%Type = Article
\bibitem[{Ohayon et~al.(2022)}]{Ohayon2022}
\bibinfo{author}{B.~Ohayon}, et~al.,
\newblock \bibinfo{title}{Precision measurement of the lamb shift in muonium},
\newblock \bibinfo{journal}{Phys. Rev. Lett.} \bibinfo{volume}{128}
  (\bibinfo{year}{2022}) \bibinfo{pages}{011802}.
  \DOIprefix\doi{10.1103/PhysRevLett.128.011802}.
%Type = Article
\bibitem[{Chu et~al.(1988)Chu, Mills, Yodh, Nagamine, Miyake, and
  Kuga}]{Chu1988Muonium1S2S}
\bibinfo{author}{S.~Chu}, \bibinfo{author}{A.~P. Mills}, \bibinfo{author}{A.~G.
  Yodh}, \bibinfo{author}{K.~Nagamine}, \bibinfo{author}{Y.~Miyake},
  \bibinfo{author}{T.~Kuga},
\newblock \bibinfo{title}{Laser excitation of the muonium 1s--2s transition},
\newblock \bibinfo{journal}{Physical Review Letters} \bibinfo{volume}{60}
  (\bibinfo{year}{1988}) \bibinfo{pages}{101--104}.
  \DOIprefix\doi{10.1103/PhysRevLett.60.101}.
%Type = Article
\bibitem[{Maas et~al.(1994)Maas, Braun, Geerds, Jungmann, Matthias, zu~Putlitz,
  Reinhard, Schwarz, Willmann, Zhang, Baird, Sandars, Woodman, Eaton, Matousek,
  Toner, Towrie, Barr, Ferguson, Persaud, Riis, Berkeland, Boshier, Hughes, and
  Woodle}]{Maas1994Muonium1S2S}
\bibinfo{author}{F.~Maas}, \bibinfo{author}{B.~Braun},
  \bibinfo{author}{H.~Geerds}, \bibinfo{author}{K.~Jungmann},
  \bibinfo{author}{B.~Matthias}, \bibinfo{author}{G.~zu~Putlitz},
  \bibinfo{author}{I.~Reinhard}, \bibinfo{author}{W.~Schwarz},
  \bibinfo{author}{L.~Willmann}, \bibinfo{author}{L.~Zhang},
  \bibinfo{author}{P.~Baird}, \bibinfo{author}{P.~Sandars},
  \bibinfo{author}{G.~Woodman}, \bibinfo{author}{G.~Eaton},
  \bibinfo{author}{P.~Matousek}, \bibinfo{author}{W.~Toner},
  \bibinfo{author}{M.~Towrie}, \bibinfo{author}{J.~Barr},
  \bibinfo{author}{A.~Ferguson}, \bibinfo{author}{M.~Persaud},
  \bibinfo{author}{E.~Riis}, \bibinfo{author}{D.~Berkeland},
  \bibinfo{author}{M.~Boshier}, \bibinfo{author}{V.~Hughes},
  \bibinfo{author}{K.~Woodle},
\newblock \bibinfo{title}{{A measurement of the 1S--2S transition frequency in
  muonium}},
\newblock \bibinfo{journal}{Phys. Lett. A} \bibinfo{volume}{187}
  (\bibinfo{year}{1994}) \bibinfo{pages}{247--254}.
  \DOIprefix\doi{10.1016/0375-9601(94)90903-2}.
%Type = Article
\bibitem[{Cortinovis et~al.(2023)Cortinovis, Ohayon, de~Sousa~Borges, Janka,
  Golovizin, Zhadnov, and Crivelli}]{Cortinovis2023}
\bibinfo{author}{I.~Cortinovis}, \bibinfo{author}{B.~Ohayon},
  \bibinfo{author}{L.~de~Sousa~Borges}, \bibinfo{author}{G.~Janka},
  \bibinfo{author}{A.~Golovizin}, \bibinfo{author}{N.~Zhadnov},
  \bibinfo{author}{P.~Crivelli},
\newblock \bibinfo{title}{Update of {muonium} {$1S$--$2S$} transition
  frequency},
\newblock \bibinfo{journal}{Eur. Phys. J. D} \bibinfo{volume}{77}
  (\bibinfo{year}{2023}) \bibinfo{pages}{66}.
  \DOIprefix\doi{10.1140/epjd/s10053-023-00639-z}.
%Type = Article
\bibitem[{Favart et~al.(1971)Favart, McIntyre, Stowell, Telegdi, DeVoe, and
  Swanson}]{PhysRevLett.27.1336}
\bibinfo{author}{D.~Favart}, \bibinfo{author}{P.~M. McIntyre},
  \bibinfo{author}{D.~Y. Stowell}, \bibinfo{author}{V.~L. Telegdi},
  \bibinfo{author}{R.~DeVoe}, \bibinfo{author}{R.~A. Swanson},
\newblock \bibinfo{title}{Precision measurement of the hyperfine interval of
  muonium by a novel technique: Ramsey resonance in zero field},
\newblock \bibinfo{journal}{Phys. Rev. Lett.} \bibinfo{volume}{27}
  (\bibinfo{year}{1971}) \bibinfo{pages}{1336--1340}.
  \DOIprefix\doi{10.1103/PhysRevLett.27.1336}.
%Type = Article
\bibitem[{Kobrak et~al.(1973)Kobrak, Swanson, Favart, Kells, Magnon, McIntyre,
  Roehrig, Stowell, Telegdi, and Eckhause}]{Kobrak1973}
\bibinfo{author}{H.~G.~E. Kobrak}, \bibinfo{author}{R.~A. Swanson},
  \bibinfo{author}{D.~Favart}, \bibinfo{author}{W.~Kells},
  \bibinfo{author}{A.~Magnon}, \bibinfo{author}{P.~M. McIntyre},
  \bibinfo{author}{J.~Roehrig}, \bibinfo{author}{D.~Y. Stowell},
  \bibinfo{author}{V.~L. Telegdi}, \bibinfo{author}{M.~Eckhause},
\newblock \bibinfo{title}{A new precision measurement of the muonium hyperfine
  interval in argon},
\newblock \bibinfo{journal}{Phys. Lett. B} \bibinfo{volume}{43}
  (\bibinfo{year}{1973}) \bibinfo{pages}{526--528}.
  \DOIprefix\doi{10.1016/0370-2693(73)90362-6}.
%Type = Article
\bibitem[{Casperson et~al.(1975)}]{Casperson1975}
\bibinfo{author}{D.~E. Casperson}, et~al.,
\newblock \bibinfo{title}{A new high precision measurement of the muonium
  hyperfine structure interval},
\newblock \bibinfo{journal}{Phys. Lett. B} \bibinfo{volume}{59}
  (\bibinfo{year}{1975}) \bibinfo{pages}{397–400}.
  \DOIprefix\doi{10.1016/0370-2693(75)90099-4}.
%Type = Article
\bibitem[{Danzmann et~al.(1989)Danzmann, Fee, and Chu}]{Danzmann:1989}
\bibinfo{author}{K.~Danzmann}, \bibinfo{author}{M.~Fee},
  \bibinfo{author}{S.~Chu},
\newblock \bibinfo{title}{{Doppler-free laser spectroscopy of positronium and
  muonium: reanalysis of the 1S--2S measurements}},
\newblock \bibinfo{journal}{Phys. Rev. A} \bibinfo{volume}{39}
  (\bibinfo{year}{1989}) \bibinfo{pages}{6072--6073}.
  \DOIprefix\doi{10.1103/PhysRevA.39.6072}.
%Type = Article
\bibitem[{Farley and Picasso(1979)}]{FarleyPicasso1979}
\bibinfo{author}{F.~J.~M. Farley}, \bibinfo{author}{E.~Picasso},
\newblock \bibinfo{title}{The muon ($g-2$) experiments},
\newblock \bibinfo{journal}{Annual Review of Nuclear and Particle Science}
  \bibinfo{volume}{29} (\bibinfo{year}{1979}) \bibinfo{pages}{243--282}.
  \DOIprefix\doi{10.1146/annurev.ns.29.120179.001331}.
%Type = Article
\bibitem[{Miller et~al.(2007)Miller, de~Rafael, and Roberts}]{Miller2007}
\bibinfo{author}{J.~P. Miller}, \bibinfo{author}{E.~de~Rafael},
  \bibinfo{author}{B.~L. Roberts},
\newblock \bibinfo{title}{Muon ($g-2$): experiment and theory},
\newblock \bibinfo{journal}{Reports on Progress in Physics}
  \bibinfo{volume}{70} (\bibinfo{year}{2007}) \bibinfo{pages}{795--881}.
  \DOIprefix\doi{10.1088/0034-4885/70/5/R03}.
%Type = Article
\bibitem[{Lamb and Retherford(1947)}]{LambRetherford1947}
\bibinfo{author}{W.~E.~J. Lamb}, \bibinfo{author}{R.~C. Retherford},
\newblock \bibinfo{title}{Fine structure of the hydrogen atom by a microwave
  method},
\newblock \bibinfo{journal}{Phys. Rev.} \bibinfo{volume}{72}
  (\bibinfo{year}{1947}) \bibinfo{pages}{241--243}.
  \DOIprefix\doi{10.1103/PhysRev.72.241}.
%Type = Article
\bibitem[{Kusch and Foley(1948)}]{KuschFoley1948}
\bibinfo{author}{P.~Kusch}, \bibinfo{author}{H.~M. Foley},
\newblock \bibinfo{title}{The magnetic moment of the electron},
\newblock \bibinfo{journal}{Phys. Rev.} \bibinfo{volume}{74}
  (\bibinfo{year}{1948}) \bibinfo{pages}{250--263}.
  \DOIprefix\doi{10.1103/PhysRev.74.250}.
%Type = Article
\bibitem[{Schwinger(1948)}]{Schwinger1948}
\bibinfo{author}{J.~Schwinger},
\newblock \bibinfo{title}{On quantum-electrodynamics and the magnetic moment of
  the electron},
\newblock \bibinfo{journal}{Phys. Rev.} \bibinfo{volume}{73}
  (\bibinfo{year}{1948}) \bibinfo{pages}{416--417}.
  \DOIprefix\doi{10.1103/PhysRev.73.416}.
%Type = Article
\bibitem[{Sommerfield(1957)}]{Sommerfield1957}
\bibinfo{author}{C.~M. Sommerfield},
\newblock \bibinfo{title}{Magnetic dipole moment of the electron},
\newblock \bibinfo{journal}{Phys. Rev.} \bibinfo{volume}{107}
  (\bibinfo{year}{1957}) \bibinfo{pages}{328--329}.
  \DOIprefix\doi{10.1103/PhysRev.107.328}.
%Type = Article
\bibitem[{Elend(1966)}]{Elend1966}
\bibinfo{author}{H.~H. Elend},
\newblock \bibinfo{title}{On the anomalous magnetic moment of the muon},
\newblock \bibinfo{journal}{Physics Letters} \bibinfo{volume}{20}
  (\bibinfo{year}{1966}) \bibinfo{pages}{682--684}.
  \DOIprefix\doi{10.1016/0031-9163(66)91171-1}.
%Type = Article
\bibitem[{Fujikawa et~al.(1972)Fujikawa, Lee, and Sanda}]{Fujikawa1972}
\bibinfo{author}{K.~Fujikawa}, \bibinfo{author}{B.~W. Lee},
  \bibinfo{author}{A.~I. Sanda},
\newblock \bibinfo{title}{Generalized renormalizable gauge formulation of
  spontaneously broken gauge theories},
\newblock \bibinfo{journal}{Physical Review D} \bibinfo{volume}{6}
  (\bibinfo{year}{1972}) \bibinfo{pages}{2923--2943}.
  \DOIprefix\doi{10.1103/PhysRevD.6.2923}.
%Type = Article
\bibitem[{Jackiw and Weinberg(1972)}]{JackiwWeinberg1972}
\bibinfo{author}{R.~Jackiw}, \bibinfo{author}{S.~Weinberg},
\newblock \bibinfo{title}{Weak-interaction corrections to the muon magnetic
  moment and to muonic-atom energy levels},
\newblock \bibinfo{journal}{Physical Review D} \bibinfo{volume}{5}
  (\bibinfo{year}{1972}) \bibinfo{pages}{2396--2398}.
  \DOIprefix\doi{10.1103/PhysRevD.5.2396}.
%Type = Article
\bibitem[{Bouchiat and Michel(1961)}]{BouchiatMichel1961}
\bibinfo{author}{C.~Bouchiat}, \bibinfo{author}{L.~Michel},
\newblock \bibinfo{title}{La r{\'e}sonance dans la diffusion
  m{\'e}son--m{\'e}son et le moment magn{\'e}tique anormal du m{\'e}son $\mu$},
\newblock \bibinfo{journal}{Journal de Physique et le Radium}
  \bibinfo{volume}{22} (\bibinfo{year}{1961}) \bibinfo{pages}{121}.
  \DOIprefix\doi{10.1051/jphysrad:01961002202012101}.
%Type = Article
\bibitem[{Gourdin and de~Rafael(1969)}]{GourdinDeRafael1969}
\bibinfo{author}{M.~Gourdin}, \bibinfo{author}{E.~de~Rafael},
\newblock \bibinfo{title}{Hadronic contributions to the muon g-factor},
\newblock \bibinfo{journal}{Nuclear Physics B} \bibinfo{volume}{10}
  (\bibinfo{year}{1969}) \bibinfo{pages}{667--674}.
  \DOIprefix\doi{10.1016/0550-3213(69)90333-2}.
%Type = Article
\bibitem[{Aoyama et~al.(2020)}]{Aoyama2020}
\bibinfo{author}{T.~Aoyama}, et~al.,
\newblock \bibinfo{title}{The anomalous magnetic moment of the muon in the
  standard model},
\newblock \bibinfo{journal}{Physics Reports} \bibinfo{volume}{887}
  (\bibinfo{year}{2020}) \bibinfo{pages}{1--166}.
  \DOIprefix\doi{10.1016/j.physrep.2020.07.006}.
%Type = Article
\bibitem[{Aliberti et~al.(2025)}]{Aliberti2025}
\bibinfo{author}{R.~Aliberti}, et~al.,
\newblock \bibinfo{title}{The anomalous magnetic moment of the muon in the
  standard model: an update},
\newblock \bibinfo{journal}{Physics Reports} \bibinfo{volume}{1143}
  (\bibinfo{year}{2025}) \bibinfo{pages}{1--158}.
  \DOIprefix\doi{10.1016/j.physrep.2025.08.002}.
%Type = Article
\bibitem[{Lees et~al.(2012)}]{BaBar2012}
\bibinfo{author}{J.~P. Lees}, et~al. (\bibinfo{collaboration}{BaBar
  Collaboration}),
\newblock \bibinfo{title}{Precise measurement of the $e^+e^- \to
  \pi^+\pi^-(\gamma)$ cross section with the initial-state radiation method at
  babar},
\newblock \bibinfo{journal}{Physical Review D} \bibinfo{volume}{86}
  (\bibinfo{year}{2012}) \bibinfo{pages}{032013}.
  \DOIprefix\doi{10.1103/PhysRevD.86.032013}.
  \href{http://arxiv.org/abs/1205.2228}{{\tt arXiv:1205.2228}}.
%Type = Article
\bibitem[{Babusci et~al.(2013)}]{KLOE2013}
\bibinfo{author}{D.~Babusci}, et~al. (\bibinfo{collaboration}{KLOE
  Collaboration}),
\newblock \bibinfo{title}{Precision measurement of $\sigma(e^+e^- \to
  \pi^+\pi^-(\gamma))$ with the kloe detector},
\newblock \bibinfo{journal}{Physics Letters B} \bibinfo{volume}{720}
  (\bibinfo{year}{2013}) \bibinfo{pages}{336--343}.
  \DOIprefix\doi{10.1016/j.physletb.2013.02.029}.
%Type = Article
\bibitem[{Ablikim et~al.(2016)}]{BESIII2016}
\bibinfo{author}{M.~Ablikim}, et~al. (\bibinfo{collaboration}{BESIII
  Collaboration}),
\newblock \bibinfo{title}{Measurement of the $e^+e^- \to \pi^+\pi^-$ cross
  section between 600 and 900 mev using initial state radiation},
\newblock \bibinfo{journal}{Physics Letters B} \bibinfo{volume}{753}
  (\bibinfo{year}{2016}) \bibinfo{pages}{629--638}.
  \DOIprefix\doi{10.1016/j.physletb.2015.11.043}.
%Type = Article
\bibitem[{Achasov et~al.(2021)}]{SND2020update}
\bibinfo{author}{M.~N. Achasov}, et~al. (\bibinfo{collaboration}{SND
  Collaboration}),
\newblock \bibinfo{title}{Measurement of the $e^+e^- \to \pi^+\pi^-$ process
  cross section with the snd detector at the vepp-2000 collider in the energy
  region $0.525<\sqrt{s}<0.883$ gev},
\newblock \bibinfo{journal}{Journal of High Energy Physics}
  \bibinfo{volume}{2021} (\bibinfo{year}{2021}) \bibinfo{pages}{113}.
  \DOIprefix\doi{10.1007/JHEP01(2021)113}.
  \href{http://arxiv.org/abs/2004.00263}{{\tt arXiv:2004.00263}}.
%Type = Article
\bibitem[{Ignatov et~al.(2024)}]{CMD32024}
\bibinfo{author}{F.~V. Ignatov}, et~al. (\bibinfo{collaboration}{CMD-3
  Collaboration}),
\newblock \bibinfo{title}{Measurement of the $e^+e^-\to\pi^+\pi^-$ cross
  section from threshold to 1.2 gev with the cmd-3 detector},
\newblock \bibinfo{journal}{Physical Review D} \bibinfo{volume}{109}
  (\bibinfo{year}{2024}) \bibinfo{pages}{112002}.
  \DOIprefix\doi{10.1103/PhysRevD.109.112002}.
%Type = Article
\bibitem[{Blum et~al.(2018)}]{Blum2018}
\bibinfo{author}{T.~Blum}, et~al.,
\newblock \bibinfo{title}{Calculation of the hadronic vacuum polarization
  contribution to the muon anomalous magnetic moment},
\newblock \bibinfo{journal}{Physical Review Letters} \bibinfo{volume}{121}
  (\bibinfo{year}{2018}) \bibinfo{pages}{022003}.
  \DOIprefix\doi{10.1103/PhysRevLett.121.022003}.
%Type = Article
\bibitem[{Borsanyi et~al.(2021)}]{Borsanyi2021}
\bibinfo{author}{S.~Borsanyi}, et~al.,
\newblock \bibinfo{title}{Leading hadronic contribution to the muon magnetic
  moment from lattice qcd},
\newblock \bibinfo{journal}{Nature} \bibinfo{volume}{593}
  (\bibinfo{year}{2021}) \bibinfo{pages}{51--55}.
  \DOIprefix\doi{10.1038/s41586-021-03418-1}.
%Type = Article
\bibitem[{Cassels et~al.(1957)Cassels, O'Keeffe, Rigby, Wetherell, and
  Wormald}]{Cassels1957}
\bibinfo{author}{J.~M. Cassels}, \bibinfo{author}{T.~W. O'Keeffe},
  \bibinfo{author}{M.~Rigby}, \bibinfo{author}{A.~M. Wetherell},
  \bibinfo{author}{J.~R. Wormald},
\newblock \bibinfo{title}{Experiments with a polarized muon beam},
\newblock \bibinfo{journal}{Proceedings of the Physical Society. Section A}
  \bibinfo{volume}{70} (\bibinfo{year}{1957}) \bibinfo{pages}{543--546}.
  \DOIprefix\doi{10.1088/0370-1298/70/7/412}.
%Type = Article
\bibitem[{Garwin et~al.(1960)Garwin, Hutchinson, Penman, and
  Shapiro}]{Garwin1960}
\bibinfo{author}{R.~L. Garwin}, \bibinfo{author}{D.~P. Hutchinson},
  \bibinfo{author}{S.~Penman}, \bibinfo{author}{G.~Shapiro},
\newblock \bibinfo{title}{Accurate determination of the $\mu^+$ magnetic
  moment},
\newblock \bibinfo{journal}{Physical Review} \bibinfo{volume}{118}
  (\bibinfo{year}{1960}) \bibinfo{pages}{271--283}.
  \DOIprefix\doi{10.1103/PhysRev.118.271}.
%Type = Article
\bibitem[{Bargmann et~al.(1959)Bargmann, Michel, and Telegdi}]{Bargmann1959}
\bibinfo{author}{V.~Bargmann}, \bibinfo{author}{L.~Michel},
  \bibinfo{author}{V.~L. Telegdi},
\newblock \bibinfo{title}{Precession of the polarization of particles moving in
  a homogeneous electromagnetic field},
\newblock \bibinfo{journal}{Physical Review Letters} \bibinfo{volume}{2}
  (\bibinfo{year}{1959}) \bibinfo{pages}{435--436}.
  \DOIprefix\doi{10.1103/PhysRevLett.2.435}.
%Type = Article
\bibitem[{Bailey et~al.(1968)Bailey, Bartl, von Bochmann, Brown, Farley,
  Jöstlein, Picasso, and Williams}]{Bailey1968}
\bibinfo{author}{J.~Bailey}, \bibinfo{author}{W.~Bartl},
  \bibinfo{author}{G.~von Bochmann}, \bibinfo{author}{R.~C.~A. Brown},
  \bibinfo{author}{F.~J.~M. Farley}, \bibinfo{author}{H.~Jöstlein},
  \bibinfo{author}{E.~Picasso}, \bibinfo{author}{R.~W. Williams},
\newblock \bibinfo{title}{{Precision measurement of the anomalous magnetic
  moment of the muon}},
\newblock \bibinfo{journal}{Phys. Lett. B} \bibinfo{volume}{28}
  (\bibinfo{year}{1968}) \bibinfo{pages}{287--290}.
  \DOIprefix\doi{10.1016/0370-2693(68)90261-X}.
%Type = Article
\bibitem[{Bailey et~al.(1979)Bailey, Borer, Combley, Drumm, Eck, Farley, Field,
  Flegel, Hattersley, Krienen, Lange, Lebée, McMillan, Petrucci, Picasso,
  Runólfsson, von Rüden, Williams, and Wojcicki}]{Bailey1979}
\bibinfo{author}{J.~Bailey}, \bibinfo{author}{K.~Borer},
  \bibinfo{author}{F.~Combley}, \bibinfo{author}{H.~Drumm},
  \bibinfo{author}{C.~Eck}, \bibinfo{author}{F.~J.~M. Farley},
  \bibinfo{author}{J.~H. Field}, \bibinfo{author}{W.~Flegel},
  \bibinfo{author}{P.~M. Hattersley}, \bibinfo{author}{F.~Krienen},
  \bibinfo{author}{F.~Lange}, \bibinfo{author}{G.~Lebée},
  \bibinfo{author}{E.~McMillan}, \bibinfo{author}{G.~Petrucci},
  \bibinfo{author}{E.~Picasso}, \bibinfo{author}{O.~Runólfsson},
  \bibinfo{author}{W.~von Rüden}, \bibinfo{author}{R.~W. Williams},
  \bibinfo{author}{S.~G. Wojcicki}
  (\bibinfo{collaboration}{CERN-Mainz-Daresbury}),
\newblock \bibinfo{title}{{Final report on the CERN muon storage ring including
  the anomalous magnetic moment and the electric dipole moment of the muon, and
  a direct test of relativistic time dilation}},
\newblock \bibinfo{journal}{Nucl. Phys. B} \bibinfo{volume}{150}
  (\bibinfo{year}{1979}) \bibinfo{pages}{1--75}.
  \DOIprefix\doi{10.1016/0550-3213(79)90292-X}.
%Type = Article
\bibitem[{Bennett et~al.(2006)}]{Bennett2006}
\bibinfo{author}{G.~W. Bennett}, et~al. (\bibinfo{collaboration}{Muon g-2
  Collaboration}),
\newblock \bibinfo{title}{Final report of the muon {E821} anomalous magnetic
  moment measurement at {BNL}},
\newblock \bibinfo{journal}{Physical Review D} \bibinfo{volume}{73}
  (\bibinfo{year}{2006}) \bibinfo{pages}{072003}.
  \DOIprefix\doi{10.1103/PhysRevD.73.072003}.
%Type = Article
\bibitem[{Aguillard et~al.(2025)}]{Muong-2:2025xyk}
\bibinfo{author}{D.~P. Aguillard}, et~al. (\bibinfo{collaboration}{Muon g-2}),
\newblock \bibinfo{title}{{Measurement of the positive muon anomalous magnetic
  moment to 127~ppb}},
\newblock \bibinfo{journal}{Phys. Rev. Lett.} \bibinfo{volume}{135}
  (\bibinfo{year}{2025}) \bibinfo{pages}{101802}.
  \DOIprefix\doi{10.1103/7clf-sm2v}.
%Type = Article
\bibitem[{Charpak et~al.(1962)Charpak, Farley, Garwin, Muller, Sens, and
  Zichichi}]{Charpak1962}
\bibinfo{author}{G.~Charpak}, \bibinfo{author}{F.~J.~M. Farley},
  \bibinfo{author}{R.~L. Garwin}, \bibinfo{author}{T.~Muller},
  \bibinfo{author}{J.~C. Sens}, \bibinfo{author}{A.~Zichichi},
\newblock \bibinfo{title}{A new measurement of the anomalous magnetic moment of
  the muon},
\newblock \bibinfo{journal}{Phys. Lett.} \bibinfo{volume}{1}
  (\bibinfo{year}{1962}) \bibinfo{pages}{16--20}.
  \DOIprefix\doi{10.1016/0031-9163(62)90263-9}.

\end{thebibliography}
\end{document}